\documentclass[pre,showpacs,10pt,groupedaddress]{revtex4}

\usepackage{epsfig}
\usepackage{graphicx}
\usepackage{amsmath}
\usepackage{amssymb}
\usepackage{amsfonts}

\begin{document}

\title{Binary collisions of charged particles in a magnetic field}
\author{H.~B.~Nersisyan}
\altaffiliation{Permanent address: Institute of Radiophysics and Electronics, 378410
Ashtarak, Armenia}
\email{hrachya@irphe.am}
\author{G.~Zwicknagel}
\email{guenter.zwicknagel@physik.uni-erlangen.de}
\affiliation{Institut f\"{u}r Theoretische Physik II, Erlangen-N\"{u}rnberg Universit\"{a}%
t, Staudtstr. 7, D-91058 Erlangen, Germany}
\date{\today}

\begin{abstract}
Binary collisions between charged particles in an external
magnetic field are considered in second-order perturbation theory,
starting from the unperturbed helical motion of the particles. The
calculations are done with the help of an improved binary
collisions treatment which is valid for any strength of the
magnetic field, where the second-order energy and velocity
transfers are represented in Fourier space for arbitrary
interaction potentials. The energy transfer is explicitly
calculated for a regularized and screened potential which is both
of finite range and non-singular at the origin, and which involves
as limiting cases the Debye (i.e., screened) and Coulomb
potential. Two distinct cases are considered in detail. (i) The
collision of two identical (e.g., electron-electron) particles;
(ii) and the collision between a magnetized electron and an
uniformly moving heavy ion. The energy transfer involves all
harmonics of the electron cyclotron motion. The validity of the
perturbation treatment is evaluated by comparing with classical
trajectory Monte--Carlo calculations which also allows to
investigate the strong collisions with large energy and velocity
transfer at low velocities. For large initial velocities on the
other hand, only small velocity transfers occur. There the
non-perturbative numerical classical trajectory Monte--Carlo
results agree excellently with the predictions of the perturbative
treatment.
\end{abstract}

\maketitle

\section{Introduction}
\label{sec:intr}

In the presence of an external magnetic field $\mathbf{B}$ the problem of
two charged particles cannot be solved in a closed form as the relative
motion and the motion of the center of mass are coupled to each other.
Therefore no theory exists for a solution of this problem that is uniformly
valid for any strength of the magnetic field and the Coulomb force between
the particles. The energy loss of ion beams and the related processes in a
magnetized plasmas which are important in many areas of physics such as
transport, heating, magnetic confinement of thermonuclear plasmas and
astrophysics are examples of physical situations where this problem arises.
Recent applications are the cooling of heavy ion beams by electrons~\cite%
{der77,der78,sor83,pot90,mes94} and the energy transfer for heavy-ion
inertial confinement fusion (ICF) (see, e.g.,~\cite{pro05} for an overview).
The classical limit of a hydrogen or Rydberg atom in a strong magnetic field
also falls in this category (see, e.g.,~\cite{has89} and references therein)
but in contrast to the free-free transitions (scattering) the total energy
is negative here.

Numerical calculations have been performed for binary collisions (BC)
between magnetized electrons~\cite{sia76,gli92} and for collisions between
magnetized electrons and ions \cite{gzwi99,zwi00,zwi02,zwi06,ner07}. In
general the total energy $E$ of the particles interacting in a magnetic
field is conserved but the relative and center of mass energies are not
conserved separately. In addition, the presence of the magnetic field breaks
the rotational symmetry of the system and as a consequence only the
component of the angular momentum $\mathbf{L}$ parallel to the magnetic
field $L_{\parallel }$ is a constant of motion. So, the constants of motion $%
E$ and $L_{\parallel }$ reduce the phase space of the relative motion. A
different situation arises for the BC between an electron and uniformly
moving heavy ion. As an ion is much heavier than an electron, its uniform
motion is only weakly perturbed by collisions with the electrons and the
magnetic field. In this case $L_{\parallel }$ is not conserved but there
exists a conserved generalized energy $K$ \cite{zwi02,ner07} involving the
energy of relative motion and a magnetic term. The apparently simple problem
of charged particle interaction in a magnetic field is in fact a problem of
considerable complexity and the additional degree of freedom of the
cyclotron orbital motion produces a chaotic system with two degrees (or one
degree for heavy ions) of freedom~\cite{gut90,sch00,bhu02,ner07}.

In this paper we consider the BC between two charged particles treating the
interaction (Coulomb) as a perturbation to their helical motions. For
electron-heavy ion collisions this has been done previously in first order
in the ion charge $Z$ and for an ion at rest \cite{gel97} and in up to $%
\mathrm{O}(Z^{2})$ for uniformly moving heavy ion~\cite{toe02,ner03}. In
Ref.~\cite{toe02} three regimes are identified, depending on the relative
size of the parameters $a$ (the cyclotron radius), $s$ (the distance of the
closest approach), and $\delta $ (the pitch of the helix). In earlier
kinetic approaches \cite{der77,der78,sor83,pot90,mes94} only two regimes
have been distinguished: Fast collisions for $s<a$, where the Coulomb
interaction is dominant and adiabatic collisions for $s>a$, where the
magnetic field is important, as the electron performs many gyrations during
the collision with the ion. The change $\Delta E_{i}$ of the energy of the
ion has been related to the square of the momentum transfer $\Delta p$,
which has been calculated up to $\mathrm{O}(Z)$. This is somewhat
unsatisfactory, as there is another $\mathrm{O}(Z^{2})$ contribution to $%
\Delta E_{i}$, in which the second-order momentum transfer enters linearly.
Moreover, for applications in plasma physics (e.g., for calculation of the
ion energy loss in a magnetized plasma) one calculates the angular averaged
energy transfer which vanishes within first-order perturbation theory due to
symmetry reasons and the ion energy change receives contribution only from
higher orders~\cite{ner03}. Indeed, the transport phenomena, etc., are of
order $\mathrm{O}(Z^{2})$ in the ion charge.

Here we focus on BC between two magnetized identical particles
(e.g. electrons) within the second order perturbation theory and
its comparison with classical trajectory Monte--Carlo (CTMC)
simulations. The present work considerably extends our earlier
studies in Refs.~\cite{ner07,ner03} where the second-order energy
transfer for an ion--electron collision was calculated with the
help of an improved BC treatment which is valid for any strength
of the magnetic field and does not require the specification of
the interaction potential. In addition, we consider here the
impact parameter-averaged energy transfer for the BC between
magnetized electron and heavy ion moving uniformly along the
magnetic field which has not been yet considered in
Refs.~\cite{ner07,ner03}, and will give new analytical expressions
which are more appropriate for an explicit calculation of the
energy loss. Physically these two distinct cases are similar
except the time-dependent center of mass cyclotron motion in the
case of two identical particles. The paper is organized as
follows. In Sec.~\ref{sec:s1} starting from the exact equation of
motion of two charged particles moving in a magnetic field we
discuss some basic results of the exact BC treatment for the
energy and velocity transfers as well as the energy conservation.
In the following Sec.~\ref{sec:s2}, we discuss the velocity and
energy transfer during BC of magnetized particles for arbitrary
magnetic fields and strengths of the two-particle interaction
potential. The equations of motion are solved in a perturbative
manner up to the second order in interaction force starting from
the unperturbed helical motion of the particles in a magnetic
field. Then in Sec.~\ref{sec:s3} we turn to the explicit
calculation of the second order energy transfer for
electron-electron collision but without any restriction on the
magnetic field $\mathbf{B}$. The obtained energy transfer involves
all cyclotron harmonics of the electron helical motion. For
further applications (e.g., in cooling of ion beams, transport
phenomena in magnetized plasmas) we consider the regularized and
screened interaction potential which is both of finite range and
less singular than the Coulomb interaction at the origin and as
the limiting cases involves the Debye (i.e., screened) and Coulomb
potentials. An exact solution for two particle collision in the
presence of infinitely strong magnetic field is considered in
Sec.~\ref{sec:imp}. This also suggests an improved perturbative
treatment for repulsive interaction and in the case of strong
magnetic field. In Sec.~\ref{sec:s4} the results of the
perturbative binary collision model are compared with CTMC
simulations in which the scattering of ensembles of magnetized
electrons are treated exactly. This comparison allows to determine
clearly the range of validity of the perturbative treatment. The
results are summarized and discussed in Sec.~\ref{sec:disc}. Some
details of the calculation of the generalized Coulomb logarithm is
described in the Appendix~\ref{sec:ap1}. The small
velocity limits of the energy transfers are derived in the Appendix~\ref%
{sec:ap2}.

\section{Binary collision formulation. General treatment}
\label{sec:s1}

For our description of binary collision (BC) we start with considering the
equations of motion for two charged particles moving in a homogeneous
magnetic field and the related conservation laws, in general. This will then
be specified to the two particular case on with we will focus in this paper.
The BC between two electrons and between an electron and an uniformly moving
heavy ion. Next the quantities of interest, the velocity transfer and the
energy transfer of particles during the binary collision, will be introduced
and discussed, before we turn to the solution of the equations of motion in
the subsequent section.

\subsection{Relative and cm motion and conservation laws}

\label{sec:s1.1}

We consider two point charges with masses $m_{1}$, $m_{2}$ and charges $%
q_{1}e $, $q_{2}e$, respectively, moving in a homogeneous magnetic field $%
\mathbf{B}=B\mathbf{b}$ and interacting with the potential $q_{1}q_{2}
e\!\!\!/^{2} U(\mathbf{r})$ with $e\!\!\!/^{2}=e^{2}/4\pi \varepsilon _{0}$.
Here $\varepsilon _{0}$ is the permittivity of the vacuum and $\mathbf{r}=%
\mathbf{r}_{1}-\mathbf{r}_{2}$ is the relative coordinate of the colliding
particles. For two isolated charged particles this interaction is given by
the Coulomb potential, i.e.~$U_{\mathrm{C}}(\mathbf{r})=1/r$. In plasma
applications the infinite range of this potential is modified by the
screening. Then $U$ may be modeled by $U_{\mathrm{D}}(\mathbf{r}%
)=e^{-r/\lambda }/r$ with a screening length $\lambda $ which can be chosen
as the Debye screening length $\lambda _{\mathrm{D}}$, see, for example \cite%
{akh75}. The quantum uncertainty principle prevents particles (for $q_1 q_2
< 0$) from falling into the center of these potentials. In a classical
picture this can be achieved by regularization of $U(\mathbf{r})$ at the
origin, taking for example $U_{\mathrm{R}}(\mathbf{r})=\left(1-e^{-r/%
\lambdabar }\right) e^{-r/\lambda }/r$, \cite{kel63,deu77}. Here $\lambdabar
$ is a parameter, which is usually related to the de Broglie wavelength.

In the presence of an external magnetic field, the Lagrangian and the
corresponding equations of motion cannot be separated into parts describing
the relative motion $[\mathbf{r}=\mathbf{r}_{1}-\mathbf{r}_{2},\mathbf{v}=%
\dot{\mathbf{r}}]$ and the motion of the center of mass (cm) $[\mathbf{R}%
=(m_{1}\mathbf{r}_{1}+m_{2}\mathbf{r}_{2})/(m_{1}+m_{2}),\mathbf{V}=\dot{%
\mathbf{R}}]$, in general (see, e.g., \cite{sia76,zwi02,toe02,ner03}).
Introducing the reduced mass $1/\mu =1/m_{1}+1/m_{2}$ the equations of
motion are
\begin{eqnarray}
{\dot{\mathbf{v}}}(t) -\mu\, e B \left( \frac{q_{1}}{m_{1}^2}+\frac{q_{2}}{%
m_{2}^2}\right)\, \left[ \mathbf{v}(t)\times \mathbf{b}\right] & = & e B
\left( \frac{q_{1}}{m_{1}}-\frac{q_{2}}{m_{2}}\right)\, \left[ \mathbf{V}%
(t)\times \mathbf{b}\right] \, + \, \frac{q_{1}q_{2}e\!\!\!/^{2}}{\mu }%
\mathbf{F}\left( \mathbf{r}(t)\right) ,  \label{eq:a1} \\
{\dot{\mathbf{V}}}(t) - e B \left(\frac{q_{1}+ q_{2}}{m_{1}+m_{2}}\right)\, %
\left[ \mathbf{V}(t)\times \mathbf{b}\right] & = & \frac{\mu\, e B}{%
m_{1}+m_{2}}\left( \frac{q_{1}}{m_{1}}-\frac{q_{2}}{m_{2}}\right)\, \left[
\mathbf{v}(t)\times \mathbf{b}\right] ,  \label{eq:a2}
\end{eqnarray}%
where $q_{1}q_{2}e\!\!\!/^{2}\mathbf{F}\left( \mathbf{r}(t)\right) $ $(%
\mathbf{F}=-\partial U/\partial \mathbf{r})$ is the force exerted by the
particle 2 on the particle 1. The coupled, nonlinear differential equations~(%
\ref{eq:a1}) and (\ref{eq:a2}) completely describe the motion of the
particles. For solving the scattering problem, they have to be integrated
numerically for a complete set of the initial conditions.

From Eqs.~(\ref{eq:a1}) and (\ref{eq:a2}) follows the conservation of the
parallel component of the cm velocity ${\mathbf{V}}(t)\cdot \mathbf{b}%
=V_{0\parallel }$ and the total energy
\begin{equation}
E = E_{\mathrm{cm}} + E_{\mathrm{r}} = \frac{(m_{1}+m_{2})V^{2}(t)}{2} +
\frac{\mu v^{2}(t)}{2}+q_{1}q_{2}e\!\!\!/^{2}U(\mathbf{r})=\mathrm{const},
\label{eq:a12e}
\end{equation}%
but since, in general, the relative and center of mass motions are coupled
the relative $E_{\mathrm{r}}$ and cm $E_{\mathrm{cm}}$ energies are not
conserved separately.

In the case of two identical particles, taking here electrons, i.e.~$%
m_{1}=m_{2}=m$, $q_{1}=q_{2}=-1$, the equations of motion considerably
simplify to
\begin{eqnarray}
{\dot{\mathbf{v}}}(t) + \omega_{c}\, \left[ \mathbf{v}(t)\times \mathbf{b}%
\right] & = & \frac{2 e\!\!\!/^{2}}{m} \, \mathbf{F}\left( \mathbf{r}%
(t)\right) ,  \label{eq:a5} \\
{\dot{\mathbf{V}}}(t) + \omega_{c} \, \left[ \mathbf{V}(t)\times \mathbf{b}%
\right] & = & 0 \, ,  \label{eq:a6}
\end{eqnarray}%
with the cyclotron frequency of the electron $\omega_{c} = eB/m$. Here the
cm-motion, Eq.~(\ref{eq:a6}), can be solved which leads to
\begin{equation}
\mathbf{V}\left( t\right) \equiv \mathbf{V}_{0}\left( t\right)
=V_{0\parallel }\mathbf{b}+V_{0\perp }\left[ \mathbf{u}_{c}\cos \left(
\omega _{c}t\right) + \left[ \mathbf{b}\times \mathbf{u}_{c}\right] \sin
\left( \omega _{c}t\right) \right] ,  \label{eq:a6-1}
\end{equation}%
where $V_{0\perp }\geqslant 0$ is the cm velocity transverse to the magnetic
field direction $\mathbf{b}$, and $\mathbf{u}_{c}=\left( \cos \varphi
_{c},\sin \varphi _{c}\right) $ is the unit vector perpendicular to $\mathbf{%
b}$ ($\varphi _{c}$ is the phase of the cm transversal motion which is fixed
by initial conditions). The velocities $V_{0\parallel }$ and $V_{0\perp }$
are related to the particles unperturbed parallel $v_{0\nu \parallel }$ and
transverse $v_{0\nu \bot }$ velocities ($\nu =1,2$)
\begin{equation}
V_{0\parallel }=\frac{v_{01\parallel }+v_{02\parallel }}{2},\qquad V_{0\perp
}\mathbf{u}_{c}=\frac{v_{01\perp }\mathbf{u}_{1}+v_{02\perp }\mathbf{u}_{2}}{%
2}  \label{eq:a6-2}
\end{equation}%
with the unit vectors $\mathbf{u}_{\nu }=\left( \cos \varphi _{\nu },\sin
\varphi _{\nu }\right) $ ($\varphi _{\nu }$ are the initial phases of the
particles) which fix the initial transverse velocities (or coordinates) of
the particles 1 and 2.

With the help of the equation of motion~(\ref{eq:a5}) and relation (\ref%
{eq:a6-1}) it can be easily proven that the relative ($E_{\mathrm{r}}$) and
cm ($E_{\mathrm{cm}}$) energies
\begin{equation}
E_{\mathrm{cm}}=m\mathbf{V}_{0}^{2}( t) =m( V_{0\parallel }^{2}+V_{0\perp
}^{2}) ,\qquad E_{\mathrm{r}}=\frac{mv^{2}}{4}+ e\!\!\!/^{2}U(\mathbf{r})
\label{eq:a6-3}
\end{equation}
are here conserved separately. They can be expressed by the particles
initial (unperturbed) velocities $\mathbf{v}_{0\nu }$ according to $E_{%
\mathrm{cm}}=\frac{m}{4}\left( \mathbf{v}_{01}+\mathbf{v}_{02}\right) ^{2}$,
$E_{\mathrm{r}}=\frac{m}{4}\left( \mathbf{v}_{01}-\mathbf{v}_{02}\right)^{2}$%
. Note that $E_{\mathrm{cm}}$ and $E_{\mathrm{r}}$ are here functions only
of the difference of the initial phases of the particles $\varphi =\varphi
_{1}-\varphi _{2}$.

In the second case of BC between an electron and an heavy ion of mass $M$
and charge $Ze$, i.e.~$q_{1}=-1, q_{2}=Z$, $m_1 = m, m_{2}=M$, and assuming $%
m/M \to 0, \mu \to m$, the equations of motion (\ref{eq:a1}) and (\ref{eq:a2}%
) turn into
\begin{eqnarray}
{\dot{\mathbf{v}}}(t)+\omega _{c}\left[ \mathbf{v}(t)\times \mathbf{b}\right]
& = & -\omega _{c}\left[ \mathbf{v}_{i}\times \mathbf{b}\right] - \frac{%
Ze\!\!\!/^{2}}{m}\mathbf{F}\left( \mathbf{r}(t)\right) ,  \label{eq:a7} \\
{\dot{\mathbf{V}}}(t) & = & 0 \, ,  \label{eq:a8}
\end{eqnarray}
where $\mathbf{v}_{i}$ is the given heavy ion velocity and $\mathbf{V}(t) =
\mathbf{v}_{i} = \mathrm{const}$. While the relative energy $E_{\mathrm{r}}$
itself is not conserved in this case, there exists nevertheless a constant
of motion
\begin{equation}
K = \frac{mv^{2}}{2}-Ze\!\!\!/^{2}U(\mathbf{r}) + m\omega _{c}\mathbf{r}%
\cdot \left[ \mathbf{v}_{i}\times \mathbf{b}\right] = E_{\mathrm{r}} +
m\omega _{c}\mathbf{r}\cdot \left[ \mathbf{v}_{i}\times \mathbf{b}\right]
\label{eq:a7b}
\end{equation}%
which can be easily checked with the help of Eq.~(\ref{eq:a7}). In contrast
to the unmagnetized case, the relative energy transfer during an
ion--electron collision is thus proportional to $\delta r_{\bot }v_{i\bot }$%
, where $\delta r_{\bot }$ and $v_{i\bot }$ are the perpendicular components
of the change of relative position and the ion velocity. Only for ions which
move along the magnetic field direction, i.e.~$(\mathbf{v}_{i\bot }=0)$,
where the magnetic term in Eq.~(\ref{eq:a7b}) vanishes, the relative energy
is conserved. This case is very similar to the case of electron-electron
collisions except that the cm velocity is constant.

\subsection{Energy loss and velocity transfer}

\label{sec:s1.2}

In the general case the rate $dE_\nu/dt$ at which the energy $E_{\nu} =
m_{\nu} v_{\nu}^2/2$ of particle $\nu$ changes during the collision with the
other particle can be obtained by multiplying the equation of motion for
particle $\nu$ by its velocity $\mathbf{v}_{\nu }(t)=\mathbf{V}(t)+\varrho
_{\nu }(\mu /m_{\nu })\mathbf{v}(t)$, where $\varrho _{1}=1$, $\varrho
_{2}=-1$. As $q_{1}q_{2}e\!\!\!/^{2}\mathbf{F}\left( \mathbf{r}\right) $ is
the force exerted by the particle 2 on the particle 1 the integration of
this rate over the whole collision yields the energy transfer (see, \cite%
{ner07} for further details)
\begin{equation}
\Delta E_{\nu } = \varrho _{\nu }q_{1}q_{2}e\!\!\!/^{2}\int_{-\infty
}^{\infty }\mathbf{V}(\tau )\cdot \mathbf{F}\left( \mathbf{r}(\tau )\right)
d\tau .  \label{eq:a10}
\end{equation}%
assuming that for $t\rightarrow \pm \infty$, $r(t)\rightarrow \infty $ and $%
U(\mathbf{r}(t))\rightarrow 0$. According to the conservation of total
energy we have $\Delta E_{1}=-\Delta E_{2}$, as it can be directly seen from
Eq.~(\ref{eq:a10}).

Alternatively the energy changes (\ref{eq:a10}) can be expressed by the
velocity transferred to particle 1 (or 2) during the collision. For this
purpose we split the velocity of the $\nu $th particle, $\mathbf{v}_{\nu
}(t) $, into $\mathbf{v}_{\nu }(t)=\mathbf{v}_{0\nu }(t)+\delta \mathbf{v}%
_{\nu }(t)$. Here $\mathbf{v}_{0\nu }(t)$ describes the free helical motion
of the particles in the magnetic field, $\dot{\mathbf{v}}_{0\nu } - (q_\nu e
B/m_\nu) \left[ \mathbf{v}_{0\nu }\times \mathbf{b}\right] =0$, while $%
\delta \mathbf{v}_{\nu }(t)$ is the related velocity change (with $\delta
\mathbf{v}_{\nu }(t\to -\infty)\to 0$). From the energy change with respect
to the free motion $\delta E_{\nu }(t) = (m_{\nu} /2) \{\left[ \mathbf{v}%
_{0\nu }(t)+\delta \mathbf{v}_{\nu }(t)\right] ^{2}-\mathbf{v}^{2}_{0\nu
}(t) \}$ the total energy transfer $\Delta E_{\nu }$ can then be expressed
through the total velocity transfer $\Delta \mathbf{v}_{\nu }=\delta \mathbf{%
v}_{\nu}(t\to\infty ) = [\mathbf{v}_{\nu}(t)-\mathbf{v}_{0\nu
}(t)]_{t\to\infty}$ and $\mathbf{v}_{0\nu}\cdot \Delta\mathbf{v}_{\nu} =%
\mathbf{v}_{0\nu}(t)\cdot \delta\mathbf{v}_{\nu} (t)\vert_{t\to\infty}$
\begin{equation}
\Delta E_{1}=-\Delta E_{2}=m_{1}\left( \mathbf{v}_{01} \cdot \Delta \mathbf{v%
}_{1}+\frac{1}{2}\Delta \mathbf{v}_{1}^{2}\right) .  \label{eq:a21}
\end{equation}%
Employing this way of calculating the energy transfer, as e.g. in \cite%
{toe02}, the potential $U(\mathbf{r})$ has to be specified already at an
early stage of the calculation. In Sec.~\ref{sec:s2} we will show that Eq.~(%
\ref{eq:a10}) allows for a more general formulation in which the cut--off at
large distances and the regularization at small distances can be treated
much easier.

So far we considered the energy transfers of the particles in the laboratory
frame. In addition the energy transfers $\Delta E_{1}$ and $\Delta E_{2}$
can be expressed by the change of the relative energy $\Delta E_{\mathrm{r}}
= \Delta (\mu \mathbf{v}^2/2)$ and the relative and cm velocity transfer $%
\Delta[\mathbf{V}\cdot\mathbf{v}]$, respectively. Since $\Delta E_{\mathrm{r}%
}=-\Delta E_{\mathrm{cm}}$ (due to the conservation of total energy), there
follows a relation between the energy transfers $\Delta E_{1}$ and $\Delta
E_{\mathrm{r}}$ given by
\begin{equation}
\Delta E_{1}=\frac{m_{2}-m_{1}}{m_{2}+m_{1}}\Delta E_{\mathrm{r}} + \mu
\Delta \left[\mathbf{V}\cdot \mathbf{v}\right] .  \label{eq:a21a}
\end{equation}%
Here, the second term is the total change of the scalar product $\mathbf{V}%
(t)\cdot \mathbf{v}(t)$. For instance, this quantity is a constant for the
free, unperturbed motion of identical particles where $\mathbf{V}%
_{0}(t)\cdot \mathbf{v}_{0}(t)= (v_{01}^{2}-v_{02}^{2})/2 =\mathrm{const}$
and is given by the initial velocities $\mathbf{v}_{01}, \mathbf{v}_{02}$.

In the case of identical particles the relative energy and the cm energy are
conserved, i.e.~$\Delta E_{r}=-\Delta E_{\mathrm{cm}}=0$, and thus $\Delta
E_{1} = - \Delta E_{2} = \mu \Delta \left[\mathbf{V}\cdot \mathbf{v}\right]$%
. We note however that the longitudinal ($\Delta E_{r\parallel}$) and
transverse ($\Delta E_{r\bot } = - \Delta E_{r\parallel}$) parts of the
energy transfer $\Delta E_{r}$ do not vanish, where
\begin{equation}
\Delta E_{r\parallel }=\frac{m}{4}\Delta v_{\parallel }\left( 2v_{r\parallel
}+\Delta v_{\parallel }\right) \neq 0.  \label{eq:a21c}
\end{equation}%
Here $v_{r\parallel }=v_{01\parallel }-v_{02\parallel }$ and $\Delta
v_{\parallel }$ are the components of $\mathbf{v}_{0}(t)$ and $\Delta
\mathbf{v}$ parallel to $\mathbf{b}$, respectively.

For the collision of an electron and a heavy ion, with $m_2 = M \gg m=m_1$
and $\mathbf{V} = \mathbf{v}_{i}$, the energy transfer to the ion $\Delta
E_{i} = \Delta E_{2} = - \Delta E_{1}$ then follows from Eq.~(\ref{eq:a21a})
as
\begin{equation}
\Delta E_{i}= -\Delta E_{r} - m\mathbf{v}_{i}\cdot \Delta \mathbf{v} = -
\frac{m}{2} \Delta \left(\mathbf{v}^2 \right) - m\mathbf{v}_{i}\cdot \Delta
\mathbf{v} .  \label{eq:a21d}
\end{equation}%
When the ion moves parallel to the magnetic field direction the relative
energy transfer in Eq.~(\ref{eq:a21d}) vanishes and $\Delta E_{i\parallel
}=-mv_{i\parallel }\Delta v_{\parallel }$, where $v_{i\parallel }$ is the
component of the ion velocity parallel to $\mathbf{b}$.

\section{Perturbative treatment. General theory}
\label{sec:s2}

\subsection{Trajectory correction}
\label{sec:s2.1}

In this section we consider the theoretical treatment of the scattering of
two identical particle, here electrons, where we seek an approximate
solution of Eq.~(\ref{eq:a5}) by assuming the interaction force between the
particles as a perturbation to the free helical motion. For the case of
ion--electron scattering the corresponding considerations and derivations
are discussed in detail in Ref.~\cite{ner07} and we thus focus on the
electron-electron case in the forthcoming discussion.

As the velocity of the cm motion is already fixed by Eq.~(\ref{eq:a6-1}), we
have to look for the solution of Eq.~(\ref{eq:a5}) for the variables $%
\mathbf{r}$ and $\mathbf{v}$ in a perturbative manner
\begin{equation}
\mathbf{r}(t)=\mathbf{r}_{0}(t)+\mathbf{r}^{(1)}(t)+\mathbf{r}%
^{(2)}(t)...,\qquad \mathbf{v}(t)=\mathbf{v}_{0}(t)+\mathbf{v}^{(1)}(t)+%
\mathbf{v}^{(2)}(t)...,  \label{eq:a22}
\end{equation}%
where $\mathbf{r}_{0}(t),\mathbf{v}_{0}(t)$ are the unperturbed
two-particles relative coordinate and velocity, respectively, $\mathbf{r}%
^{(n)}(t),\mathbf{v}^{(n)}(t)\propto q^{2n}\mathbf{F}_{n-1}$ ($n=1,2,...$)
are the $n$th order perturbations of $\mathbf{r}(t)$ and $\mathbf{v}(t)$,
which are proportional to $q^{2n}$ (for electrons $q=-1$). $\mathbf{F}%
_{n}(t) $ is the $n$th order correction to the two-particle interaction
force. Using the expansion (\ref{eq:a22}) for the $n$th order corrections $%
\mathbf{F}_{n}$ we obtain
\begin{equation}
\mathbf{F}\left( \mathbf{r}(t)\right) =\mathbf{F}_{0}\left( \mathbf{r}%
_{0}(t)\right) +\mathbf{F}_{1}\left( \mathbf{r}_{0}(t),\mathbf{r}%
_{1}(t)\right) +...,  \label{eq:a23}
\end{equation}%
where
\begin{equation}
\mathbf{F}_{0}\left( \mathbf{r}_{0}(t)\right) =\mathbf{F}\left( \mathbf{r}%
_{0}(t)\right) =-i\int d\mathbf{k}U(\mathbf{k})\mathbf{k}e^{i\mathbf{k}\cdot
\mathbf{r}_{0}(t)},  \label{eq:a24}
\end{equation}%
\begin{equation}
\mathbf{F}_{1}\left( \mathbf{r}_{0}(t),\mathbf{r}_{1}(t)\right) =\left.
\left( \mathbf{r}_{1}(t)\cdot \frac{\partial }{\partial \mathbf{r}}\right)
\mathbf{F}(\mathbf{r})\right\vert _{\mathbf{r}=\mathbf{r}_{0}(t)}=\int d%
\mathbf{k}U(\mathbf{k})\mathbf{k}\left[ \mathbf{k}\cdot \mathbf{r}_{1}(t)%
\right] e^{i\mathbf{k}\cdot \mathbf{r}_{0}(t)}.  \label{eq:a25}
\end{equation}%
In Eqs.~(\ref{eq:a24}) and (\ref{eq:a25}), we have introduced the
two-particle interaction potential $U(\mathbf{r})$ through $\mathbf{F}(%
\mathbf{r})=-\partial U(\mathbf{r})/\partial \mathbf{r}$ and the force
corrections have been written using a Fourier transformation in space.

We start with the zero-order unperturbed helical motion of two electrons in
the relative frame
\begin{eqnarray}
&&\mathbf{r}_{0}(t) =\mathbf{R}_{0}+\mathbf{b}v_{r\parallel }t+a\left[
\mathbf{u}_{r}\sin \left( \omega _{c}t\right) - \left[ \mathbf{b}\times
\mathbf{u}_{r}\right] \cos \left( \omega _{c}t\right) \right] ,
\label{eq:a27} \\
&&\mathbf{v}_{0}(t) =\dot{\mathbf{r}}_{0}(t)\ ,  \notag
\end{eqnarray}%
where $\mathbf{u}_{r}=\left( \cos \varphi _{r},\sin \varphi _{r}\right) $ ($%
\varphi _{r}$ is the initial phase of the relative motion) is the unit
vector perpendicular to the magnetic field, $v_{r\parallel }=v_{01\parallel
}-v_{02\parallel }$ and $v_{0\bot }\mathbf{u}_{r}=v_{01\perp }\mathbf{u}%
_{1}-v_{02\perp }\mathbf{u}_{2}$ (with $v_{0\bot }\geqslant 0$) are the
unperturbed relative velocity components parallel and perpendicular to $%
\mathbf{b}$, respectively. Here $v_{r\parallel }\mathbf{b}$ is the relative
velocity of the guiding centers of two particles, and $a=v_{0\bot }/\omega
_{c}$ is the relative cyclotron radius. It should be noted that in Eq.~(\ref%
{eq:a27}) the variables $\mathbf{u}_{r}$ and $\mathbf{R}_{0}$ are
independent and are defined by the initial conditions. Explicitly the
relative cyclotron radius $a$ and the phase $\varphi _{r}$ are connected to
the particles cyclotron radii $a_{1}$, $a_{2}$ and phases $\varphi _{1}$, $%
\varphi _{2}$, and $\varphi = \varphi _{1} - \varphi _{2}$, according to
\begin{equation}
e^{i\varphi _{r}}=\frac{a_{1}e^{i\varphi _{1}}-a_{2}e^{i\varphi _{2}}}{a}%
,\quad a^{2}=a_{1}^{2}+a_{2}^{2}-2a_{1}a_{2}\cos \varphi .  \label{eq:a27-1}
\end{equation}

The equation for the first-order relative velocity correction is given by
\begin{equation}
{\dot{\mathbf{v}}}^{(1)}(t)+ \omega _{c}[{\mathbf{v}}^{(1)}(t)\times \mathbf{%
b}]=\frac{2q^{2}e\!\!\!/^{2}}{m}\mathbf{F}\left( \mathbf{r}_{0}(t)\right)
\label{eq:a28}
\end{equation}%
with the solution
\begin{eqnarray}
&&\mathbf{r}^{(1)}(t) = \frac{2q^{2}e\!\!\!/^{2}}{m}\left\{ \mathbf{b}%
P_{\parallel }(t)+\mathrm{Re}\left[ \mathbf{P}_{\bot }(t)-\mathbf{b}\left(
\mathbf{b}\cdot \mathbf{P}_{\bot }(t)\right) -i \left[ \mathbf{b}\times
\mathbf{P}_{\bot }(t)\right] \right] \right\} .  \label{eq:a30} \\
&&\mathbf{v}^{(1)}(t) = \dot{\mathbf{r}}^{(1)}(t) .  \notag
\end{eqnarray}%
Here we have assumed that all corrections vanish at $t\rightarrow -\infty $
and have introduced the following abbreviations
\begin{eqnarray}
P_{\parallel }(t)=-i\int d\mathbf{k} U(\mathbf{k}) (\mathbf{k} \cdot\mathbf{b%
}) \int_{-\infty }^{t}d\tau \left( t-\tau \right) e^{i\mathbf{k}\cdot
\mathbf{r}_{0}(\tau)} ,  \label{eq:a32} \\
\mathbf{P}_{\bot}(t)=-\frac{1}{\omega _{c}}\int d\mathbf{k} U(\mathbf{k})
\mathbf{k} \int_{-\infty }^{t}d\tau [ e^{i\omega _{c}(t-\tau )}-1] e^{i%
\mathbf{k}\cdot \mathbf{r}_{0}(\tau)} ,  \notag
\end{eqnarray}%
wherein $\mathbf{r}_{0}(t)$ is given by Eq.~(\ref{eq:a27}) and the Fourier
representation of the interaction force, Eq.~(\ref{eq:a24}), has been used.

\subsection{First and second order energy transfers}

\label{sec:s2.2}

The total energy change $\Delta E_{1}$ of the first particle during
collision with the particle 2 is given by Eq.~(\ref{eq:a10}). Insertion of
Eq.~(\ref{eq:a23}) into the general expression (\ref{eq:a10}) yields
\begin{equation}
\Delta E_{1}=\Delta E_{1}^{(1)}+\Delta E_{1}^{(2)}+...,  \label{eq:a36}
\end{equation}%
where
\begin{equation}
\Delta E_{1}^{(1)}=q^{2}e\!\!\!/^{2}\int_{-\infty }^{\infty }dt\mathbf{V}%
_{0}\left( t\right) \cdot \mathbf{F}\left( \mathbf{r}_{0}(t)\right) ,\quad
\Delta E_{1}^{(2)}=q^{2}e\!\!\!/^{2}\int_{-\infty }^{\infty }dt\mathbf{V}%
_{0}\left( t\right) \cdot \mathbf{F}_{1}\left( \mathbf{r}_{0}(t),\mathbf{r}%
_{1}(t)\right)  \label{eq:a37}
\end{equation}%
are the first- and second order energy transfer, respectively.

We now introduce the variable $\mathbf{s}=\mathbf{R}_{0\bot }$ which is the
component of $\mathbf{R}_{0}$ perpendicular to the relative velocity vector $%
v_{r\parallel }\mathbf{b}$. From Eq.~(\ref{eq:a27}) we can see that $\mathbf{%
s}$ is the distance of closest approach for the guiding centers of the two
particles' helical motion. For practical applications the energy change is
now given by the average of $\Delta E_{1}$ with respect to the initial
phases of the particles $\varphi _{1}$, $\varphi _{2}$ and the azimuthal
angle $\vartheta _{\mathbf{s}}$ of $\mathbf{s}$. Such an averaged quantity $%
f $ will be abbreviated by $\langle f\rangle $ in the forthcoming
considerations.

We start with calculating the first order longitudinal relative velocity
transfer, $\Delta v_{\parallel }^{(1)}$, which also contributes to the
second order relative energy transfer according to Eq.~(\ref{eq:a21c}). This
quantity can be easily extracted from the parallel component of the
first-order relative velocity correction, $\mathbf{b} \cdot \mathbf{v}%
^{(1)}(t)$, in the limit $t\rightarrow \infty$, i.e. after completion of the
interaction. Thus, using the Fourier series of the exponential function $%
e^{iz\sin\theta}$ \cite{gra80}, we obtain from Eqs.~(\ref{eq:a30}) and (\ref%
{eq:a32})
\begin{equation}
\Delta v_{\parallel }^{(1)}=-\frac{4\pi iq^{2}e\!\!\!/^{2}}{m}\int d\mathbf{k%
} U(\mathbf{k})k_{\parallel } e^{i\mathbf{k}\cdot \mathbf{R}_{0}}
\sum_{n=-\infty}^{\infty} e^{in (\varphi_{r} -\theta)} J_{n}(k_{\bot} a)
\delta\left(\zeta_{n} (\mathbf{k}) \right) .  \label{eq:a38b}
\end{equation}
Here $J_{n}$ are the Bessel functions of the $n$th order, $\zeta_{n} (%
\mathbf{k})=n\omega_{c}+k_{\parallel} v_{r\parallel}$, $k_{\parallel } =%
\mathbf{k}\cdot \mathbf{b}$ and $k_{\bot}$ are the components of $\mathbf{k}$
parallel and transverse to $\mathbf{b}$, respectively, $\tan\theta
=k_{y}/k_{x}$. Note that the relative cyclotron radius $a$ as well as the
velocity of the cm motion $V_{0\bot }$ depend on $\varphi = \varphi_1 -
\varphi_2$ (see Eq.~(\ref{eq:a6-2}) and (\ref{eq:a27-1})). Performing the
averages with respect to the initial phases $\varphi _{1}$, $\varphi_{2}$
and the azimuthal angle $\vartheta _{\mathbf{s}}$ results in $\langle \Delta
v_{\parallel }^{(1)}\rangle = 0$ for spherically symmetric interaction
potentials ($U(\mathbf{r})=U(r)$ and $U(\mathbf{k})=U(k)$) (see, e.g., Refs.
\cite{ner07,ner03}) due to symmetry.

The first-order energy transfer is obtained by substituting Eqs. (\ref%
{eq:a6-1}), (\ref{eq:a24}) and (\ref{eq:a27}) into the first one of Eq.~(\ref%
{eq:a37}). Similarly as the first-order relative velocity correction $\Delta
v_{\parallel }^{(1)}$ (\ref{eq:a38b}) the first-order energy change also
vanishes if averaged with respect to the initial phases of the particles $%
\varphi _{1}$, $\varphi _{2}$ and $\vartheta _{\mathbf{s}}$.

Thus the energy change receives a contribution only from higher orders, and
we next evaluate the second order energy transfer $\Delta E_{1}^{(2)}$ by
inserting Eqs.~(\ref{eq:a6-1}), (\ref{eq:a25}), (\ref{eq:a27}), (\ref{eq:a30}%
) and (\ref{eq:a32}) into the second equation of Eq.~(\ref{eq:a37}). This
quantity is then averaged with respect to the initial phases of the
particles $\varphi _{1}$, $\varphi _{2}$ and the azimuthal angle $\vartheta
_{\mathbf{s}}$ of the impact parameter $\mathbf{s}$. The obtained angular
integrals are easily evaluated using the Fourier series of the exponential
function. After averaging the energy transfer $\Delta E_{1}^{(2)}$ with
respect to $\varphi _{1}$ and $\varphi _{2}$ the remaining part will depend
on $\delta \left( \left( \mathbf{k}+\mathbf{k}^{\prime }\right) \cdot
\mathbf{b}\right) $, i.e. the component of $\mathbf{k}+\mathbf{k}^{\prime }$
along the magnetic field $\mathbf{b}$. Thus this $\delta $-function enforces
$\mathbf{k}+\mathbf{k}^{\prime }$ to lie in the plane transverse to $\mathbf{%
b}$ so that $e^{i\left( \mathbf{k+k}^{\prime }\right) \cdot \mathbf{R}%
_{0}}\delta \left( \left( \mathbf{k}+\mathbf{k}^{\prime }\right) \cdot
\mathbf{b}\right) =e^{i\mathbf{Q}\cdot \mathbf{s}}\delta \left( \left(
\mathbf{k}+\mathbf{k}^{\prime }\right) \cdot \mathbf{b}\right) $, where $%
\mathbf{Q}=\mathbf{k}_{\bot }+\mathbf{k}_{\bot }^{\prime }$. In addition,
instead of phase $\varphi _{1}$ we introduce a new variable $\varphi$
according to $\varphi _{1} =\varphi +\varphi _{2}$. Then after performing $%
\varphi _{2}$ -integration there remains average with respect to the phase
difference $\varphi$. The result of the angular averaging finally reads
\begin{equation}
\langle \Delta E_{1}^{(2)}\rangle =\int_{0}^{2\pi }\frac{d\varphi }{2\pi }%
\mathcal{E}\left( \varphi ,s\right) ,  \label{eq:a42}
\end{equation}%
where
\begin{eqnarray}
\mathcal{E}\left( \varphi ,s\right) &=&-\frac{2\pi iq^{4}e\!\!\!/^{4}}{%
m\left\vert v_{r\parallel }\right\vert }\int d\mathbf{k}d\mathbf{k}^{\prime
}U(\mathbf{k})U(\mathbf{k}^{\prime })J_{0}\left( Qs\right) \delta
(k_{\parallel }^{\prime }+k_{\parallel })\sum_{n=-\infty }^{\infty }\left(
-1\right) ^{n}e^{in \left(\theta -\theta ^{\prime } \right) }
\label{eq:a42a} \\
&&\times J_{n}\left( k_{\bot }^{\prime }a\right) J_{n}\left( k_{\bot
}a\right) G_{n}\left( \mathbf{k},\mathbf{k}^{\prime }\right) \left(
V_{0\parallel }k_{\parallel }-n\omega _{c}\frac{V_{0\bot }}{v_{0\bot }}\cos
\vartheta \right) ,  \notag
\end{eqnarray}%
\begin{equation}
G_{n}\left( \mathbf{k},\mathbf{k}^{\prime }\right) =-\frac{2k_{\parallel}
k^{\prime}_{\parallel}}{\left( \zeta _{n}(\mathbf{k}^{\prime })-i0\right)
^{2}}-\frac{ k_{\bot} k^{\prime}_{\bot} e^{i(\theta ^{\prime}-\theta)} }{%
\left( \zeta _{n}(\mathbf{k}^{\prime })-i0\right) \left( \zeta _{n-1}(%
\mathbf{k}^{\prime })-i0\right) }-\frac{k_{\bot} k^{\prime}_{\bot}
e^{-i(\theta ^{\prime}-\theta)}}{\left( \zeta _{n}(\mathbf{k}^{\prime
})-i0\right) \left( \zeta _{n+1}(\mathbf{k}^{\prime })-i0\right) } .
\label{eq:a41}
\end{equation}
The phase factor $\vartheta =\varphi _{r}-\varphi_{c}$ in Eq.~(\ref{eq:a42a}%
) can be evaluated explicitly by using Eqs.~(\ref{eq:a6-2}) and (\ref%
{eq:a27-1}). Introducing the phase difference of the particles $\varphi
=\varphi _{1}-\varphi _{2}$ it reads
\begin{equation}
e^{i\left( \varphi _{r}-\varphi _{c}\right) }\equiv e^{i\vartheta }=\frac{%
a_{1}^{2}-a_{2}^{2}+2ia_{1}a_{2}\sin \varphi }{\sqrt{\left(
a_{1}^{2}-a_{2}^{2}\right) ^{2}+4a_{1}^{2}a_{2}^{2}\sin ^{2}\varphi }}.
\label{eq:a38a}
\end{equation}
The series representation (\ref{eq:a42a}) of the second-order energy
transfer is valid for any strength of the magnetic field.

For most applications it is also useful to integrate the $\varphi
_{1},\varphi _{2},\vartheta _{\mathbf{s}}$-averaged energy transfer, $%
\langle \Delta E_{1}^{(2)}\rangle $, with respect to the impact parameters $%
s $ in the full 2D space. We thus introduce a generalized cross-section \cite%
{gzwi99,zwi00,ner03,ner07} through the relation
\begin{equation}
\sigma =\int_{0}^{\infty }\langle \Delta E_{1}^{(2)}\rangle
sds=\int_{0}^{2\pi }\frac{d\varphi }{2\pi }\overline{\sigma }\left( \varphi
\right) =\int_{0}^{2\pi }\frac{d\varphi }{2\pi }\int_{0}^{\infty }\mathcal{E}%
\left( \varphi ,s\right) sds\,.  \label{eq:a43}
\end{equation}%
As $\overline{\sigma }(\varphi )$ results from the $s$-integration of the
energy transfer $\mathcal{E}(\varphi ,s)$ (\ref{eq:a42a}) one obtains an
expression for $\overline{\sigma }(\varphi )$ which represents an infinite
sum over Bessel functions. Moreover, assuming regularized interaction and
performing $\mathbf{k}$-integration in $\overline{\sigma }(\varphi )$ yields
an infinite sum over modified Bessel functions (see, e.g., an example for
ion--electron collision in Ref.~\cite{ner07}). For arbitrary axially
symmetric interaction potential similar expression is derived in Appendix~%
\ref{sec:ap2} (see Eq.~(\ref{eq:apb1})). However, for practical applications
it is much more convenient to use an equivalent integral representation of
the effective cross-section which does not involve any special function.
This expression can be derived from the Bessel-function representation of
the cross-section $\overline{\sigma }(\varphi )$ using the integral
representation of the Dirac $\delta $-function as well as the summation
formula for $\sum_{n}e^{in\varphi }J_{n}^{2}\left( a\right) $ \cite{gra80}.
The energy transfer $\overline{\sigma }(\varphi )$ after lengthy but
straightforward calculations then reads
\begin{equation}
\overline{\sigma }(\varphi )=\overline{\sigma }_{\parallel }(\varphi )+%
\overline{\sigma }_{\bot }(\varphi ) ,  \label{eq:cross}
\end{equation}
with
\begin{eqnarray}
\overline{\sigma }_{\parallel }(\varphi ) &=&-\frac{2\left( 2\pi \right)
^{2}q^{4}e\!\!\!/^{4}V_{0\parallel }}{m\omega _{c}^{2}v_{r\parallel }}%
\int_{0}^{\infty }tdt\int d\mathbf{k}\left\vert U(\mathbf{k})\right\vert ^{2}
\label{eq:a43b} \\
&&\times \left( k_{\parallel }^{2}+k_{\bot }^{2}\frac{\sin t}{t}\right)
k_{\parallel }\sin \left( k_{\parallel }\delta t\right) J_{0}\left( 2k_{\bot
}a\sin \frac{t}{2}\right) ,  \notag
\end{eqnarray}
\begin{eqnarray}
\overline{\sigma }_{\bot }(\varphi ) &=&-\frac{2\left( 2\pi \right)
^{2}q^{4}e\!\!\!/^{4}V_{0\bot }}{m\omega _{c}^{2}\left\vert v_{r\parallel
}\right\vert }\cos \vartheta \int_{0}^{\infty }tdt\int d\mathbf{k}\left\vert
U(\mathbf{k})\right\vert ^{2}  \label{eq:a43bb} \\
&&\times \left( k_{\parallel }^{2}+k_{\bot }^{2}\frac{\sin t}{t}\right)
k_{\bot }\cos \left( k_{\parallel }\delta t\right) \cos \left( \frac{t}{2}%
\right) J_{1}\left( 2k_{\bot }a\sin \frac{t}{2}\right) ,  \notag
\end{eqnarray}%
where $\delta =|v_{r\parallel }|/\omega _{c}$ is the relative pitch of the
particles helices, divided by $2\pi $. In Eq.~(\ref{eq:cross}) the
cross-section $\overline{\sigma }(\varphi )$ has been splitted into two
parts which correspond to the cm motion along ($\overline{\sigma }%
_{\parallel}(\varphi )$) and transverse ($\overline{\sigma }_{\perp}(\varphi
)$) to the magnetic field.

An expression similar to Eqs.~(\ref{eq:a42}) and (\ref{eq:a42a}) has been
obtained in \cite{ner03,ner07} for electron-heavy ion (no cyclotron motion)
collision where the direction $\mathbf{b}$ of the magnetic field and the
direction $\mathbf{n}_{r}=\mathbf{v}_{r}/v_{r}$ of the relative velocity $%
\mathbf{v}_{r}=v_{e\parallel }\mathbf{b}-\mathbf{v}_{i}$ ($v_{e\parallel }$
is the component of electron velocity parallel to the magnetic field) of the
electron guiding center is singled out in the argument of the $\delta $%
-function and the summand of the $n$-summation. This prevents a closed
evaluation of the energy transfer for ion--electron collision for arbitrary
direction of the ion motion with respect to the magnetic field. But if the
ion moves along the magnetic field and the guiding center $v_{r\parallel }%
\mathbf{b}$ has no component in transverse direction, as it is always the
case for a collision of two identical gyrating particles, the energy
transfer in such ion--electron collisions can be evaluated in the same
manner as already discussed in context with deriving Eqs.~(\ref{eq:cross})-(%
\ref{eq:a43bb}) in a straightforward manner.

For that case of ion--electron collisions with constant $\mathbf{v}_{i} =
v_{i\parallel} \mathbf{b}$ the energy loss of the ion is given by $\Delta
E_{i}= -m v_{i\parallel} \Delta v_{\parallel}$, see Eq.~(\ref{eq:a21d}),
because the relative energy transfer $\Delta E_{r}$ here vanishes according
to Eq. (\ref{eq:a7b}). The velocity transfer $\langle \Delta v_{\parallel
}^{(2)}\rangle $ required for $\langle \Delta E_{i}^{(2)}\rangle$ can be
extracted from Eqs.~(\ref{eq:a42}) and (\ref{eq:a42a}). For identical
particles this is the term in Eq.~(\ref{eq:a42a}) which is proportional to $%
V_{0\parallel }$ (see, e.g., Eq.~(\ref{eq:a21a})), i.e.
\begin{equation}
\langle \Delta v_{\parallel }^{(2)}\rangle =\int_{0}^{2\pi }\frac{d\varphi }{%
2\pi }\mathcal{U}_{1}\left( \varphi ,s\right)  \label{eq:a42b}
\end{equation}%
with
\begin{eqnarray}
\mathcal{U}_{1}\left( \varphi ,s\right) &=&-\frac{4\pi iq^{4}e\!\!\!/^{4}}{%
m^{2}\left\vert v_{r\parallel }\right\vert }\int d\mathbf{k}d\mathbf{k}%
^{\prime }U(\mathbf{k})U(\mathbf{k}^{\prime })k_{\parallel }J_{0}\left(
Qs\right) \delta (k_{\parallel }^{\prime }+k_{\parallel })  \label{eq:a42c}
\\
&&\times \sum_{n=-\infty }^{\infty }\left( -1\right) ^{n}e^{in \left(\theta
-\theta ^{\prime } \right) }J_{n}\left( k_{\bot }a\right) J_{n}\left(
k_{\bot }^{\prime }a\right) G_{n}\left( \mathbf{k},\mathbf{k}^{\prime
}\right) .  \notag
\end{eqnarray}
For the present case of these specific ion--electron collisions the
quantities $q^4$ and $\mu =m/2$ in Eq.~(\ref{eq:a42c}) have to be replaced
with $Z^2$ and $m$, respectively. Thus the $s$-integrated $\langle \Delta
v_{\parallel }^{(2)}\rangle$ for ion--electron collisions corresponds to
Eq.~(\ref{eq:a43b}), i.e.~more precisely to
\begin{equation}
\int_{0}^{\infty }\langle \Delta v_{\parallel }^{(2)} \rangle sds= \frac{Z^2%
}{q^4} \frac{1}{2mV_{0\parallel}} \int_{0}^{2\pi }\frac{d\varphi }{2\pi } \
\overline{\sigma}_{\parallel}(\varphi) .  \label{eq:a43_v2}
\end{equation}%
It should be emphasized that for ion--electron collision the phase dependent
transversal relative velocity $v_{0\bot }(\varphi)$ is replaced by the
electron transversal velocity $v_{e\bot }$ and the relative cyclotron radius
$v_{0\bot}/\omega _{c}$ for identical particles is replaced by $v_{e\bot
}/\omega_{c} $. Thus the integrands in Eqs.~(\ref{eq:a42}) and (\ref%
{eq:a43_v2}) (and similarly in other angular-averaged quantities) do not
depend on the phase $\varphi $ ($\mathcal{E}(\varphi ,s)=\mathcal{E}(s)$)
and, therefore, we have $-\langle \Delta E_{1}^{(2)}\rangle =-\mathcal{E}%
(s)\rightarrow \langle \Delta E_{i}^{(2)}\rangle $ for the second order
energy transfer to the ion $\langle \Delta E_{i}^{(2)}\rangle $.

\subsection{Second order relative energy transfers}

\label{sec:s2.3}

We now turn back to the case of identical particles and consider the
evaluation of the angular-averaged relative energy transfer in longitudinal
direction, $\langle \Delta E_{r\parallel }^{(2)}\rangle =-\langle \Delta
E_{r\bot }^{(2)}\rangle $. According to Eq.~(\ref{eq:a21c}) the longitudinal
relative energy transfer $\langle \Delta E_{r\parallel }^{(2)}\rangle$
involves both $\langle \Delta v_{\parallel }^{(2)}\rangle$ (\ref{eq:a42b})
and the the angular-averaged value of $(\Delta v_{\parallel }^{(1)})^{2}$.
The latter is calculated from Eq.~(\ref{eq:a38b}). The angular-averaging
procedure is the same as for deriving Eqs.~(\ref{eq:a42}) and (\ref{eq:a42a}%
) and yields
\begin{equation}
\langle (\Delta v_{\parallel }^{(1)})^{2}\rangle =\int_{0}^{2\pi }\frac{%
d\varphi }{2\pi }\mathcal{U}_{2}(\varphi ,s)  \label{eq:a42d}
\end{equation}%
with
\begin{eqnarray}
\mathcal{U}_{2}(\varphi ,s) &=&\frac{4\left( 2\pi
\right)^{2}q^{4}e\!\!\!/^{4}} {m^{2} \left\vert v_{r\parallel }\right\vert }%
\int d\mathbf{k}d\mathbf{k}^{\prime }U(\mathbf{k})U(\mathbf{k}^{\prime
})k_{\parallel }^{2}J_{0}\left( Qs\right) \delta (k_{\parallel }^{\prime
}+k_{\parallel })  \label{eq:a42e} \\
&&\times \sum_{n=-\infty }^{\infty }\left( -1\right) ^{n}e^{in\varsigma
\left( \theta -\theta ^{\prime }\right) }J_{n}\left( k_{\bot }a\right)
J_{n}\left( k_{\bot }^{\prime }a\right) \delta \left( \zeta _{n}(\mathbf{k}%
)\right) .  \notag
\end{eqnarray}%
The relative energy change in the longitudinal direction is then calculated
using Eqs.~(\ref{eq:a21c}), (\ref{eq:a42b}) and (\ref{eq:a42d})
\begin{equation}
\langle \Delta E_{r\bot }^{(2)}\rangle \equiv -\langle \Delta E_{r\parallel
}^{(2)}\rangle =\int_{0}^{2\pi }\frac{d\varphi }{2\pi }\mathcal{E}_{r\bot
}\left( \varphi ,s\right)  \label{eq:rel1}
\end{equation}%
with
\begin{equation}
\mathcal{E}_{r\bot }\left( \varphi ,s\right) \equiv -\mathcal{E}_{r\parallel
}\left( \varphi ,s\right) =-\frac{m}{4}\left[ \mathcal{U}_{2}(\varphi
,s)+2v_{r\parallel }\mathcal{U}_{1}(\varphi ,s)\right] .  \label{eq:rel2}
\end{equation}%
For further applications (see Sec.~\ref{sec:imp}) also the angular averaged
square of the first order relative energy transfer in the transversal
direction, $\langle (\Delta E_{r\bot }^{(1)})^{2}\rangle $ is needed, which
is given by
\begin{equation}
\langle (\Delta E_{r\bot }^{(1)})^{2}\rangle =\frac{m^{2}}{4}v_{r\parallel
}^{2}\langle (\Delta v_{\parallel }^{(1)})^{2}\rangle =\frac{m^{2}}{4}%
v_{r\parallel }^{2}\int_{0}^{2\pi }\frac{d\varphi }{2\pi }\mathcal{U}%
_{2}(\varphi ,s).  \label{eq:rel3}
\end{equation}

Similarly to the definition and derivation of Eqs.~(\ref{eq:a43}) and (\ref%
{eq:cross}) we define the related cross-sections by
\begin{eqnarray}
\sigma _{r\bot } &=&\int_{0}^{\infty }\langle \Delta E_{r\bot }^{(2)}\rangle
sds=\int_{0}^{2\pi }\frac{d\varphi }{2\pi }\overline{\sigma }_{r\bot }\left(
\varphi \right) ,  \label{eq:rel4} \\
\sigma _{r1}^{2} &=&\int_{0}^{\infty }\langle (\Delta E_{r\bot
}^{(1)})^{2}\rangle sds=\int_{0}^{2\pi }\frac{d\varphi }{2\pi }\overline{%
\sigma }_{r1}\left( \varphi \right)  \label{eq:rel5}
\end{eqnarray}%
with
\begin{eqnarray}
\overline{\sigma }_{r\bot }\left( \varphi \right) &=&\int_{0}^{\infty }%
\mathcal{E}_{r\bot }\left( \varphi ,s\right) sds=\frac{2\left( 2\pi \right)
^{2}q^{4}e\!\!\!/^{4}}{m\left\vert v_{r\parallel }\right\vert \omega _{c}}%
\int_{0}^{\infty }dt\int d\mathbf{k}\left\vert U(\mathbf{k})\right\vert
^{2}k_{\parallel }J_{0}\left( 2k_{\bot }a\sin \frac{t}{2}\right)
\label{eq:rel6} \\
&&\times \left[ \delta t\left( k_{\parallel }^{2}+k_{\bot }^{2}\frac{\sin t}{%
t}\right) \sin \left( k_{\parallel }\delta t\right) -k_{\parallel }\cos
\left( k_{\parallel }\delta t\right) \right] ,  \notag
\end{eqnarray}

\begin{eqnarray}
\overline{\sigma }_{r1}\left( \varphi \right) &=&\frac{m^{2}}{4}%
v_{r\parallel }^{2}\int_{0}^{\infty }\mathcal{U}_{2}(\varphi ,s)sds
\label{eq:rel7} \\
&=&2\left( 2\pi \right) ^{2}q^{4}e\!\!\!/^{4}\delta \int_{0}^{\infty }dt\int
d\mathbf{k}\left\vert U(\mathbf{k})\right\vert ^{2}k_{\parallel }^{2}\cos
\left( k_{\parallel }\delta t\right) J_{0}\left( 2k_{\bot }a\sin \frac{t}{2}%
\right) .  \notag
\end{eqnarray}

\section{The energy transfer for the screened and regularized potential}
\label{sec:s3}

For the Coulomb interaction $U(k)=U_{\mathrm{C}}(k)$, the full 2D
integration over the $\mathbf{s}$-space results in a logarithmic divergence
of the $\mathbf{k}$-integration in Eqs.~(\ref{eq:a43b}), (\ref{eq:a43bb}), (%
\ref{eq:rel6}) and (\ref{eq:rel7}). To cure this cutoff parameters $k_{\min }
$ and $k_{\max }$ must be introduced, see \cite{ner03,ner07} for details.
But the averaged energy transfer, Eq.~(\ref{eq:a42}) with Eq.~(\ref{eq:a42a}%
), can be evaluated without further approximation for any axially symmetric
interaction potential, $U(\mathbf{k})=U(|k_{\parallel }|,k_{\perp })$. In
this case the averaged energy transfer can be represented as the sum of all
cyclotron harmonics as it has been done for ion--electron interaction in
Ref.~\cite{ner07}.

To continue the interaction must be specified. In the following we consider
throughout the regularized screened potential $U(\mathbf{r})=U_{\mathrm{R}}(%
\mathbf{r})$ introduced in Sec.~\ref{sec:s1.1} with
\begin{equation}
U_{\mathrm{R}}(\mathbf{r})=\left(1-e^{-r/\lambdabar }\right) \frac{%
e^{-r/\lambda}}{r} \ , \quad U_{\mathrm{R}}(k_{\parallel },k_{\perp })=\frac{%
2}{\left( 2\pi \right) ^{2}}\left( \frac{1}{k_{\perp }^{2}+\kappa ^{2}}-%
\frac{1}{k_{\perp }^{2}+\chi ^{2}}\right) ,  \label{eq:a48}
\end{equation}%
where $\kappa ^{2}=k_{\parallel }^{2}+\lambda ^{-2}$, $\chi
^{2}=k_{\parallel }^{2}+d^{-2}a$, $d^{-1}=\lambda ^{-1}+\lambdabar ^{-1}$,
for which the $\mathbf{k}$-integrations involved in Eqs.~(\ref{eq:a43b}), (%
\ref{eq:a43bb}), (\ref{eq:rel6}) and (\ref{eq:rel7}) converge.

\subsection{Second order energy transfer}

\label{sec:s3.1}

A simple but important particular case is that of vanishing cyclotron radius
($v_{01\bot }=v_{02\bot }=0$), i.e. when initially the electrons move along
the magnetic field. From Eqs.~(\ref{eq:a38b}), (\ref{eq:a42d}) and (\ref%
{eq:a42e}), and using the regularized potential (\ref{eq:a48}), it is
straightforward to show that in this case $\Delta v^{(1)}_{\parallel} =0$
and $\mathcal{U}_{2}(\varphi ,s) =0$. Substituting Eq.~(\ref{eq:a48}) into
Eqs.~(\ref{eq:a42a}), (\ref{eq:a42c}) and (\ref{eq:rel2}) and setting $a=0$
we obtain after $\mathbf{k}$, $\mathbf{k}^{\prime}$-integrations
\begin{equation}
\mathcal{E}\left( \varphi ,s\right) =-\frac{V_{0\parallel}}{v_{r\parallel}}%
\mathcal{E}_{r\bot}\left( \varphi ,s\right) =-\frac{4q^{4}e\!\!\!/^{4}V_{0%
\parallel }}{mv_{r\parallel }^{3}s^{2}}\left[ (\kappa _{1}s)K_{1}(\kappa
_{1}s)-(\chi _{1}s)K_{1}(\chi _{1}s)\right] ^{2} ,  \label{eq:a59a}
\end{equation}%
where $\kappa _{1}^{2}=\delta ^{-2}+\lambda^{-2}$, $\chi _{1}^{2}=\delta
^{-2}+d^{-2}$ and $K_{n}$ are the modified Bessel functions. From Eq.~(\ref%
{eq:a59a}) it is seen that the modified Bessel functions guarantee the
convergence of the impact parameter integrated energy transfer at small $s$.

Eq.~(\ref{eq:a59a}) is the leading term of the expansion of the the energy
transfers with respect to small cyclotron radius $a$ and was obtained in the
limit $a\rightarrow 0$ where the particles move initially along their
guiding center trajectories. Moreover, for $\omega _{c}\rightarrow \infty $
also the relative pitch $\delta \rightarrow 0$ and these trajectories are
rectilinear along the lines of the magnetic field. The particles just pass
each other along a straight line and for symmetry reasons the velocity and
energy transfers vanish. For a finite $\omega _{c}$ corresponding to a
finite pitch the contribution of the leading term in Eq.~(\ref{eq:a59a})
describes the perturbation of the guiding centers trajectories. By the
reason of symmetry the next order term must be quadratic $\sim a^{2}$ and
accounts for the finite relative cyclotron motion of the particles.

As we discussed in Sec.~\ref{sec:s2.2} the energy transfer (\ref{eq:a42})
must be integrated with respect to the impact parameters $s$ for practical
applications. For general interaction potential this has been done in Sec.~%
\ref{sec:s2.2}, Eqs.~(\ref{eq:a43})-(\ref{eq:a43bb}). In general for a study
of the convergence of the $s$-integrated energy transfers we note that the
case with $s=a$ is most critical for the convergence of the cross-sections.
This is intuitively clear as the gyrating particles at $\left\vert
a_{1}-a_{2}\right\vert <s<a_{1}+a_{2}$ may hit each other on such a
trajectory. This should not matter for the potential (\ref{eq:a48}), which
has been regularized near the origin for exactly that purpose. On the other
hand, the energy transfer for the un-regularized potentials $U_{\mathrm{C}}$
and $U_{\mathrm{D}}$ diverges for $s=a$ (see Ref.~\cite{ner07} for some
explicit examples).

For the present case of the regularized and screened interaction potential
Eq.~(\ref{eq:a48}), i.e.~substituting this potential into Eqs.~(\ref{eq:a43b}%
) and (\ref{eq:a43bb}), the impact parameter integrated cross sections are
\begin{eqnarray}
\overline{\sigma }_{\parallel }(\varphi ) &=&-\frac{2q^{4}e\!\!\!/^{4}V_{0%
\parallel }}{m\omega _{c}^{3}\lambda ^{3}}\frac{\left\vert v_{r\parallel
}\right\vert }{v_{r\parallel }}\int_{0}^{\infty }\frac{t^{2}dt}{R^{3}\left(
t\right) }\left\{ e^{-R\left( t\right) }\left[ \mathcal{F}_{1}\left( R\left(
t\right) ,\xi t,t\right) +\frac{4}{\varkappa ^{2}-1}\mathcal{F}_{2}\left(
R\left( t\right) ,\xi t,t\right) \right] \right.  \label{eq:a63} \\
&&\left. +e^{-\varkappa R\left( t\right) }\left[ \mathcal{F}_{1}\left(
\varkappa R\left( t\right) ,\varkappa \xi t,t\right) -\frac{4\varkappa ^{2}}{%
\varkappa ^{2}-1}\mathcal{F}_{2}\left( \varkappa R\left( t\right) ,\varkappa
\xi t,t\right) \right] \right\} ,  \notag
\end{eqnarray}
\begin{eqnarray}
\overline{\sigma }_{\bot }(\varphi ) &=&\frac{q^{4}e\!\!\!/^{4}}{m\omega
_{c}^{3}\lambda ^{3}}\frac{v_{02\perp }^{2}-v_{01\perp }^{2}}{\left\vert
v_{r\parallel }\right\vert }\int_{0}^{\infty }\frac{t\sin tdt}{R^{3}\left(
t\right) }\left\{ e^{-R\left( t\right) }\left[ \mathcal{F}_{3}\left( R\left(
t\right) ,\xi t,t\right) +\frac{4}{\varkappa ^{2}-1}\mathcal{F}_{4}\left(
R\left( t\right) ,\xi t,t\right) \right] \right.  \label{eq:a63x} \\
&&\left. +e^{-\varkappa R\left( t\right) }\left[ \mathcal{F}_{3}\left(
\varkappa R\left( t\right) ,\varkappa \xi t,t\right) -\frac{4\varkappa ^{2}}{%
\varkappa ^{2}-1}\mathcal{F}_{4}\left( \varkappa R\left( t\right) ,\varkappa
\xi t,t\right) \right] \right\} .  \notag
\end{eqnarray}%
Here $\xi =\delta /\lambda =\left\vert v_{r\parallel }\right\vert /\omega
_{c}\lambda $, $R^{2}\left( t\right) =\xi ^{2}t^{2}+4(a^{2}/\lambda
^{2})\sin ^{2}(t/2)$, $\varkappa =\lambda /d=1+\lambda /\lambdabar $, and
\begin{equation}
\mathcal{F}_{1}\left( R,\zeta ,t\right) =2+2R-R^{2}+\left( 1-\frac{\sin t}{t}%
\right) \left[ R^{2}+R+1-\frac{\zeta ^{2}}{R^{2}}\left( R^{2}+3R+3\right) %
\right] ,  \label{eq:a64}
\end{equation}%
\begin{equation}
\mathcal{F}_{2}\left( R,\zeta ,t\right) =R+1-\frac{1}{R^{2}}\left( 1-\frac{%
\sin t}{t}\right) \left[ R^{3}+4R^{2}+9R+9-\frac{\zeta ^{2}}{R^{2}}\left(
R^{3}+6R^{2}+15R+15\right) \right] ,  \label{eq:a65}
\end{equation}%
\begin{equation}
\mathcal{F}_{3}\left( R,\zeta ,t\right) =2+2R-R^{2}+\left( 1-\frac{\sin t}{t}%
\right) \left[ R^{2}-R-1-\frac{\zeta ^{2}}{R^{2}}\left( R^{2}+3R+3\right) %
\right] ,  \label{eq:a66}
\end{equation}%
\begin{equation}
\mathcal{F}_{4}\left( R,\zeta ,t\right) =R+1-\frac{1}{R^{2}}\left( 1-\frac{%
\sin t}{t}\right) \left[ R^{3}+2R^{2}+3R+3-\frac{\zeta ^{2}}{R^{2}}\left(
R^{3}+6R^{2}+15R+15\right) \right] .  \label{eq:a67}
\end{equation}%
Note that the transversal cross-section $\overline{\sigma }_{\bot }(\varphi )
$ vanish at $a_{1}=a_{2}$ and the energy transfer occurs only due to the cm
motion along the magnetic field $\overline{\sigma }_{\parallel }(\varphi )$.
We also note in particular the dependence of the sign of the transversal
cross-section Eq.~\eqref{eq:a63x} on $a_{1}-a_{2}$.

As already discussed in Sec.~\ref{sec:s2.2} the energy transfer in the case
of electron-heavy ion collisions is given by the $s$-integrated $\langle
\Delta v_{\parallel }^{(2)}\rangle$, see Eq.~(\ref{eq:a43_v2}), now with $%
\overline{\sigma}_{\parallel}(\varphi)$ from Eq.~(\ref{eq:a63}).

\subsection{Second order relative energy transfers}

\label{sec:s3.2}

Similarly, substituting Eq.~(\ref{eq:a48}) into Eqs.~(\ref{eq:rel6}) and (%
\ref{eq:rel7}), we obtain for the relative cross-sections $\overline{\sigma }%
_{r\bot }\left( \varphi \right) $ and $\overline{\sigma }_{r1}\left( \varphi
\right)$\
\begin{eqnarray}
\overline{\sigma }_{r\bot }\left( \varphi \right) &=&\frac{2q^{4}e\!\!\!/^{4}%
}{m\left\vert v_{r\parallel }\right\vert \omega _{c}\lambda }%
\int_{0}^{\infty }\frac{dt}{R^{3}\left( t\right) }\left\{ e^{-R\left(
t\right) }\left[ \mathcal{Y}_{1}\left( R\left( t\right) ,\xi t,t\right) +%
\frac{4}{\varkappa ^{2}-1}\mathcal{Y}_{2}\left( R\left( t\right) ,\xi
t,t\right) \right] \right.  \label{eq:rel8} \\
&&\left. +\frac{1}{\varkappa ^{2}}e^{-\varkappa R\left( t\right) }\left[
\mathcal{Y}_{1}\left( \varkappa R\left( t\right) ,\varkappa \xi t,t\right) -%
\frac{4\varkappa ^{2}}{\varkappa ^{2}-1}\mathcal{Y}_{2}\left( \varkappa
R\left( t\right) ,\varkappa \xi t,t\right) \right] \right\} ,  \notag
\end{eqnarray}%
\begin{eqnarray}
\overline{\sigma }_{r1}\left( \varphi \right) &=&2q^{4}e\!\!\!/^{4}\xi
\int_{0}^{\infty }\frac{dt}{R\left( t\right) }\left\{ e^{-R\left( t\right) }%
\left[ \mathcal{P}_{1}\left( R\left( t\right) ,\xi t\right) -\frac{4}{%
\varkappa ^{2}-1}\mathcal{P}_{2}\left( R\left( t\right) ,\xi t\right) \right]
\right.  \label{eq:rel9} \\
&&\left. +e^{-\varkappa R\left( t\right) }\left[ \mathcal{P}_{1}\left(
\varkappa R\left( t\right) ,\varkappa \xi t\right) +\frac{4\varkappa ^{2}}{%
\varkappa ^{2}-1}\mathcal{P}_{2}\left( \varkappa R\left( t\right) ,\varkappa
\xi t\right) \right] \right\} ,  \notag
\end{eqnarray}%
where
\begin{equation}
\mathcal{P}_{1}\left( R,\zeta \right) =1-\frac{\zeta ^{2}}{R^{2}}\left(
R+1\right) ,\qquad \mathcal{P}_{2}\left( R,\zeta \right) =\frac{1}{R^{2}}%
\left[ R+1-\frac{\zeta ^{2}}{R^{2}}\left( R^{2}+3R+3\right) \right] ,
\label{eq:rel10}
\end{equation}%
\begin{eqnarray}
\mathcal{Y}_{1}\left( R,\zeta ,t\right) &=&-R^{2}+\zeta ^{2}\left(
3R+3-R^{2}\right) +\zeta ^{2}\left( 1-\frac{\sin t}{t}\right) \left[
R^{2}+R+1-\frac{\zeta ^{2}}{R^{2}}\left( R^{2}+3R+3\right) \right] ,
\label{eq:rel11} \\
\mathcal{Y}_{2}\left( R,\zeta ,t\right) &=&\left( \zeta ^{2}+1\right) \left(
R+1\right) -\frac{\zeta ^{2}}{R^{2}}\left( R^{2}+3R+3\right)
\label{eq:rel12} \\
&&-\frac{\zeta ^{2}}{R^{2}}\left( 1-\frac{\sin t}{t}\right) \left[
R^{3}+4R^{2}+9R+9-\frac{\zeta ^{2}}{R^{2}}\left( R^{3}+6R^{2}+15R+15\right) %
\right] .  \notag
\end{eqnarray}%
It must be emphasized that in contrast to the quantity $\overline{\sigma }%
(\varphi )$ the relative cross-sections $\overline{\sigma }_{r\bot }(\varphi
)$ and $\overline{\sigma }_{r1} (\varphi )$ do not depend on the cm velocity
components $V_{0\parallel }$ and $V_{0\bot }$.

Next we also consider the cross-sections $\overline{\sigma }(\varphi )$, $%
\overline{\sigma }_{r\bot }(\varphi )$ and $\overline{\sigma }_{r1}(\varphi )
$ for vanishing cyclotron radius, $a\rightarrow 0$. In this limit $\overline{%
\sigma }_{\bot }(\varphi )=0$ and  $\overline{\sigma }_{r1}(\varphi )=0$ as
can be easily seen from Eq.~(\ref{eq:rel7}) as well as by explicitly
evaluating Eq.~(\ref{eq:rel9}) with $R\left( t\right) =\xi t$. The
integration of Eq.~(\ref{eq:a59a}) with respect to the impact parameter (see
Appendix~\ref{sec:ap1} for details) yields
\begin{equation}
\overline{\sigma }\left( \varphi \right) =-\frac{V_{0\parallel }}{%
v_{r\parallel }}\overline{\sigma }_{r\bot }\left( \varphi \right) =-\frac{%
4q^{4}e\!\!\!/^{4}V_{0\parallel }}{mv_{r\parallel }^{3}}\left[ \frac{\xi
^{2}\left( \varkappa ^{2}+1\right) +2}{2\xi ^{2}\left( \varkappa
^{2}-1\right) }\ln \frac{\xi ^{2}\varkappa ^{2}+1}{\xi ^{2}+1}-1\right] .
\label{eq:a69}
\end{equation}%
Note that $\overline{\sigma }(\varphi )$ at $a\rightarrow 0$ can be
alternatively evaluated from Eq.~(\ref{eq:a63}). In this limit the
expression for $R(t)$ is simplified to $R(t) =\xi t$. After performing the $t
$-integration we again arrive at Eq.~(\ref{eq:a69}). In addition Eq.~(\ref%
{eq:a69}) is approximately valid also for finite cyclotron radius $a$,
assuming that the longitudinal velocity $v_{r\parallel }$ is larger than the
transversal ones, $v_{0\perp }$ and $V_{0\perp }$. Indeed, in this case $%
R(t) \simeq \xi t$ since $\delta \gg a$ and in Eqs.~(\ref{eq:a63}) and (\ref%
{eq:a63x}) the transversal cross-section $\overline{\sigma }_{\bot}(\varphi )
$ can be neglected compared to the longitudinal one. This indicates that in
the high velocity limit with $v_{r\parallel }\gg v_{0\perp },$ $V_{0\perp }$
the transversal motion of the particles as well as its cm transversal motion
are not important and can be neglected. This high velocity limit of the
cross-section $\overline{\sigma }_{r1}\left( \varphi \right) $ is obtained
from Eq.~(\ref{eq:rel9}). Since only the contribution of small $t$ is
important the function $R\left( t\right) $ is approximated by $R\left(
t\right) \simeq t(\xi ^{2}+a^{2}/\lambda ^{2})^{1/2}\simeq \xi
t(1+a^{2}/2\delta ^{2})$ (here we keep the small term $\sim a^{2}/\delta ^{2}
$ because $\overline{\sigma }_{r1}\left( \varphi \right) $ vanishes at $%
a\rightarrow 0$). Using this result for $R\left( t\right) $ from Eqs.~(\ref%
{eq:a63x}), (\ref{eq:rel9}) and (\ref{eq:a69}) in the high velocity limit we
obtain within the leading term approximation for the cross-sections
\begin{eqnarray}
&&\overline{\sigma }_{r1}\left( \varphi \right) \simeq 2q^{4}e\!\!\!/^{4}%
\frac{v_{0\bot }^{2}}{v_{r\parallel }^{2}}\Lambda (\varkappa ),
\label{eq:a69a} \\
&&\overline{\sigma }_{\parallel}(\varphi ) \simeq -\frac{V_{0\parallel }}{%
v_{r\parallel }}\overline{\sigma }_{r\bot }\left( \varphi \right) \simeq -%
\frac{4q^{4}e\!\!\!/^{4}V_{0\parallel }}{mv_{r\parallel }^{3}}\Lambda
(\varkappa ) ,  \label{eq:a69b} \\
&&\overline{\sigma}_{\bot}(\varphi ) \simeq \frac{2q^{4}e\!\!\!/^{4}}{%
mv_{r\parallel}^{4}}\left( v_{02\perp}^{2}-v_{01\perp}^{2}\right) \Lambda
(\varkappa) \, .  \label{eq:a69c}
\end{eqnarray}%
Here $\Lambda (\varkappa )$\ is a generalized Coulomb logarithm and is given
by Eq.~(\ref{eq:ap5}). Note that $\overline{\sigma }(\varphi)$ and $%
\overline{\sigma }_{r\bot }(\varphi )$ are isotropic, i.e. do not depend on $%
\varphi $ while $\overline{\sigma }_{r1}(\varphi )$ contains a term which is
proportional to $\cos \varphi $. Also, in the high velocity limit the
cross-sections do not depend on the magnetic field strength and decays as $%
\overline{\sigma }_{\parallel}(\varphi ) \sim v_{r\parallel }^{-3}$, $%
\overline{\sigma }_{\bot}(\varphi ) \sim v_{r\parallel }^{-4}$,  $\overline{%
\sigma }_{r\bot }(\varphi ) \sim v_{r\parallel }^{-2}$ and $\overline{\sigma
}_{r1}(\varphi ) \sim v_{r\parallel }^{-2}$.

Finally we briefly turn to the case of small relative velocity, $%
v_{r\parallel}\ll v_{0\bot}$, $V_{0\bot}$. It should be emphasized that the
integral representations of the cross-sections, Eqs.~(\ref{eq:a63}), (\ref%
{eq:rel8}) and (\ref{eq:rel9}), are not adopted for evaluation of these
quantities at small velocities. For this purpose it is much more convenient
to use an alternative Bessel-function representation of the cross-sections
as shown in Appendix~\ref{sec:ap2}. In addition, it is expected that the
limit of small $v_{r\parallel}$ is the most critical regime for a violation
of the perturbation theory employed here. Therefore explicit analytical
expressions in this limit can be useful for an improvement of the
perturbation theory by comparing the analytical results with numerical
simulations, see Sec.~\ref{sec:s4}.

\section{Improved treatment for the repulsive interaction}
\label{sec:imp}

The magnetic field drastically changes the scattering problem of two charged
particles as discussed in Sec.~\ref{sec:intr}. One important consequence of
a strong magnetic field is the backscattering of the particles from the
repulsive potential barrier even at a finite impact parameter $s$ (let us
recall that in the case of Rutherford scattering this occurs at either
vanishing impact parameters, $s\rightarrow 0$, or at vanishing relative
velocity). For instance, in a strong magnetic field with $a\ll \lambda $, $s$
and in the case of attractive interaction the velocity and energy transfers
are very small and for symmetry reasons vanish with increasing $B$, see the
similar discussion in Sec.~\ref{sec:s3}. However, in the case of repulsive
interaction the magnetic field together with the interaction potential forms
a potential barrier because of the particles motion is effectively
one-dimensional. Here two possible scattering regimes must be clearly
distinguished. To this end, we consider an exactly solvable model for two
interacting particles moving in the presence of an infinitely strong
magnetic field on rectilinear trajectories along the field with vanishing
cyclotron radii. Introducing the relative coordinate $\zeta
=z_{1}(t)-z_{2}(t)$ the relative energy conservation (note that the cm
energy is also conserved since $\mathbf{V}=\mathbf{b}V_{\parallel }(t)=%
\mathbf{b}V_{0\parallel }=\mathrm{const}$) for the regularized Yukawa
potential can be written in the form:
\begin{equation}
{\dot{\zeta}}^{2}=v_{r\parallel }^{2}\left[ 1-\varsigma _{1}\varsigma _{2}%
\frac{s_{0}}{r}\left( 1-e^{-r/\lambdabar }\right) e^{-r/\lambda }\right] ,
\label{eq:apa1}
\end{equation}%
where $\varsigma_{1}=\left\vert q_{1}\right\vert /q_{1}$, $%
\varsigma_{2}=\left\vert q_{2}\right\vert /q_{2}$, $r^{2}(t)=\zeta
^{2}(t)+s^{2}$, $s_{0}=2\left\vert q_{1}q_{2}\right\vert e\!\!\!/^{2}/\mu
v_{r\parallel }^{2}$. Here $\mu $ is the reduced mass, $s$ is the impact
parameter and $v_{r\parallel }$ is the initial longitudinal relative
velocity of the particles. Since the relative energy is conserved there is
no relative energy transfer and $\Delta E_{r}=0$. Then the energy transfer
of the particle 1 is related to the relative velocity transfer $\Delta
v_{\parallel }={\dot{\zeta}(+\infty )-\dot{\zeta}(-\infty )}$ by $\Delta
E_{1}=\mu V_{0\parallel }\Delta v_{\parallel }$, see Eq.~(\ref{eq:a21a}).
For reasons of symmetry, no velocity can be transferred from particle 2 to
the particle 1 if the interaction is attractive ($q_{1}q_{2} < 0$), see,
e.g., Eq.~(\ref{eq:apa1}). This may also be true for repulsive case ($%
q_{1}q_{2} > 0$). For instance, at $s_{0}<\lambdabar $ (or equivalently $%
x=v_{r\parallel }/v_{s}>\left( 2/\nu \right) ^{1/2}$ with $\nu =\lambdabar
/\lambda $ and $v_{s}^{2}=\left\vert q_{1}q_{2}\right\vert e\!\!\!/^{2}/\mu
\lambda $) and arbitrary $s$ as well as at $s_{0}>\lambdabar $ (or
equivalently $x<\left( 2/\nu \right) ^{1/2}$) and $s>s_{m}=\lambda \eta
_{m}(x)$, where $\eta _{m}(x)$\ is the root of the transcendental equation
\begin{equation}
\frac{1}{\eta _{m}}\left( 1-e^{-\eta _{m}/\nu }\right) e^{-\eta _{m}}=\frac{%
x^{2}}{2},  \label{eq:apa2}
\end{equation}%
and the energy of two particles relative motion is larger than the energy of
the potential barrier which again yields vanishing velocity and energy
transfers for repulsive interaction. Thus for $q_{1}q_{2} >0$\ the energy
transfer occurs at $s_{0}>\lambdabar $ and $s<s_{m}$. In this case the
velocity transfer is $\Delta v_{\parallel }=-2v_{r\parallel }$ which
corresponds to a reversion of the initial motion, i.e. to a backscattering
event. Then the energy transfer is
\begin{equation}
\Delta E_{1}=-\Delta E_{2}=-2\mu V_{0\parallel }v_{r\parallel } \Theta
\left(q_{1}q_{2}\right) \Theta (v_{c}^{2}-v_{r\parallel }^{2})\Theta \left(
s_{m}-s\right) ,  \label{eq:apa2a}
\end{equation}%
where $\Theta (z)$ is the Heavyside function and $v_{c}^{2}=2\left\vert
q_{1}q_{2}\right\vert e\!\!\!/^{2}/\mu \lambdabar $.

Consider now the energy-velocity transfers integrated with respect to the
impact parameters $s$. The result reads
\begin{equation}
-\frac{1}{\lambda ^{2}}\int_{0}^{\infty }\frac{\Delta E_{1}(s)}{\mu
V_{0\parallel }v_{s}}sds=-\frac{1}{\lambda ^{2}}\int_{0}^{\infty }\frac{%
\Delta v_{\parallel }(s)}{v_{s}}sds=x\eta _{m}^{2}\left( x\right) .
\label{eq:apa3}
\end{equation}%
Here $\eta _{m}$ is a function of $x=v_{r\parallel }/v_{s}$. (Note that this
function vanishes at $x>\left( 2/\nu \right) ^{1/2}$, where the
transcendental equation (\ref{eq:apa2}) has no solution). Consider two
limiting cases. At $v_{r\parallel }/v_{s}\lesssim \left( 2/\nu \right)
^{1/2} $ we obtain from Eq.~(\ref{eq:apa2})
\begin{equation}
\eta _{m}\left( x\right) \simeq \frac{\left( 2\nu \right) ^{3/2}}{2\nu +1}%
\left[ \left( \frac{2}{\nu }\right) ^{1/2}-x\right]  \label{eq:apa4}
\end{equation}%
and the $s$-integrated velocity transfer vanishes as $\sim \lbrack \left(
2/\nu \right) ^{1/2}-x]^{2}$. At $v_{r\parallel }/v_{s}\rightarrow 0$, Eq.~(%
\ref{eq:apa2}) yields
\begin{equation}
\eta _{m}\left( x\right) \simeq \ln \left[ \frac{2/x^{2}}{\ln \left(
2/x^{2}\right) }\right] .  \label{eq:apa5}
\end{equation}%
and the $s$-integrated velocity transfer vanishes like
\begin{equation}
-\frac{1}{\lambda ^{2}}\int_{0}^{\infty }\frac{\Delta v_{\parallel }(s)}{%
v_{s}}sds\simeq x\ln ^{2}\left[ \frac{2/x^{2}}{\ln \left( 2/x^{2}\right) }%
\right] \rightarrow 0.  \label{eq:apa6}
\end{equation}

From this simple example it is clear that the perturbative treatment
developed in the previous sections is not applicable for repulsive
interaction and in the presence of strong magnetic field when the hard
collisions like backscattering events may occur. Nevertheless the second
order treatment for the repulsive case can be improved when the energy
transfers due to the hard collisions are involved in the theory as the
leading terms. The simple example considered above suggests that this can be
done by explicitly using the energy conservation, see, e.g. Eq.~(\ref%
{eq:apa1}). For simplicity consider the case when the relative energy of
particles is conserved, $\Delta E_{r}=0$. Then from Eq.~(\ref{eq:a21c}) (in
the general case $m/2$ must be replaced here by the reduced mass $\mu $) and
the obvious relation $\Delta E_{r\parallel }=-\Delta E_{r\bot }$ we obtain a
quadratic equation for the relative velocity transfer $\Delta v_{\parallel }$
which has two real solutions
\begin{equation}
\Delta v_{\parallel }^{\mathrm{I}}=-v_{r\parallel }\left( 1-\sqrt{1-\phi }%
\right) ,\qquad \Delta v_{\parallel }^{\mathrm{II}}=-v_{r\parallel }\left( 1+%
\sqrt{1-\phi }\right)  \label{eq:apa8}
\end{equation}%
with $\phi =2\Delta E_{r\bot }/\mu v_{r\parallel }^{2}$. The velocity
transfers $\Delta v_{\parallel }^{\mathrm{I}}$ and $\Delta v_{\parallel }^{%
\mathrm{II}}$\ correspond to two different scattering regimes. In
particular, in the limit of strong magnetic field the transversal energy
transfer is small, $\phi \ll 1$, because of the transversal motion is
strongly hindered and
\begin{equation}
\Delta v_{\parallel }^{\mathrm{I}}\simeq -v_{r\parallel }\Phi ,\qquad \Delta
v_{\parallel }^{\mathrm{II}}\simeq -v_{r\parallel }\left( 2-\Phi \right) ,
\label{eq:apa11}
\end{equation}%
where $\Phi =\frac{1}{2}\phi +\frac{1}{8}\phi ^{2}$. In the limit of
vanishing $\phi $, $\Delta v_{\parallel }^{\mathrm{I}}\rightarrow 0$ while $%
\Delta v_{\parallel }^{\mathrm{II}}\rightarrow -2v_{r\parallel }$ which
correspond to over the barrier and backscattering events, respectively.
Thus, in contrast to the Rutherford classical scattering the magnetic field
may form two scattering channels, I and II. For instance, the scattering
with an attractive interaction is realized in the regime I while in the
repulsive case the scattering may occur in both regimes I and II depending
on the relation between the initial relative energy, the height of the
potential barrier and the strength of the magnetic field. In the case of the
infinitely strong magnetic field the boundary between I and II is fixed by
the arguments of the Heavyside functions in Eq.~(\ref{eq:apa2a}). According
to Eq.~(\ref{eq:apa8}) the velocity transfers $\Delta v_{\parallel }^{%
\mathrm{I}}$ and $\Delta v_{\parallel }^{\mathrm{II}}$ obey an exact
relation $\Delta v_{\parallel }^{\mathrm{I}}+\Delta v_{\parallel }^{\mathrm{%
II}}=-2v_{r\parallel }$.

We further consider the kinematically allowed range for the regimes I and
II. The dimensionless transversal energy transfer $\phi $\ can range $-\phi
_{\max }\leqslant \phi \leqslant 1$, where $\phi _{\max }=v_{0\bot
}^{2}/v_{r\parallel }^{2}$, $v_{0\bot }$ is the initial transversal relative
velocity. Here $\phi =-\phi _{\max }$\ ($\phi =1$) corresponds to a complete
transfer of the initial relative transversal (longitudinal) motion to the
parallel (transversal) one. Therefore the quantities $\Delta v_{\parallel }^{%
\mathrm{I}}$ and $\Delta v_{\parallel }^{\mathrm{II}}$ are restricted in the
domains
\begin{equation}
-v_{r\parallel }\leqslant \Delta v_{\parallel }^{\mathrm{I}}\leqslant
v_{r\parallel }\frac{\phi _{\max }}{\sqrt{1+\phi _{\max }}+1},
\label{eq:apa9}
\end{equation}%
\begin{equation}
-v_{r\parallel }\left( 1+\sqrt{1+\phi _{\max }}\right) \leqslant \Delta
v_{\parallel }^{\mathrm{II}}\leqslant -v_{r\parallel },  \label{eq:apa10}
\end{equation}%
where the boundary between I and II is fixed by $\Delta v_{\parallel
}=-v_{r\parallel }$.

The discussion above indicates that the second order perturbative treatment
for the repulsive interaction in the regime II can be improved if instead of
the standard approximation developed in the previous sections, the second
relation in Eq.~(\ref{eq:apa11}) is used. In particular, in the presence of
a strong magnetic field the modified second order velocity transfer reads
\begin{equation}
\Delta v_{\parallel }^{(2)}\simeq -2v_{r\parallel }+v_{r\parallel }\left[
\frac{\Delta E_{r\bot }^{(2)}}{\mu v_{r\parallel }^{2}}+\frac{1}{2}\left(
\frac{\Delta E_{r\bot }^{(1)}}{\mu v_{r\parallel }^{2}}\right) ^{2}\right] .
\label{eq:apa12}
\end{equation}%
Here $\Delta E_{r\bot }^{(1)}$ and $\Delta E_{r\bot }^{(2)}$\ are the first
and second order relative transversal energy transfers, respectively (see,
e.g., Eqs.~(\ref{eq:rel1})-(\ref{eq:rel3}) for collisions of two identical
particles). Note that the second term in Eq.~(\ref{eq:apa12}) within the
square brackets enters here by the opposite sign compared to the standard
perturbative treatment where the first (backscattering) term in Eq.~(\ref%
{eq:apa12}) is not involved. Since we assumed the conservation of the
relative energy of the particles these results are valid both for BC of two
identical particles and for ion--electron collisions (with $v_{i\bot }=0$).

\section{Comparison with simulations}
\label{sec:s4}

\subsection{Classical trajectory Monte--Carlo (CTMC) simulations}
\label{sec:ctmc}

A fully numerical treatment is required for applications beyond the
perturbative regime and for checking the validity of the perturbative
approach outlined above. In the present case of binary ion--electron or
electron-electron collisions in a magnetic field and with the effective
interaction $U_{\mathrm{R}}(r)$ (\ref{eq:a48}) the numerical evaluation of
the BC energy loss is very complicated, but can be successfully investigated
by classical trajectory Monte--Carlo (CTMC) simulations \cite%
{gzwi99,zwi00,zwi02}. In the CTMC method \cite{abr66} the
trajectories for the relative motion between the ion and an
electron are calculated by a numerical integration of the
equations of motion (i.e.~Eq.~(\ref{eq:a5}) for electron-electron
collisions and Eq.~(\ref{eq:a7}) for ion--electron collisions,
respectively), starting with initial conditions for the parallel
${v}_{r \parallel}$ and the transverse $\mathbf{v}_{0 \perp}$
relative velocity. The initial positions are chosen to correspond
to a certain impact parameter $\mathbf{s}$ and are located outside
the interaction zone, which is -- employing a screened interaction
like (\ref{eq:a48}) -- defined as a sphere of several screening
lengths $\lambda$ about the ion. The numerical calculation stops
after the electron has left this interaction zone, that is, when
the collision is completed. Deducing the velocity changes from the
initial and final velocities ${v}_{\parallel}$, $\mathbf{v}_{\perp}$, ${V}%
_{\parallel}$, $\mathbf{V}_{\perp}$ yields the energy transfer $\Delta E_{1}$
(\ref{eq:a21a}) (with $\Delta E_{\mathrm{r}} = 0$) or $\Delta E_{i}$ (\ref%
{eq:a21d}). The required accuracy is achieved by using a modified
Velocity--Verlet algorithm which has been specifically designed for particle
propagation in a (strong) magnetic field \cite{spr99,zwi08}, and by adapting
continuously the actual time--step by monitoring the constant of motion $E_{%
\mathrm{r}}$ (\ref{eq:a6-3}) or $K$ (\ref{eq:a7b}). The resulting relative
deviations of $E_{\mathrm{r}}$ or $K$ are of the order of $10^{-6}-10^{-5}$.

The desired average over the initial phases, i.e.~over the orientation of
the transverse relative velocity $\mathbf{v}_{0 \perp}$ and the impact
parameter $\mathbf{s}$ is performed by a Monte--Carlo sampling \cite%
{bin97,fis99} of a large number of trajectories with different initial
values. The actual number of computed trajectories is adjusted by monitoring
the convergence of the averaging procedure. Around $10^5 - 10^{6}$
trajectories are typically needed for the energy transfer for one set of
initial relative and cm velocities and at a given magnetic field.

\subsection{Results}
\label{sec:results}

For the forthcoming discussion we put the equation of the relative motion of
two particles in a more appropriate dimensionless form by scaling lengths in
units of the screening length $\lambda $ and velocities in units of a
characteristic velocity $v_{s}$ defined by
\begin{equation}
v_{s}^{2}=\frac{\vert q_{1}q_{2}\vert e\!\!\!/^{2}}{\mu \lambda }.
\label{eq:a70}
\end{equation}%
Let us recall that $q_1 =q_2 =-1$, $\mu =m/2$ and $q_1 =-1$, $q_2 =Z$, $\mu
=m$ for electron-electron and ion--electron collisions, respectively. This
velocity $v_{s}$ gives a measure for the strength of the Coulomb interaction
with respect to the (initial) kinetic energy of relative motion $\mu
v_{r}^{2}/2$. For $v_{r}<v_{s}$ the kinetic energy is small compared to the
characteristic potential energy $\vert q_1 q_2 \vert e\!\!\!/^{2}/\lambda $
in a screened Coulomb potential and we expect to be in a non-perturbative
regime. A perturbative treatment on the other hand should be applicable for $%
v_{r}\gg v_{s}$.

\begin{figure}[tbp]
\includegraphics[width=8cm]{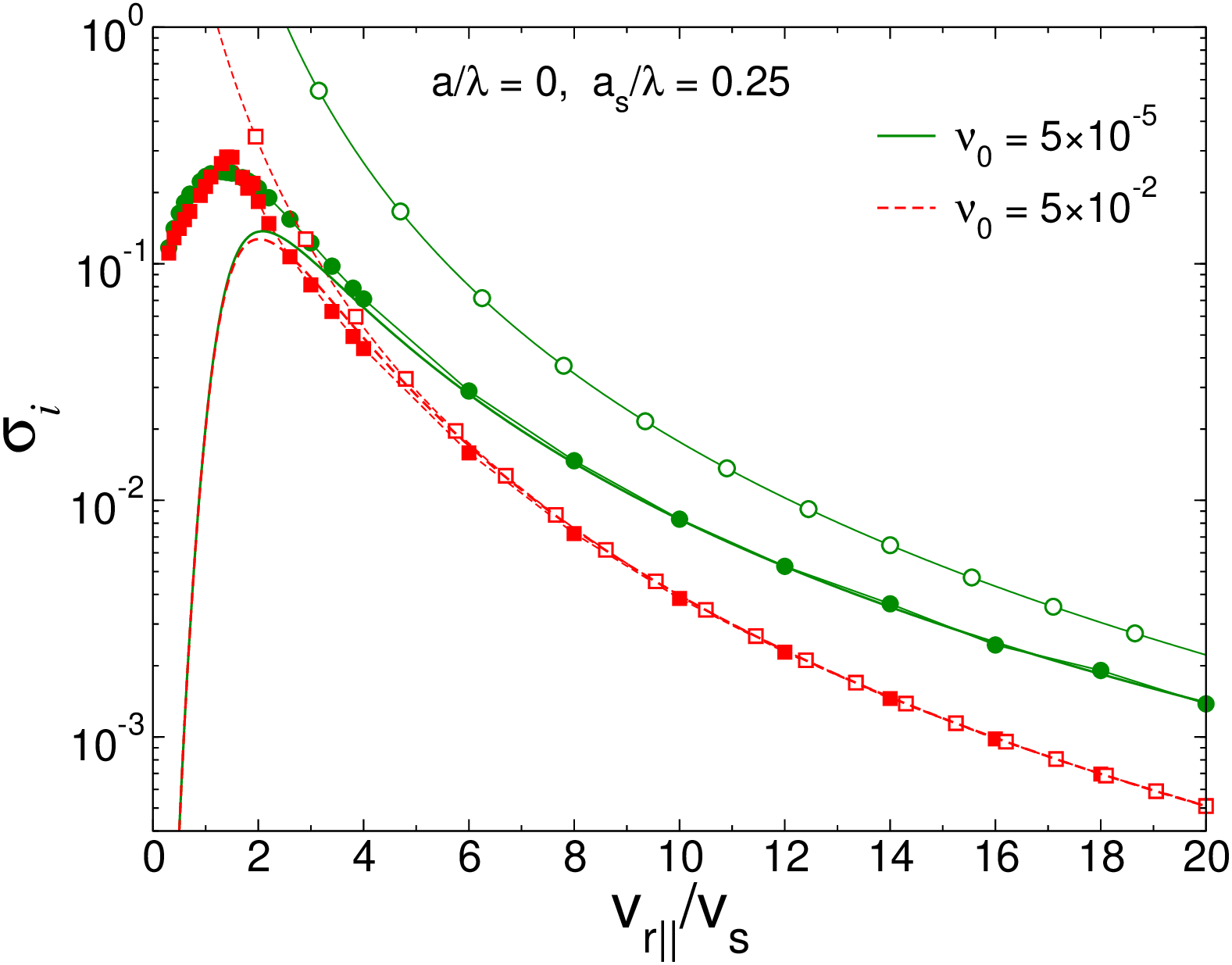} %
\includegraphics[width=8cm]{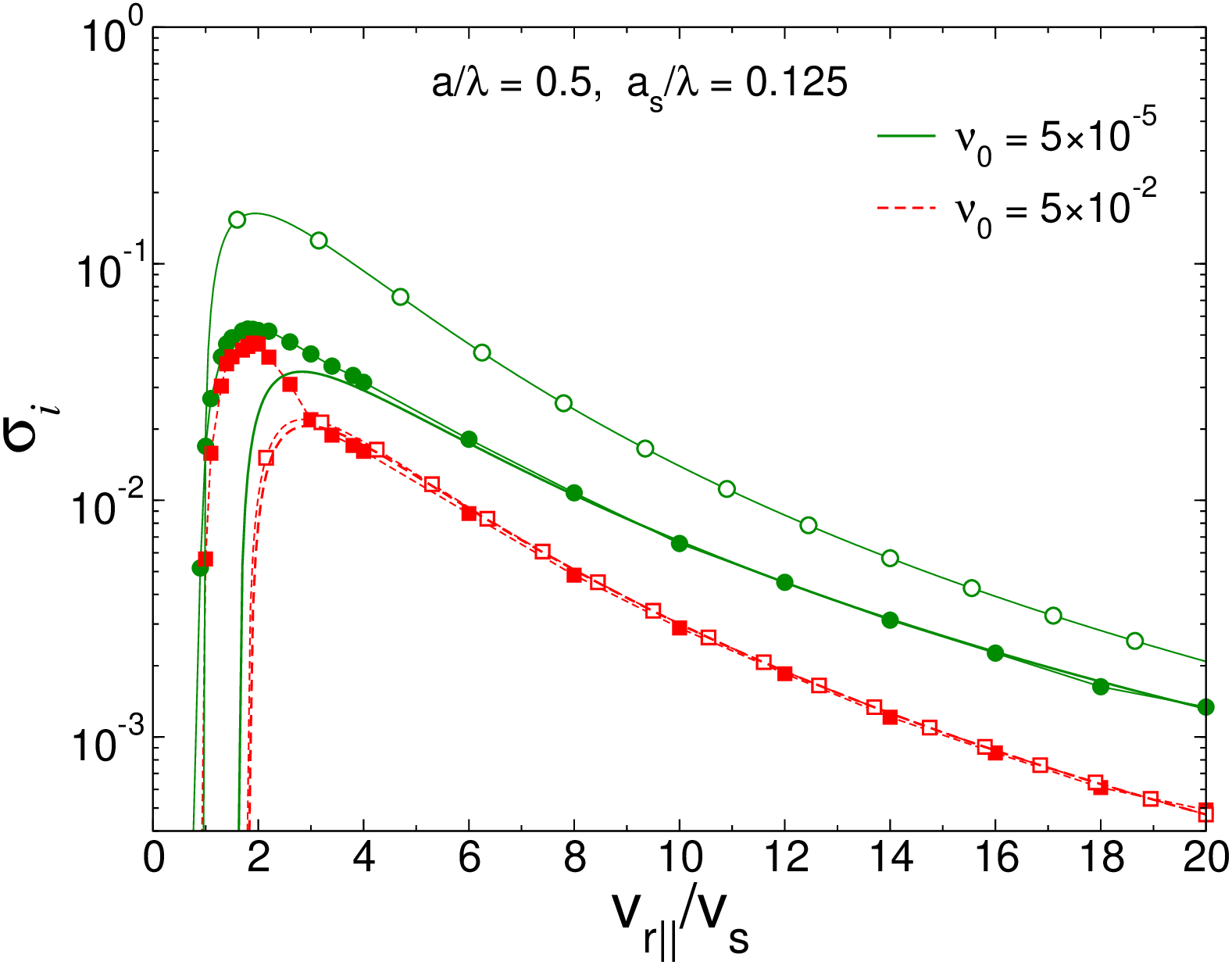}
\caption{(Color online) The cross-section $\sigma_{i}$
Eq.~(\ref{eq:sig3}) for ion--electron (with $Z>0$) collisions in
terms of the
dimensionless relative velocity component $v_{r\parallel}/v_{s}$ at $a=0$, $%
a_{s}/\lambda =0.25$ (left panel) and $a/\lambda =0.5$, $%
a_{s}/\lambda =0.125$ (right panel). The solid and dashed curves
were obtained for $\nu_{0}=\lambdabar_{0}/\lambda %
=5\times10^{-5}$ and $\nu_{0}=5\times 10^{-2}$, respectively. The
curves with filled symbols are from CTMC simulations. The curves
with open
symbols represent the second order perturbative treatment with constant $%
\lambdabar=\lambdabar_{0}$. The results of the second order
perturbative treatment with a modified (dynamical) cutoff
parameter $\lambdabar (v_{r\parallel})$ (see
Eq.~(\ref{eq:lambda_ml})) are given by the curves without
symbols.}
\label{fig:kap1}
\end{figure}

The scaled version of equations (\ref{eq:a5}) or (\ref{eq:a7}) only depends
on the two dimensionless parameters $a_s/\lambda$ and $\lambdabar/\lambda$,
and the initial conditions with positions scaled in $\lambda$ and velocities
in $v_s$. Here $a_s = v_s/\omega_{c}$ is the cyclotron radius for $v_{\perp}
= v_s$ and the parameter $a_s/\lambda \propto v_s/B$ represents a measure
for the strength of the magnetic field compared to the strength of the
Coulomb interaction (which is $\propto v_s^2$, see (\ref{eq:a70})). The
ratio $\lambdabar/\lambda$ describes the amount of softening of the screened
interaction at $r\to 0$ with $q_{1}q_{2}e\!\!\!/ ^2 U_{\mathrm{R}}(r\to 0)
\to {q_{1}q_{2}e\!\!\!/ ^2}/\lambdabar$.

\begin{figure}[tbp]
\includegraphics[width=8cm]{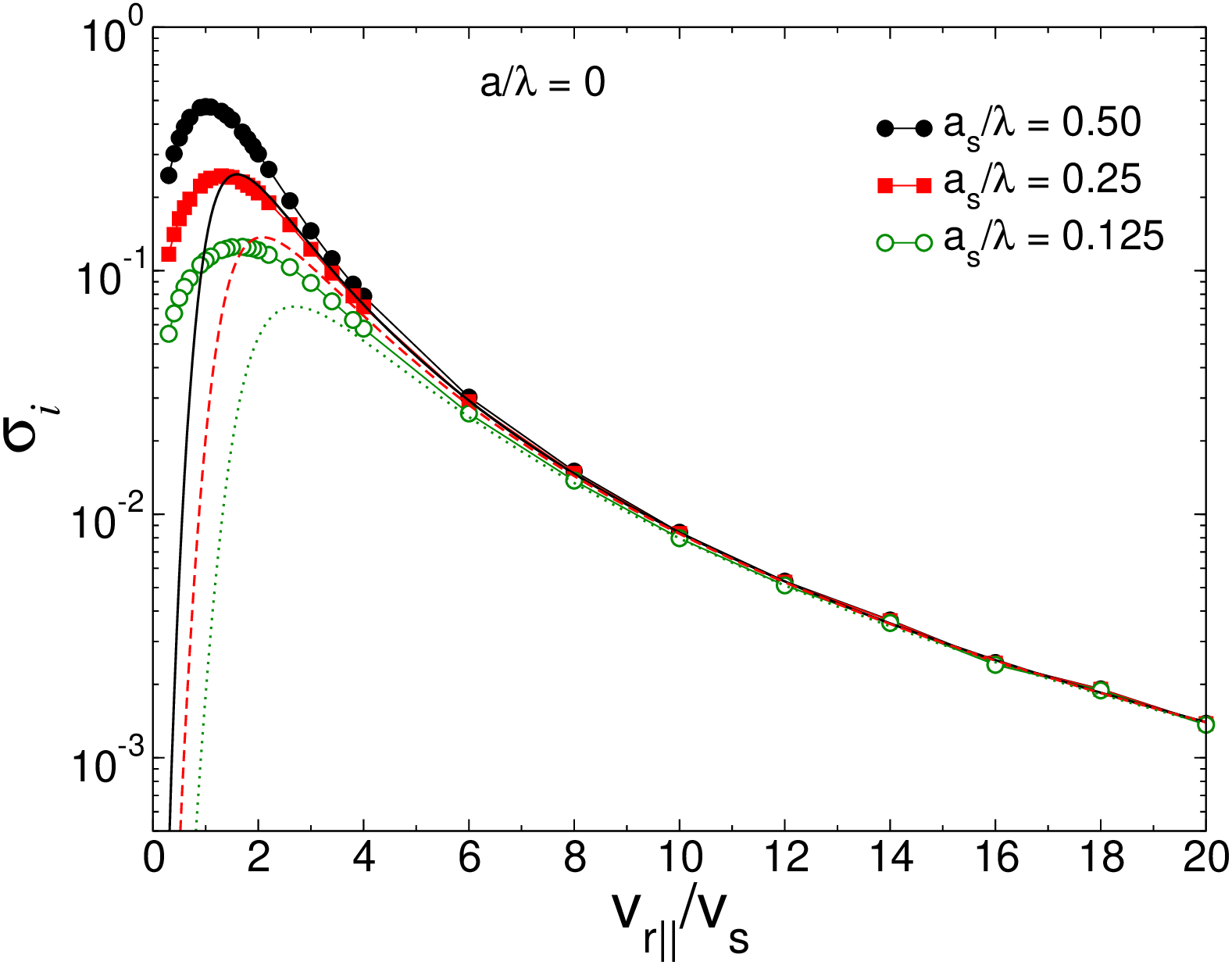} %
\includegraphics[width=8cm]{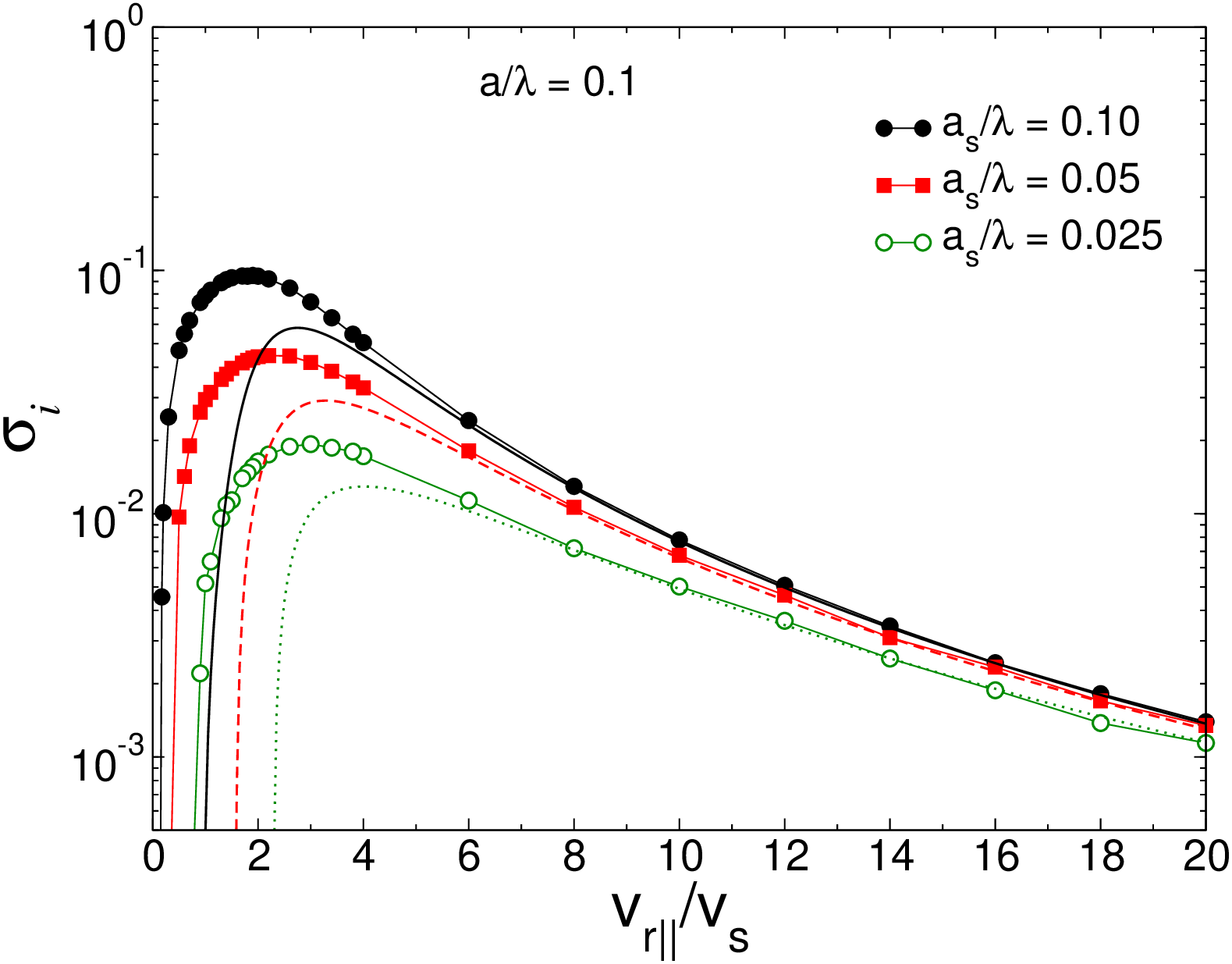}
\caption{(Color online) The cross-section $\sigma_{i}$
Eq.~(\ref{eq:sig3}) for ion--electron (with $Z>0$) collision in
terms of dimensionless relative velocity component
$v_{r\parallel}/v_{s}$ at $a=0$ (left panel) and $a/\lambda =0.1$
(right panel). The curves with and without symbols correspond to
CTMC simulations and the second order
perturbative treatment, respectively. Left panel, $a_{s}/\lambda %
=0.5 $ (solid curve), $a_{s}/\lambda =0.25$ (dashed curve), $a_{s}/%
\lambda =0.125$ (dotted curve). Right panel, $a_{s}/ \lambda %
=0.1$ (solid curve), $a_{s}/\lambda =0.05$ (dashed curve), $a_{s}/
\lambda =0.025$ (dotted curve).} \label{fig:1}
\end{figure}
\begin{figure}[tbp]
\includegraphics[width=8cm]{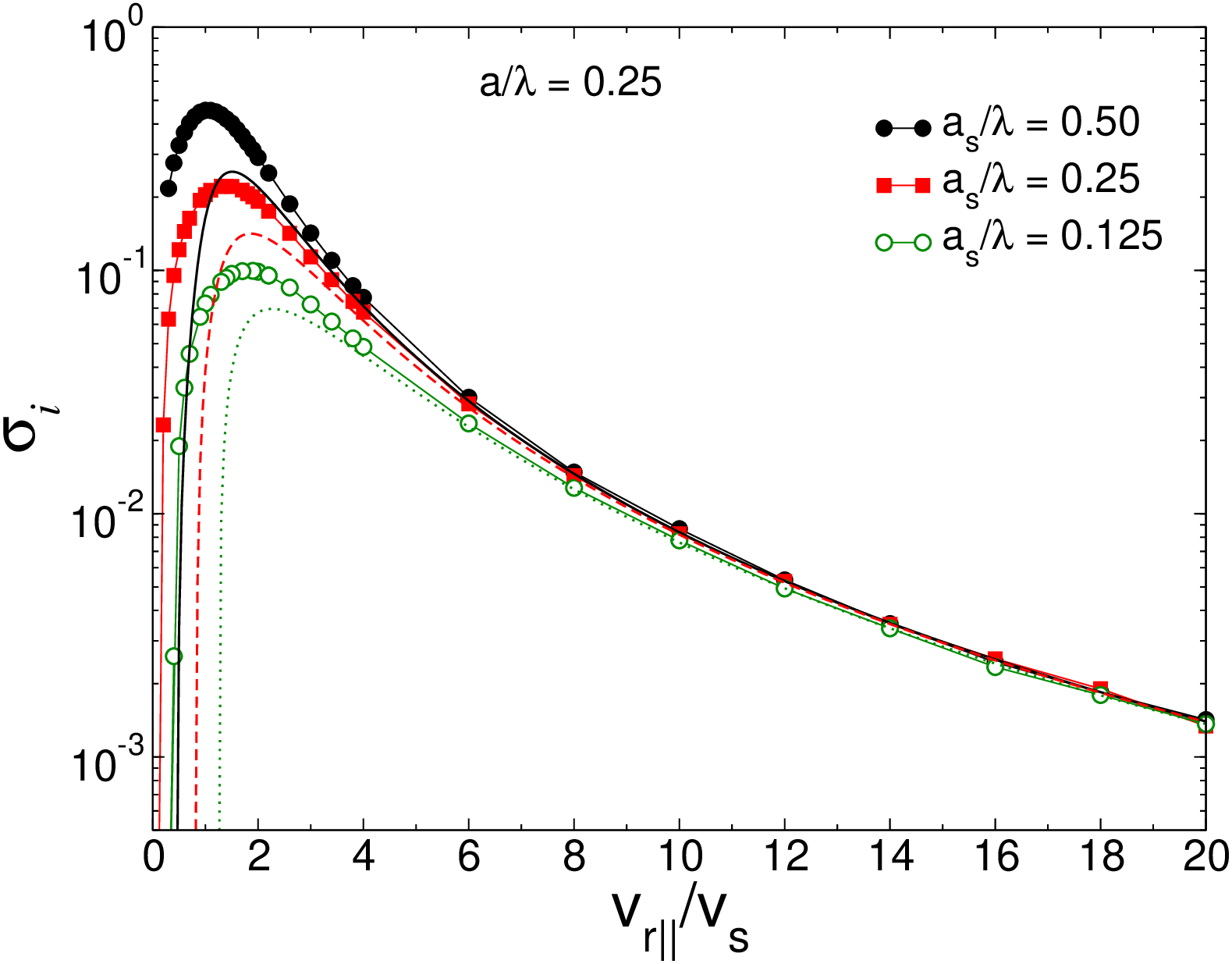} %
\includegraphics[width=8cm]{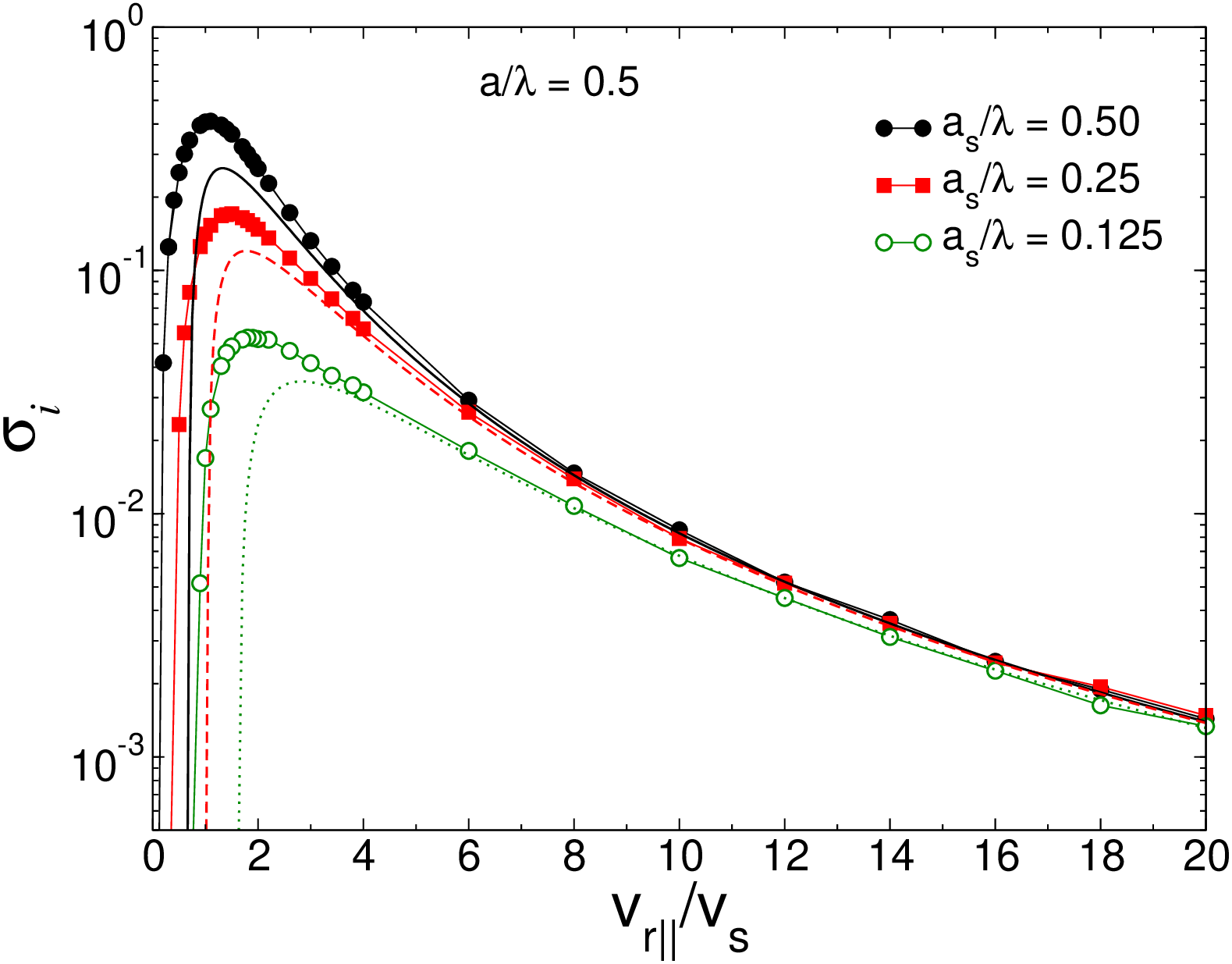}
\caption{(Color online) Same as in Fig.~\ref{fig:1} but for $a/%
\lambda=0.25$ (left panel) and $a/\lambda =0.5$ (right
panel). In left and right panels, $a_{s}/\lambda =0.5 $ (solid), $%
a_{s}/\lambda =0.25$ (dashed), $a_{s}/\lambda =0.125$ (dotted).}
\label{fig:2}
\end{figure}

In the analytical perturbative approach we thus apply the same scaling of
length and velocities and introduce for electron-electron collisions the
dimensionless cross-sections
\begin{eqnarray}
\sigma _{\parallel } =-\frac{1}{m V_{0\parallel }v_{s}\lambda ^{2}}%
\int_{0}^{2\pi }\frac{d\varphi }{2\pi }\overline{\sigma }_{\parallel }\left(
\varphi \right) ,  \label{eq:sig1} \\
\sigma _{\bot } =-\frac{1}{m v_{s}^{2}\lambda ^{2}}\int_{0}^{2\pi }\frac{%
d\varphi }{2\pi }\overline{\sigma }_{\bot }\left( \varphi \right) .
\label{eq:sig2}
\end{eqnarray}
The corresponding dimensionless cross-section for ion--electron interaction
is given by
\begin{equation}
\sigma _{i} =\frac{1}{mv_{i\parallel} v_{s} \lambda^{2}}
\int_{0}^{\infty} \langle\Delta E_{i} \rangle sds =-\frac{1}{v_{s}
\lambda^{2}} \int_{0}^{\infty} \langle\Delta v_{\parallel} \rangle
sds .  \label{eq:sig3}
\end{equation}
It should be noted that in the regimes where the cross-section $\overline{%
\sigma}_{\parallel}(\varphi)$ is not strongly sensitive with respect to the
initial phases $\varphi$ the quantity $\sigma _{i}$ in Eq.~(\ref{eq:sig3})
is approximately given by $\sigma_{i}\simeq 2\sigma _{\parallel}$ (see also
the relation (\ref{eq:a43_v2})) in the units introduced above. An example of
such regime has been considered in Sec.~\ref{sec:s3.2} where $\overline{%
\sigma}_{\parallel}(\varphi)$ in the high-velocity limit and in leading
order is independent on $\varphi$, see Eq.~(\ref{eq:a69b}). One can expect
that the approximate relation between the ion--electron and
electron-electron cross-sections is valid within second order perturbative
treatment with arbitrary $Z$ and within exact CTMC simulations, where,
however, the ion--electron interaction must be repulsive ($Z<0$).

\begin{figure}[tbp]
\includegraphics[width=8cm]{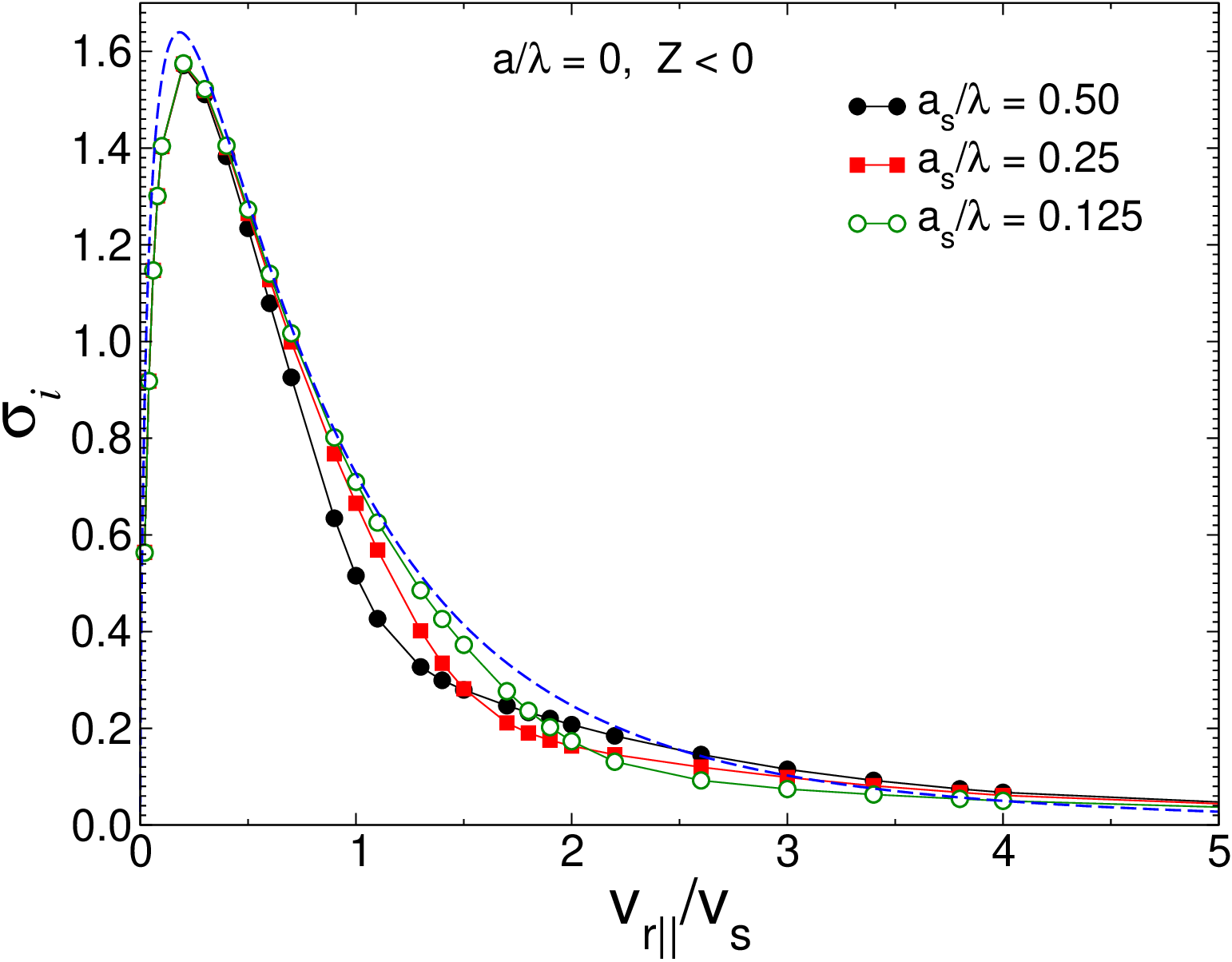} %
\includegraphics[width=8cm]{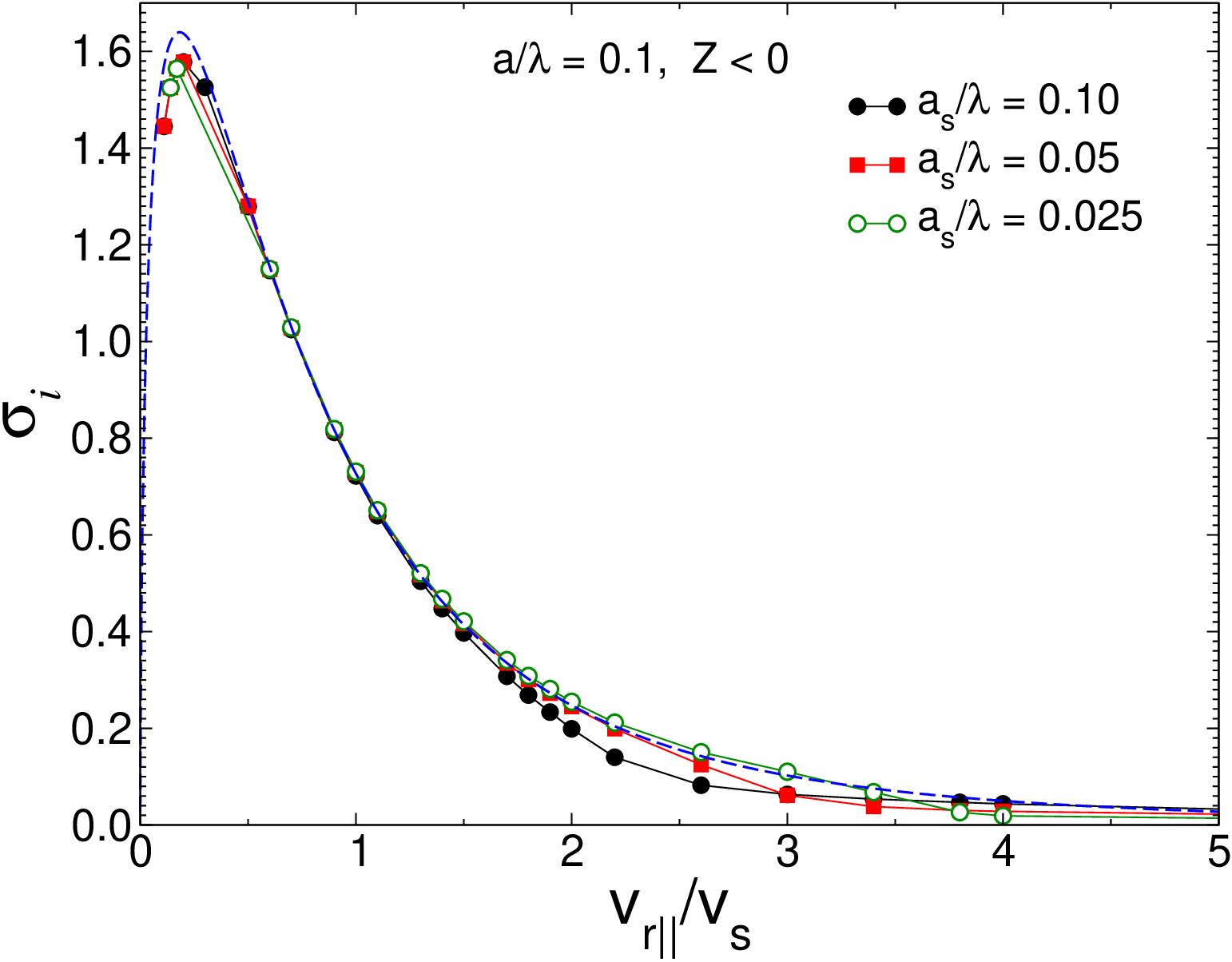}
\caption{(Color online) The cross-section $\sigma_{i}$
Eq.~(\ref{eq:sig3}) for ion--electron collision with $Z<0$ at
$a=0$ (left panel) and $a/\lambda =0.1$ (right panel). The curves
with symbols correspond to CTMC simulations. Left panel,
$a_{s}/\lambda =0.5 $
(filled circles), $a_{s}/\lambda =0.25$ (squares), $a_{s}/%
\lambda =0.125$ (open circles). Right panel, $a_{s}/\lambda =0.1$
(filled circles), $a_{s}/\lambda =0.05$ (squares), $a_{s}/%
\lambda =0.025$ (open circles). The dashed curves are obtained employing
Eq.~(\ref{eq:apa3}) with constant cutoff parameter $\nu =%
\nu_{0} =5\times 10^{-5}$.}
\label{fig:3}
\end{figure}
\begin{figure}[tbp]
\includegraphics[width=8cm]{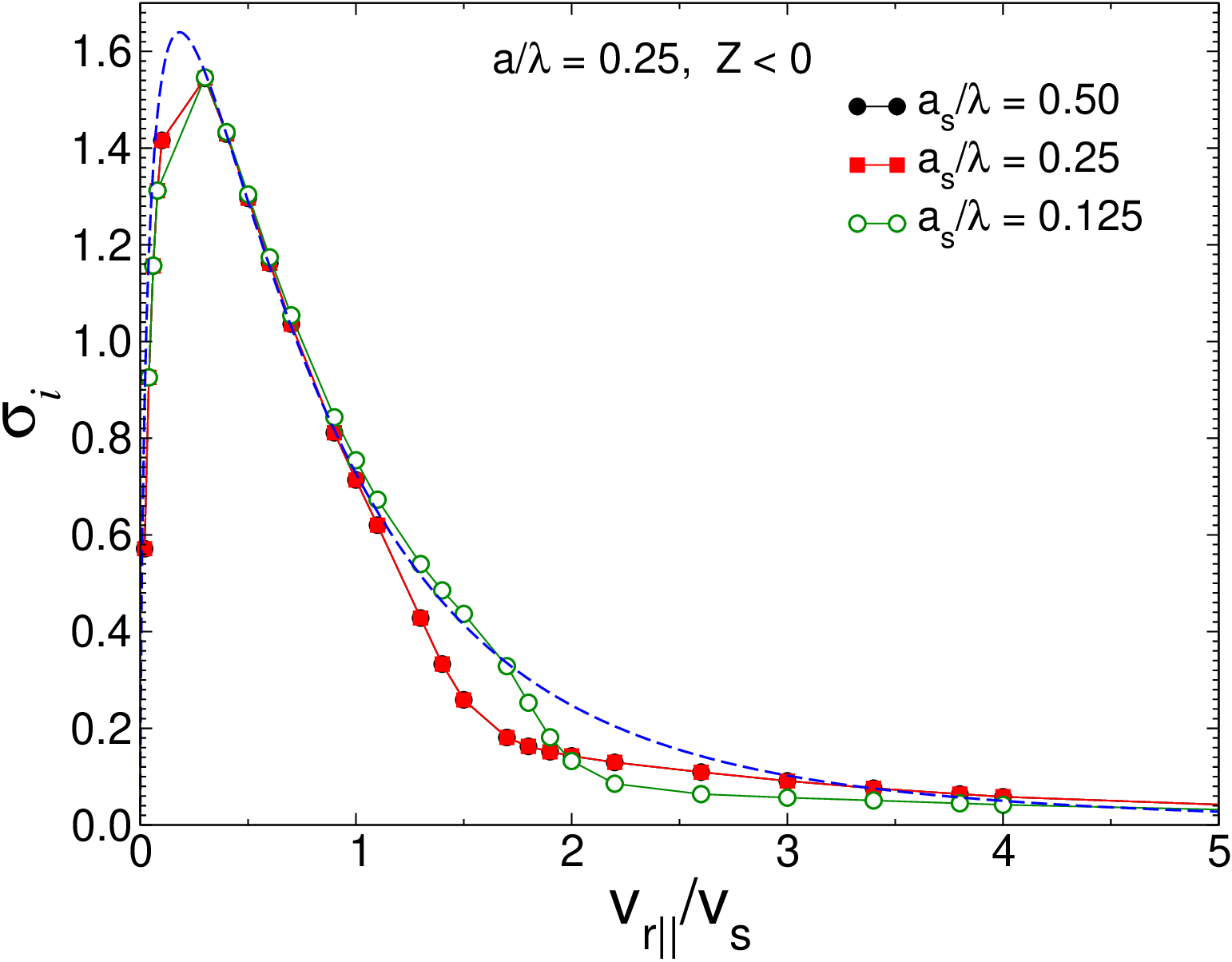} %
\includegraphics[width=8cm]{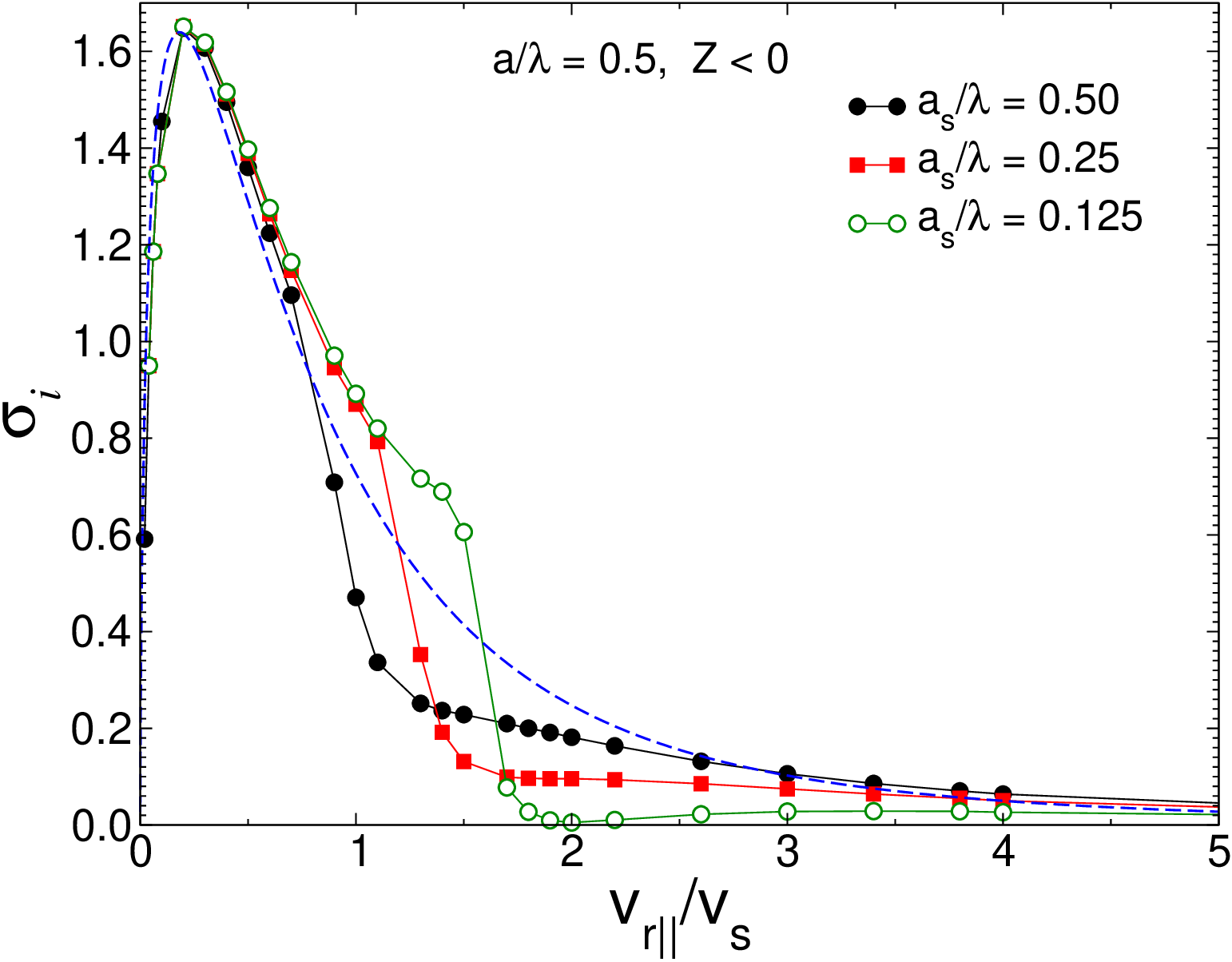}
\caption{(Color online) Same as in Fig.~\ref{fig:3} but for $a/%
\lambda =0.25$ (left panel) and $a/\lambda =0.5$ (right panel). In
left and right panels, $a_{s}/\lambda =0.5 $ (filled
circles), $a_{s}/\lambda =0.25$ (squares), $a_{s}/\lambda %
=0.125$ (open circles).}
\label{fig:4}
\end{figure}

Next we specify the cutoff parameter $\lambdabar$ which is a measure of
softening of the interaction potential at short distances. As we discussed
in previous sections the regularization in the potential (\ref{eq:a48}) is
sufficient to guarantee the existence of the $s$-integrated energy
transfers, see e.g. Eq.~(\ref{eq:a43}), but there remains the problem of
treating hard collisions. For a perturbation treatment the change in
relative velocity must be small compared to $v_{r}$ and this condition is
increasingly difficult to fulfill in the regime $v_{r}\to 0$. This suggests
a physically reasonable procedure: The potential must be softened near the
origin. In fact the parameter $\lambdabar$ which describes the effects of
quantum diffraction should be related to the de Broglie wavelength which is
inversely proportional to $v_{r}$. Here within classical picture of
collisions we employ in a perturbative treatment the dynamical cutoff
parameter $\varkappa (v_{r\parallel})=1+\lambda /\lambdabar (v_{r\parallel})$%
, where
\begin{equation}
\lambdabar^{2}(v_{r\parallel})=Cb^{2}_{0}(v_{r\parallel})+\lambdabar^{2}_{0}
, \quad b_{0}(v_{r\parallel})=\frac{\vert q_{1}q_{2}\vert e\!\!\!/^{2}}{\mu
(v^{2}_{r\parallel}+v^{2}_{0\perp})} .  \label{eq:lambda_ml}
\end{equation}
In the case of ion--electron collision $\vert q_{1}q_{2}\vert$, $\mu$ and $%
v_{0\perp}$ in Eq.~(\ref{eq:lambda_ml}) have to be replaced by $\vert Z\vert$%
, $m$ and $v_{e\perp}$, respectively. Here $\lambdabar_{0}$ is some constant
cutoff parameter, and $b_{0}(v_{r\parallel})$ is the distance of closest
approach of two charged particles in the absence of a magnetic field. Also
in Eq.~(\ref{eq:lambda_ml}) we have introduced an additional fitting
parameter $C$. For determining $C$ we consider the second order transport
cross-section $\sigma_{\mathrm{tr}}=-(2\pi
/\mu v_{r\parallel}V_{0\parallel})\sigma $, where $\sigma$ is given by Eq.~(%
\ref{eq:a43}). For the regularized potential and in high-velocity limit $%
\sigma $ is given by Eq.~(\ref{eq:a69b}), where the generalized Coulomb
logarithm $\Lambda (\varkappa)$ and the cutoff $\varkappa=\varkappa
(v_{r\parallel})$ are determined by Eqs.~(\ref{eq:ap5}) and (\ref%
{eq:lambda_ml}), respectively. Setting $\lambdabar_{0}=0$ the obtained
high-velocity transport cross-section is then compared with an exact
asymptotic expression derived in Ref.~\cite{hah71} for the Yukawa-type (i.e.
with $\lambdabar \to 0$) interaction potential which yields $C=e^{2\gamma
-1}/4\simeq 0.292$, where $\gamma$ is Euler's constant. As will be shown
below the second order cross-sections with dynamical cutoff parameter (\ref%
{eq:lambda_ml}) excellently agree with CTMC simulations at high velocities.
The CTMC simulations have been carried out with constant $\lambdabar
=\lambdabar_{0}\ll \lambda$ (or $\varkappa = \varkappa_{0}\simeq
\lambda/\lambdabar_{0} \gg 1$), that is, the interaction is almost Coulomb
at short distances. As an example in Fig.~\ref{fig:kap1} we compare the
cross-sections $\sigma_{i}$ obtained with CTMC simulations (curves with
filled symbols) and within second order perturbative treatment either with
constant $\lambdabar =\lambdabar_{0}$ (open symbols) or dynamical cutoff
parameters (\ref{eq:lambda_ml}) (curves without symbols). Two distinct cases
are considered here with $\nu_{0}=\lambdabar_{0}/\lambda =5\times10^{-5}$
and $\nu_{0}=5\times 10^{-2}$. Also $a=0$ (no cyclotron motion at the
initial state), $a_{s}/\lambda=0.25$ and $a/\lambda=0.5$, $%
a_{s}/\lambda=0.125$ in left and right panels, respectively. It is seen that
with respect to the 'softness' of the regularized potential the agreement
between CTMC and the perturbative treatment improves, keeping all other
parameters fixed, for increasing $\lambdabar /\lambda$, that is a weaker
interaction potential $U_{\mathrm{R}}$.

\begin{figure}[tbp]
\includegraphics[width=8cm]{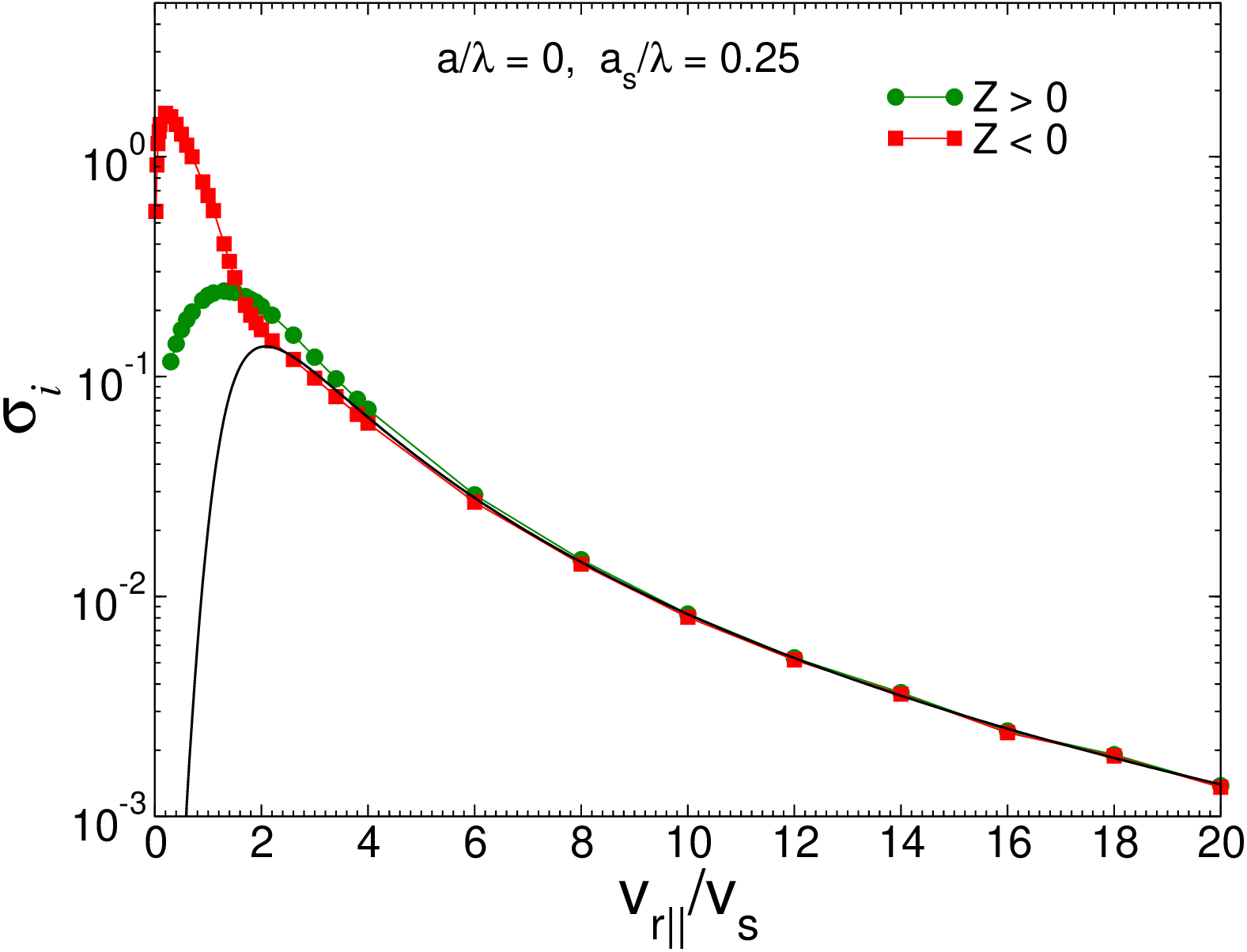} %
\includegraphics[width=8cm]{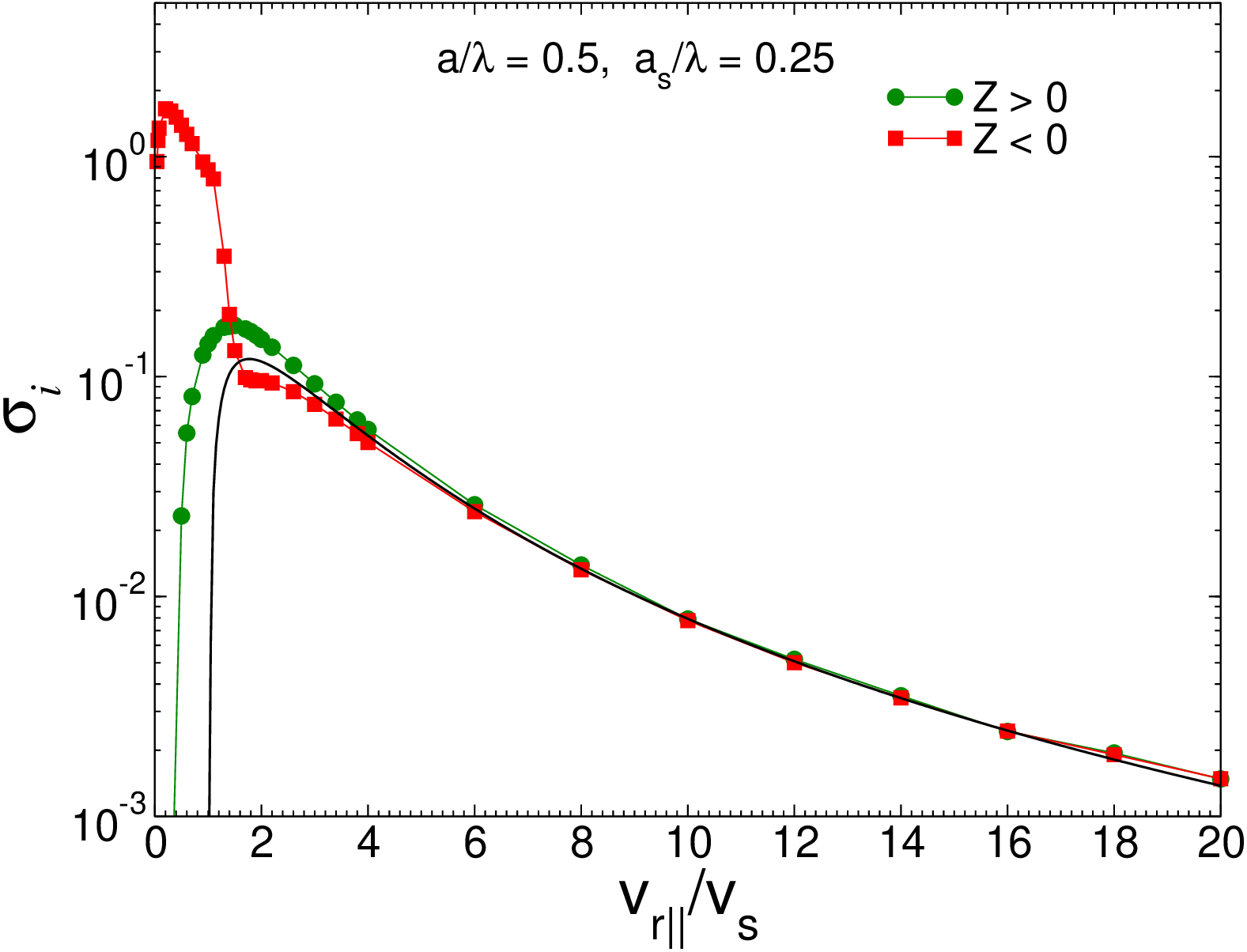}
\caption{(Color online) Comparison of the CTMC simulations for the
attractive (circles) and repulsive (squares) ion--electron interactions at $%
a_{s}/\lambda =0.25$. The solid curves correspond to the second
order theory with modified cutoff parameter. In left and right
panels $a=0$ and $a/\lambda =0.5$, respectively.}
\label{fig:at_vs_rep}
\end{figure}
\begin{figure}[tbp]
\includegraphics[width=8cm]{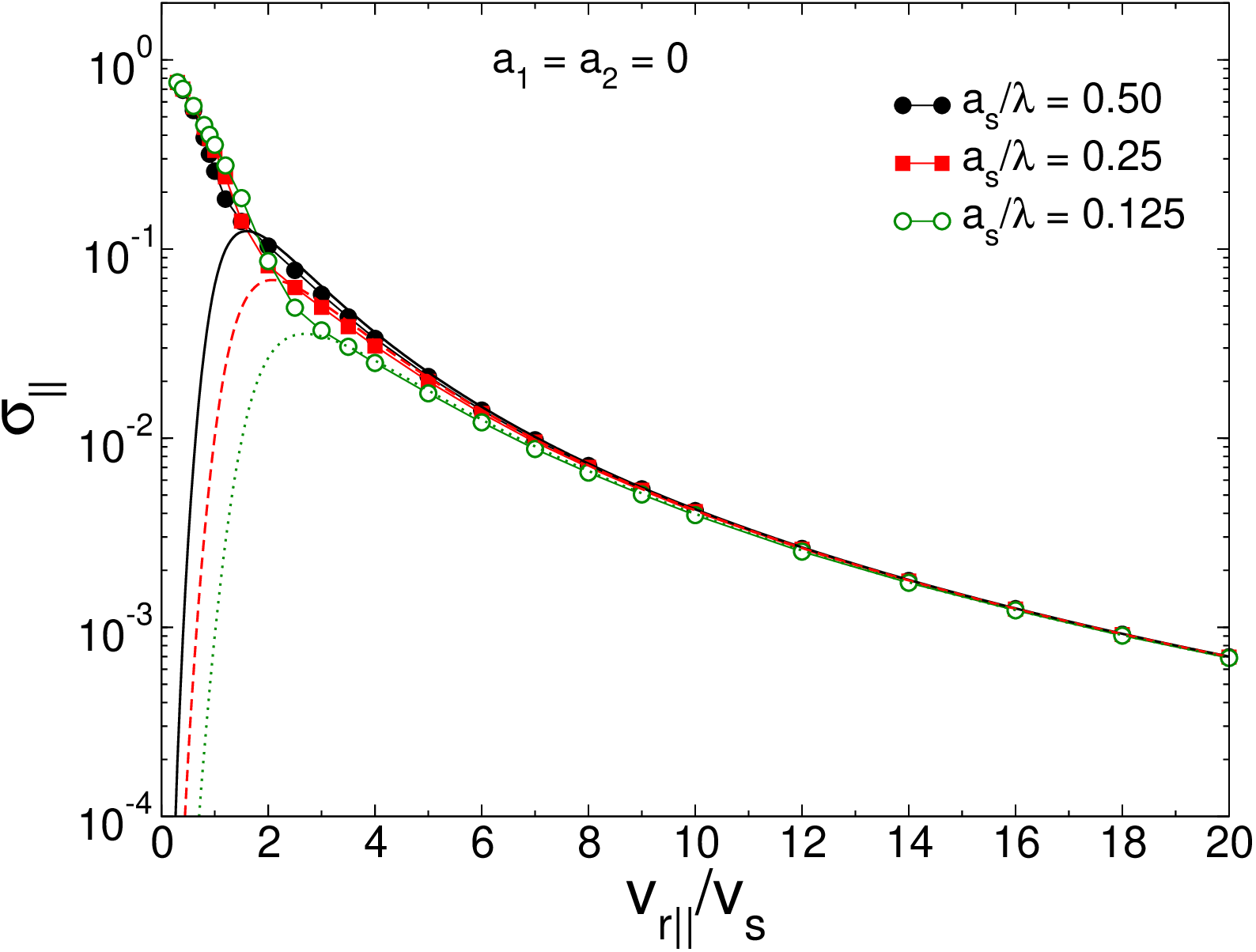}
\includegraphics[width=8cm]{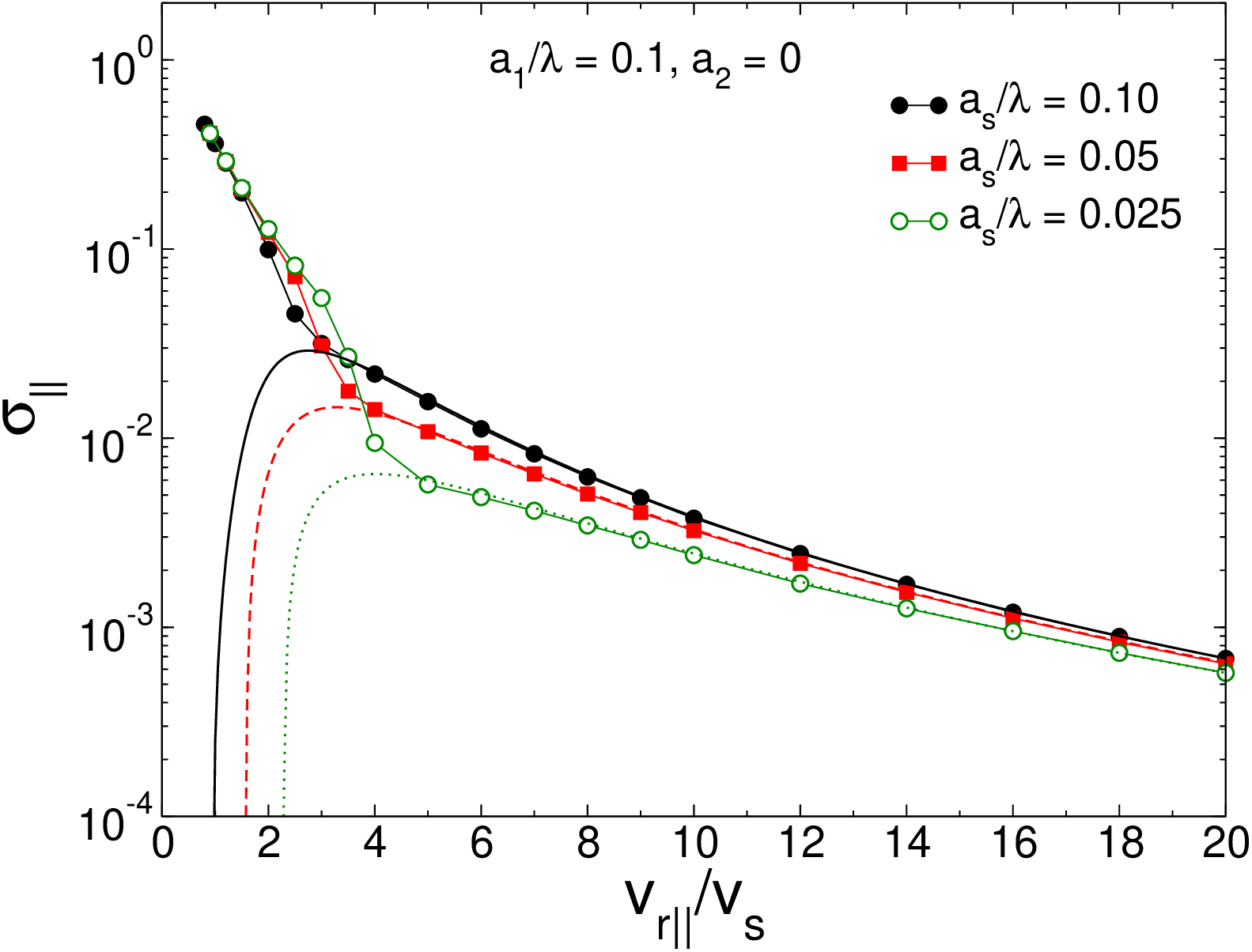}
\caption{(Color online) The cross-section $\sigma_{\parallel}$ Eq.~(%
\ref{eq:sig1}) for electron-electron collision in terms of
dimensionless relative velocity component $v_{r\parallel}/v_{s}$
for $a_{2}=0 $ and at $a_{1}=0$ (left panel) and $a_{1}/\lambda
=0.1$ (right panel). The curves with and without symbols
correspond to CTMC simulations
and the second order perturbative treatment, respectively. Left panel, $%
a_{s}/\lambda =0.5 $ (solid), $a_{s}/\lambda =0.25$ (dashed),
$a_{s}/\lambda =0.125$ (dotted). Right panel, $a_{s}/
\lambda =0.1$ (solid), $a_{s}/\lambda =0.05$ (dashed), $%
a_{s}/\lambda =0.025$ (dotted).}
\label{fig:5}
\end{figure}
\begin{figure}[tbp]
\includegraphics[width=8cm]{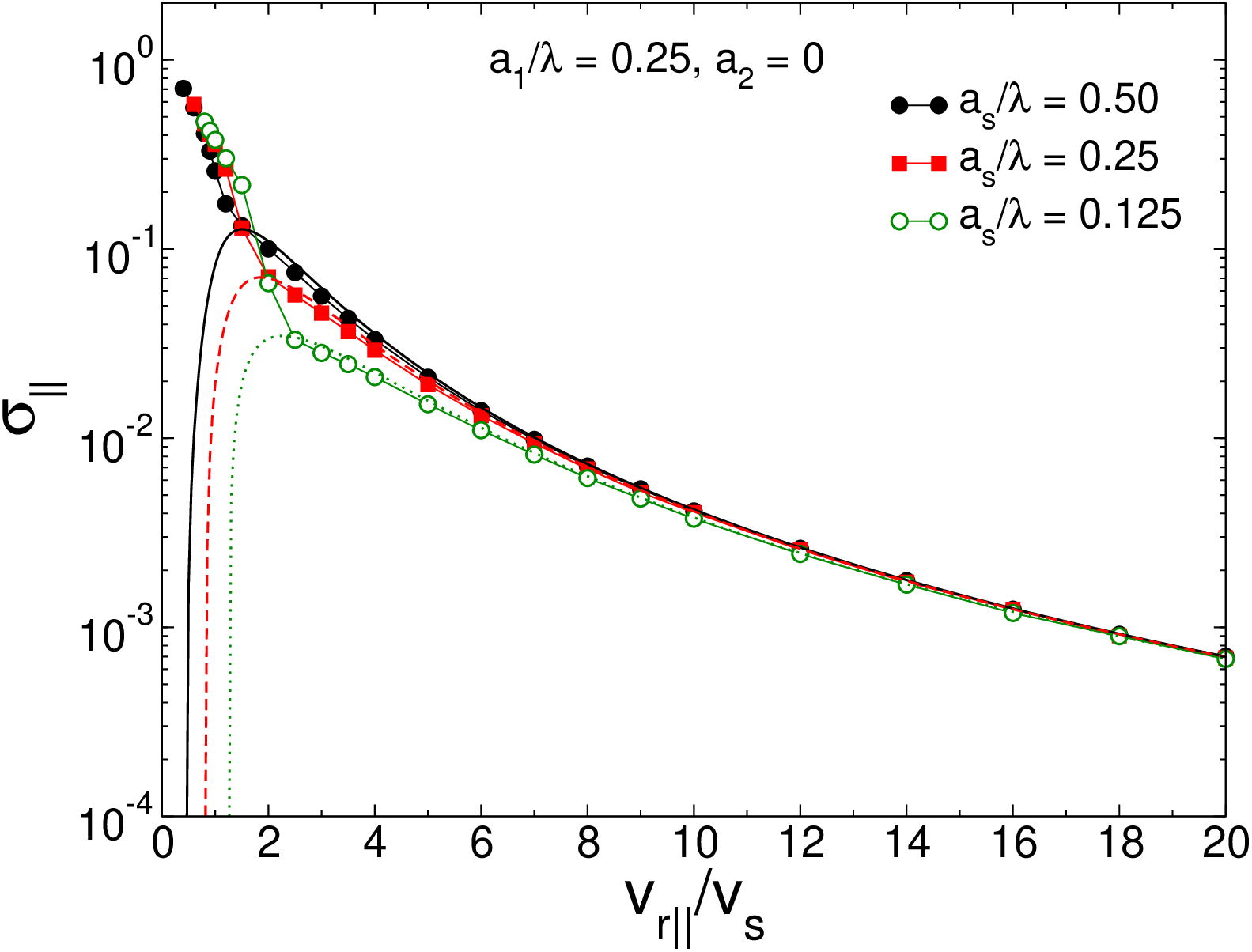}
\includegraphics[width=8cm]{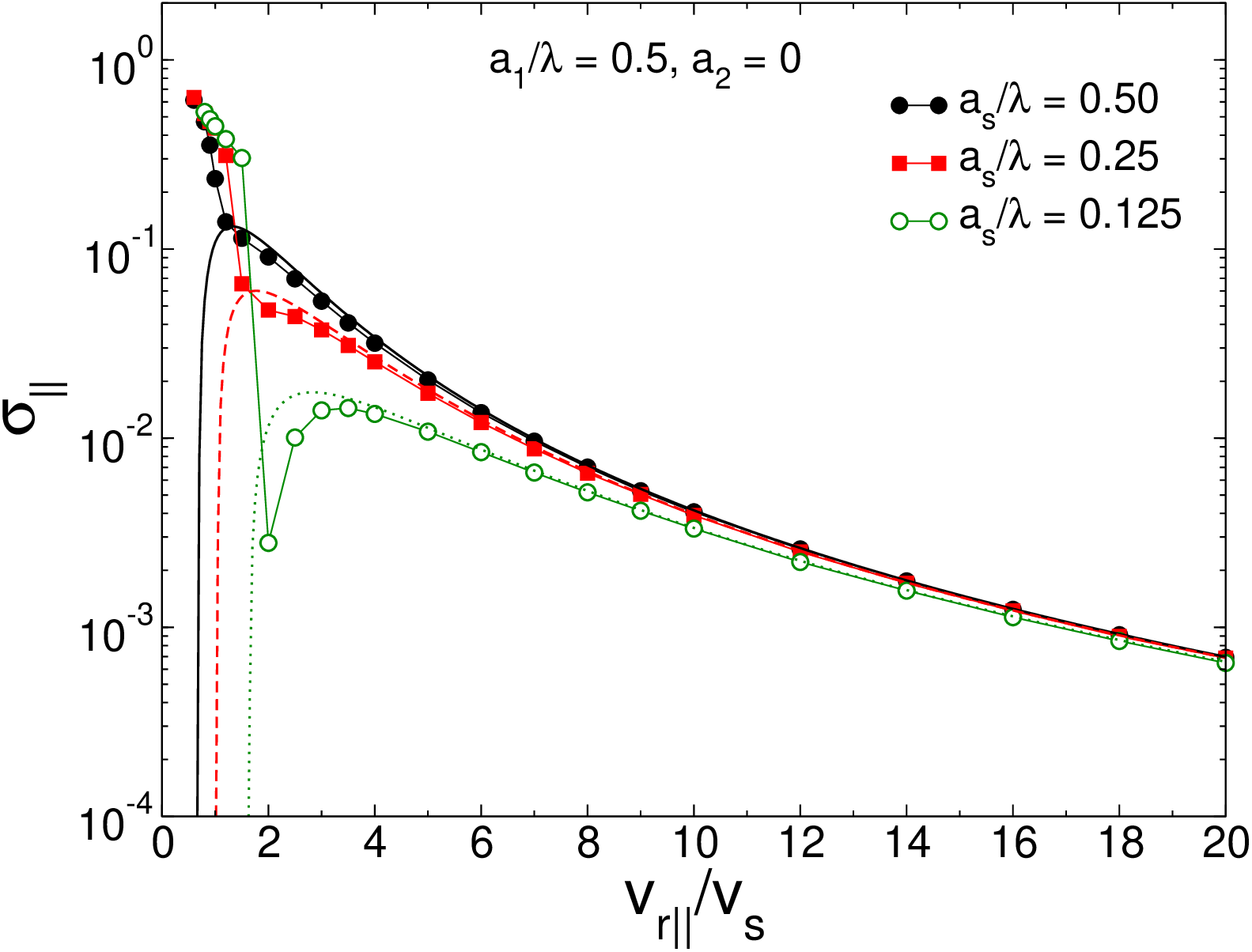}
\caption{(Color online) Same as in Fig.~\ref{fig:5} but for $a_{1}/%
\lambda=0.25$ (left panel) and $a_{1}/\lambda =0.5$ (right
panel). In left and right panels, $a_{s}/\lambda =0.5 $ (solid), $%
a_{s}/\lambda =0.25$ (dashed), $a_{s}/\lambda =0.125$ (dotted).}
\label{fig:6}
\end{figure}
\begin{figure}[tbp]
\includegraphics[width=8cm]{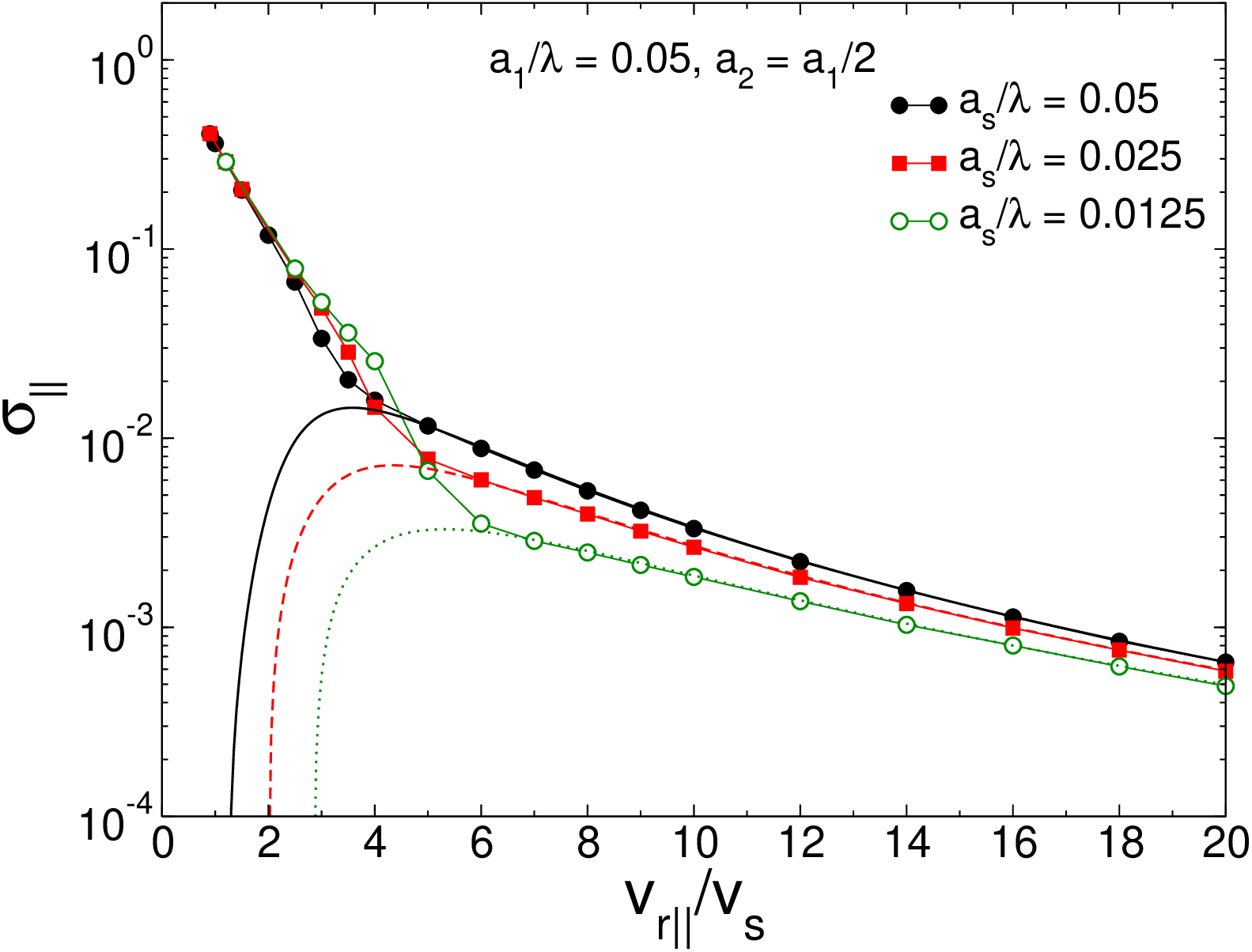} %
\includegraphics[width=8cm]{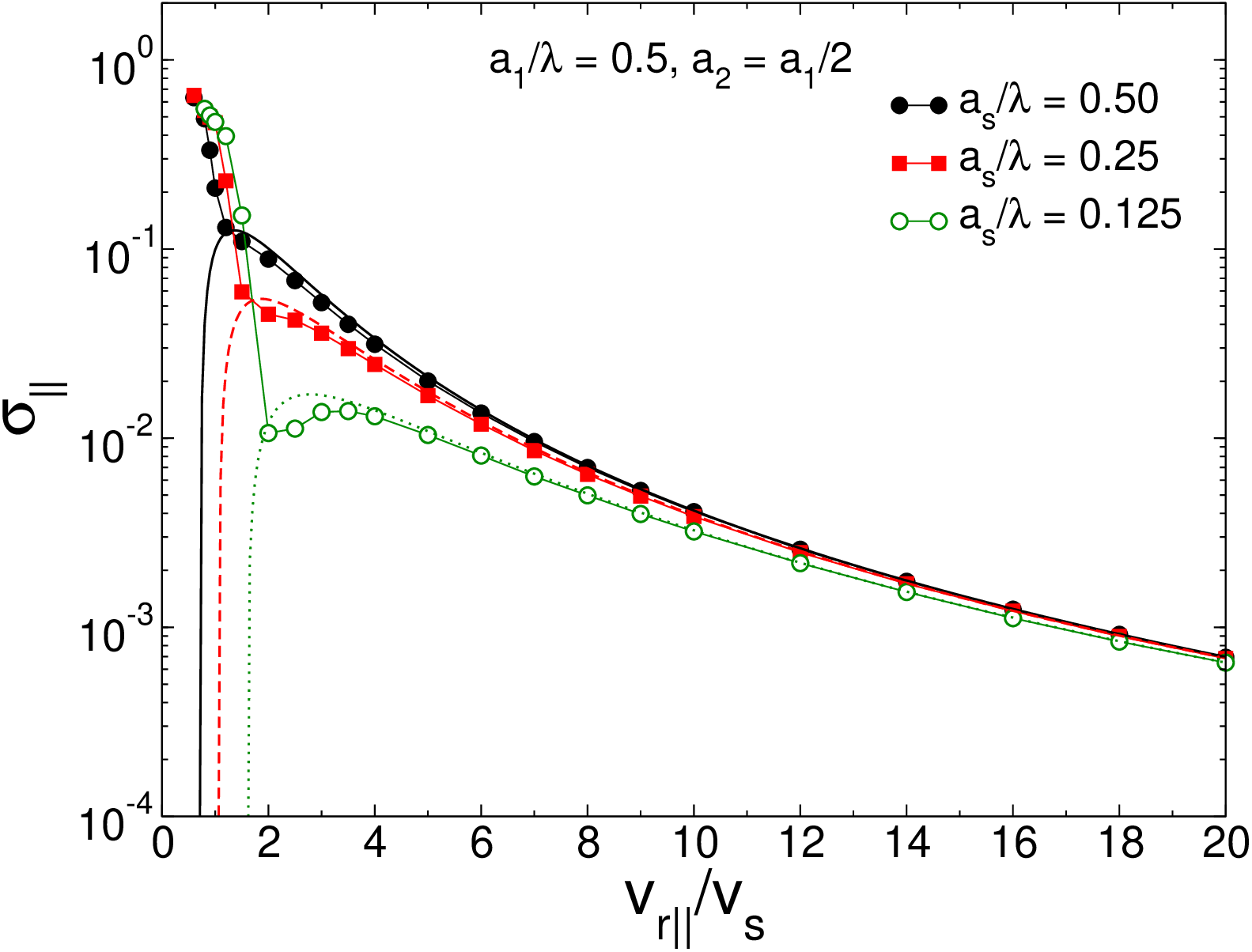}
\caption{(Color online) Same as in Fig.~\ref{fig:5} but for $%
a_{2}=a_{1}/2$ with $a_{1}/\lambda=0.05$ (left panel) and $a_{1}/%
\lambda =0.5$ (right panel). Left panel, $a_{s}/\lambda =0.05 $
(solid curve), $a_{s}/\lambda =0.025$ (dashed curve), $a_{s}/
\lambda =0.0125$ (dotted curve). Right panel, $a_{s}/\lambda %
=0.5$ (solid), $a_{s}/\lambda =0.25$ (dashed), $a_{s}/%
\lambda =0.125$ (dotted).}
\label{fig:7}
\end{figure}

\begin{figure}[tbp]
\includegraphics[width=8cm]{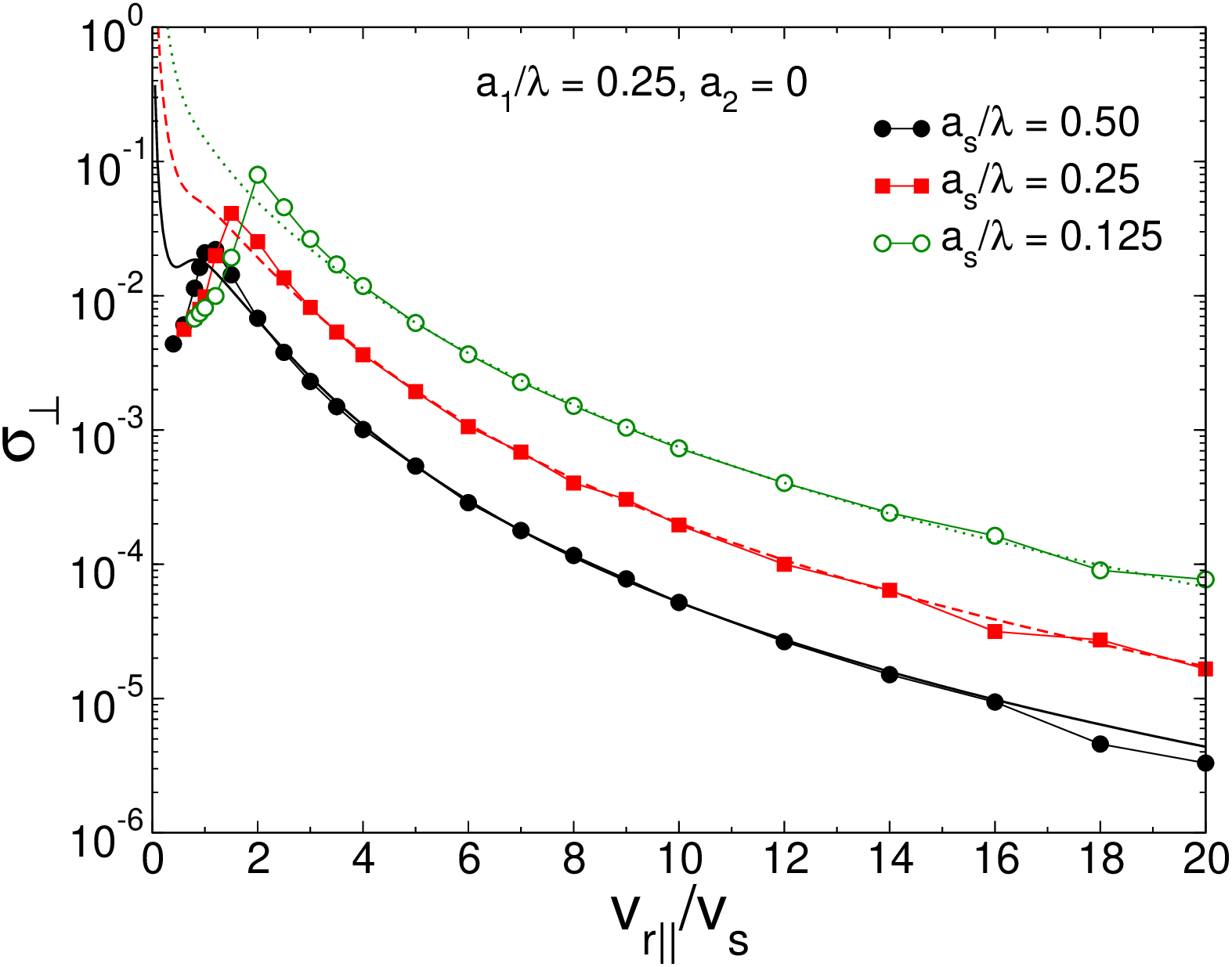}
\includegraphics[width=8cm]{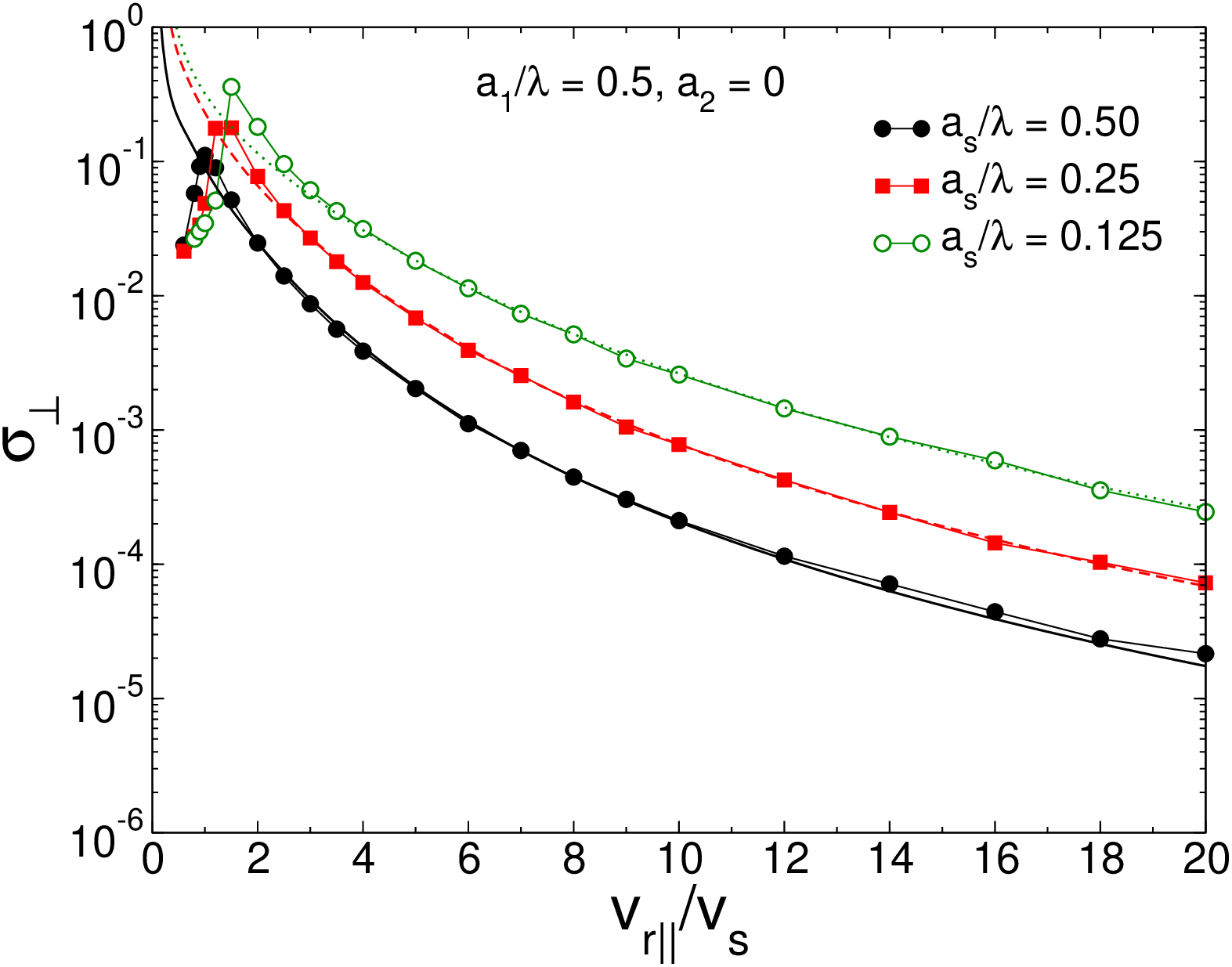}
\caption{(Color online) The cross-section $\sigma_{\bot}$ Eq.~(%
\ref{eq:sig2}) for electron-electron collision in terms of
dimensionless relative velocity component $v_{r\parallel}/v_{s}$
for $a_{2}=0
$ and at $a_{1}/\lambda=0.25$ (left panel) and $a_{1}/%
\lambda =0.5$ (right panel). The curves with and without symbols
correspond to CTMC simulations and the second order perturbative
treatment, respectively. In left and right panels, $a_{s}/\lambda
=0.5 $ (solid curve), $a_{s}/\lambda =0.25$ (dashed),
$a_{s}/\lambda =0.125 $ (dotted).} \label{fig:8}
\end{figure}

\begin{figure}[tbp]
\includegraphics[width=8cm]{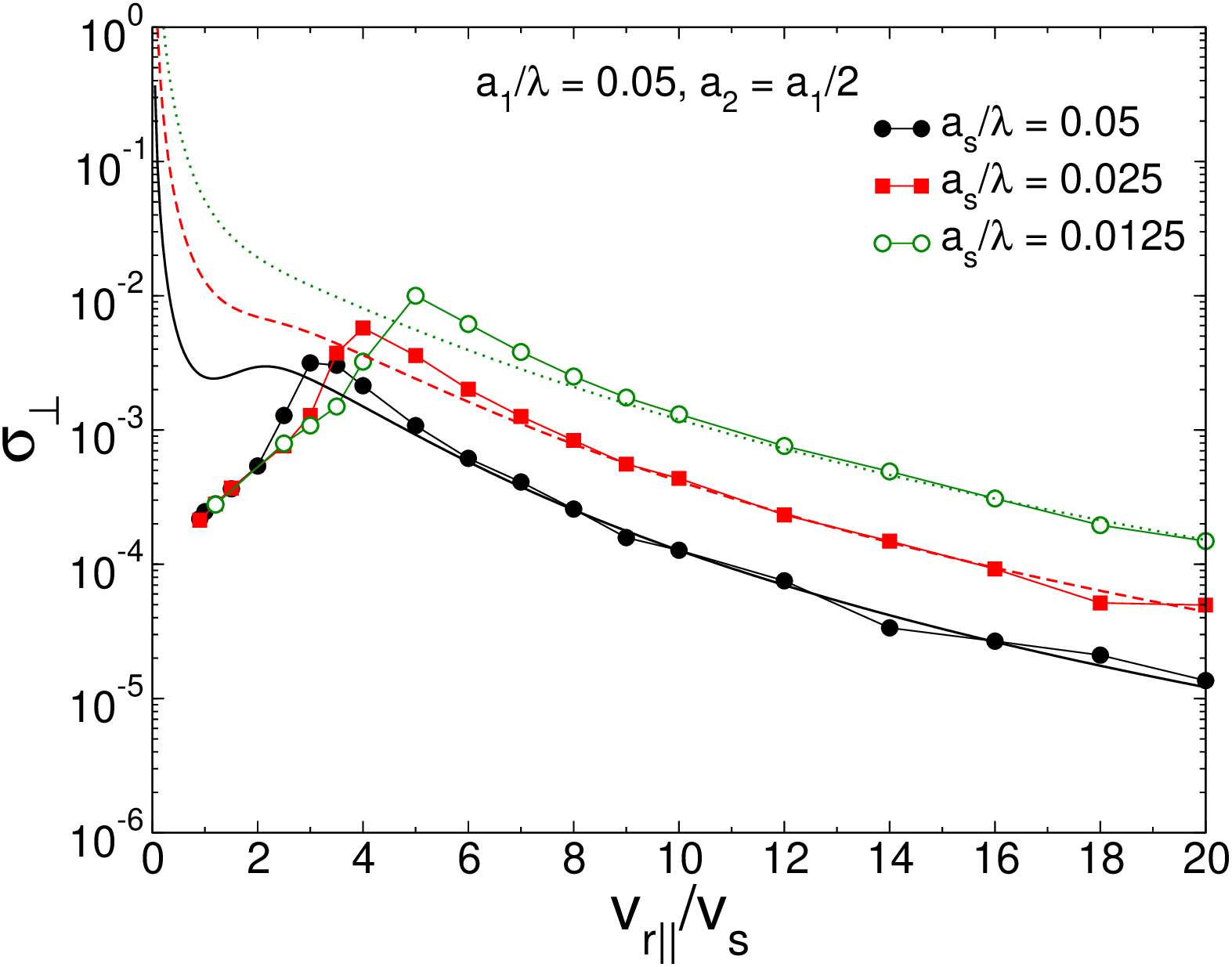} %
\includegraphics[width=8cm]{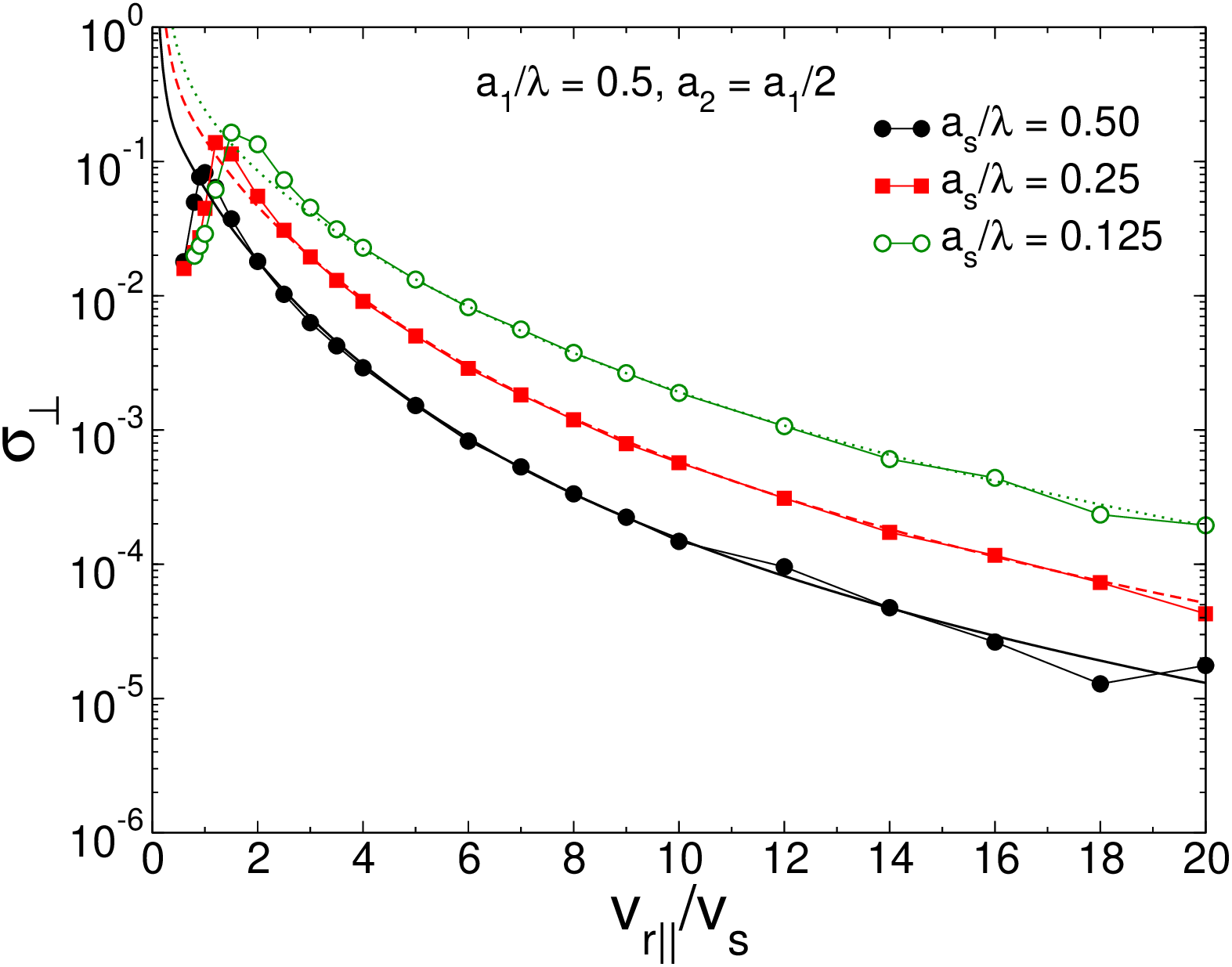}
\caption{(Color online) Same as in Fig.~\ref{fig:8} but for $%
a_{2}=a_{1}/2$ and at $a_{1}/\lambda=0.05$ (left panel) and $a_{1}/%
\lambda =0.5$ (right panel). Left panel, $a_{s}/\lambda %
=0.05 $ (solid curve), $a_{s}/\lambda =0.025$ (dashed), $a_{s}/%
\lambda =0.0125$ (dotted). Right panel, $a_{s}/\lambda =0.5 $
(solid), $a_{s}/\lambda =0.25$ (dashed), $a_{s}/\lambda %
=0.125$ (dotted).}
\label{fig:9}
\end{figure}

Systematic investigations and comparisons of the cross-sections determined
by the CTMC simulations and the second order perturbative treatment Eqs.~(%
\ref{eq:a63}), (\ref{eq:a63x}), (\ref{eq:sig1})--(\ref{eq:sig3}) with the
dynamical cutoff parameter $\lambdabar(v_{r\parallel})$ (\ref{eq:lambda_ml})
are presented in Figs.~\ref{fig:1}--\ref{fig:9}. We have also compared the
CTMC results with the simple (exact) model given by Eq.~(\ref{eq:apa3}) for
a repulsive ion--electron interactions, Figs.~\ref{fig:3} and \ref{fig:4}.
Shown are $\sigma_{i}$ for attractive (with $Z>0$) and repulsive (with $Z<0$%
) interactions (Figs.~\ref{fig:1}--\ref{fig:at_vs_rep}), $\sigma_{\parallel}$
(Figs.~\ref{fig:5}--\ref{fig:7}), and $\sigma_{\perp}$ (Figs.~\ref{fig:8}
and \ref{fig:9}) as functions of $v_{r\parallel}/v_{s}$ for fixed cyclotron
radii $a$, $a_{1}$ and $a_{2}$ and varying the strength of the magnetic
field $a_{s}/\lambda =B_{s}/B$ with $B_{s} =mv_{s}/e\lambda$. For each pair
of fixed $a_{i}$ ($i=1,2$) and $a_{s}$ the transversal relative velocity is
determined by $v_{0i\perp}/v_{s} =a_{i}/a_{s}$ (or $v_{e\perp}/v_{s} =a/a_{s}
$ for ion--electron collisions). The CTMC results are indicated by the
curves with filled or open symbols and the corresponding second order
predictions are given by the curves without symbols. Both the CTMC and
second order calculations have been done for a regularized potential $U_{%
\mathrm{R}}$ with $\nu_{0}=5\times 10^{-5}$. Note also the logarithmic
scales for the cross-sections in Figs.~\ref{fig:1}, \ref{fig:2} and \ref%
{fig:at_vs_rep}--\ref{fig:9}.

First we discuss some general observations of the behavior of the
cross--sections. In Figs.~\ref{fig:1}, \ref{fig:2} and \ref{fig:at_vs_rep}--%
\ref{fig:9} we can clearly observe that in the regimes of large relative
velocities $v_{r\parallel}/v_{s} \gtrsim \kappa$ (where typically $%
2\leqslant\kappa \leqslant 4$) the second order perturbative treatment
agrees perfectly with the numerical CTMC results. In addition in the limit
of very large velocities $v_{r\parallel}/v_{s} \gg 1$ the cross--sections $%
\sigma_{i}$ and $\sigma_{\parallel}$ calculated either within perturbation
theory or CTMC method with different strength of the magnetic field and
transversal velocities converge to the same value. This behavior agrees with
the predictions of the asymptotic expression (\ref{eq:a69b}) which is
independent on $B$ and $v_{01\perp}$, $v_{02\perp}$ (or $v_{e\perp}$). It
can also be seen that the smaller the transversal velocities, the better is
the convergence to the regime of Eq.~(\ref{eq:a69b}). At large relative
velocities the second order and CTMC cross-sections $\sigma_{\perp}$ shown
in Figs.~\ref{fig:8} and \ref{fig:9} agree with Eq.~(\ref{eq:sig2}) with the
asymptotic expression (\ref{eq:a69c}). Since at $v_{r\parallel}/v_{s} \gg 1$
the quantity $\sigma_{\perp}$ is not affected by the magnetic field but
behaves as $\sigma_{\perp}\sim v^{2}_{01\perp},v^{2}_{02\perp}$ for fixed
electron cyclotron radii it will be larger for smaller $a_{s}$ (more
precisely $\sigma_{\perp}\sim a^{-2}_{s}$) as shown in Figs.~\ref{fig:8} and %
\ref{fig:9}.

At small velocities with $v_{r\parallel}/v_{s} \lesssim 1$ the second order
treatment considerably deviates from CTMC simulations, see Figs.~\ref%
{fig:kap1}--\ref{fig:2} and Figs.~\ref{fig:at_vs_rep}--\ref{fig:9}. Here the
second order cross--sections are given by approximate expressions (\ref%
{eq:apb7}) and (\ref{eq:apb16}) where the parameter $\varkappa$ at small
relative velocities is given by $\varkappa \simeq \varkappa
(0)=1+\lambda/\lambdabar (0)$ and $\lambdabar (0)$ is the dynamical cutoff $%
\lambdabar (v_{r\parallel})$ Eq.~(\ref{eq:lambda_ml}) at $v_{r\parallel}=0$.
Note that at finite cyclotron radii of the particles the quantity $%
\lambdabar (0)$ is a constant depending on the value of $\nu_{0}$ and the
transversal velocities. However, for vanishing cyclotron radii (as, e.g. in
the left panels of Figs.~\ref{fig:kap1}, \ref{fig:1}, \ref{fig:at_vs_rep}
and \ref{fig:5}) the cutoff parameter (\ref{eq:lambda_ml}) at small
velocities behaves as $\lambdabar/\lambda \sim (v_{s}/v_{r\parallel})^{2}$.
The quantity $(\varkappa -1)^{2}$ involved in Eqs.~(\ref{eq:apb7}) and (\ref%
{eq:apb16}) falls as $\sim (v_{r\parallel}/v_{s})^{4}$. This results in a
strong self--cutting at small velocities. Thus employing the cutoff (\ref%
{eq:lambda_ml}) the second order cross--section $\sigma_{\parallel}$ is
strongly reduced and decreases as $\sigma_{\parallel}\sim v^{5}_{r\parallel}$
and $\sigma_{\parallel} \sim v^{6}_{r\parallel}$ at $a=0$ and $a\neq 0$,
respectively. The transversal cross--section $\sigma_{\perp}$, i.e. the
second term in Eq.~(\ref{eq:apb7}), does not contain a term $(\varkappa
-1)^{2}$ and diverges as $\sigma_{\perp} \sim v^{-2}_{r\parallel}$, see
Figs.~\ref{fig:8} and \ref{fig:9}. In this small velocity regime the second
order perturbative treatment is clearly invalid and a non--perturbative
description is required.


In Figs.~\ref{fig:3} and \ref{fig:4} we demonstrate the cross-section $%
\sigma_{i}$ for repulsive ($Z<0$) ion--electron interaction. The
dashed curves represent the cross--sections obtained from simple
model considered in
Sec.~\ref{sec:imp}, see Eq.~(\ref{eq:apa3}), with constant cutoff parameter $%
\nu_{0}$. Let us recall that the model (\ref{eq:apa3}) completely ignores
the cyclotron motion of the particles. It is seen that in Figs.~\ref{fig:3}
and \ref{fig:4} (left panel) the agreement with CTMC simulations is quite
satisfactory even for finite cyclotron radius $a$ and magnetic field.
However, with increasing $a$ the CTMC simulations show a more involved
picture as e.g. in Fig.~\ref{fig:4} (right panel) than the predictions of a
simple model (\ref{eq:apa3}). And the deviations from CTMC becomes more
pronounced with increasing electron transversal velocity $v_{e\perp}$ and
magnetic field, see e.g. open symbols in the right panel of Fig.~\ref{fig:4}%
. In the CTMC simulations the cross--section shrinks strongly at $%
v_{r\parallel}\gtrsim v_{s}$ with increasing relative velocity. This feature
can be explain on the basis of a simple model discussed in Sec.~\ref{sec:imp}%
. At strong but finite magnetic field the hard collisions like
backscattering events may also occur but the total relative energy $\mu
v^{2}_{r\parallel}/2$ in Eq.~(\ref{eq:apa1}) will be replaced here by the
total energy $\mu (v^{2}_ {r\parallel}+v^{2}_{e\perp})/2$. This will reduce
the domain of the backscattering events (regime II introduced in Sec.~\ref%
{sec:imp}). With increasing $v_{r\parallel}$ this domain will be further
shrunk and finally the scattering may occur only in the regime I where a
strong magnetic field may strongly reduce the energy transfer. (Let us
recall that the ion moves along the magnetic field. In this case the energy
transfer vanishes with increasing $B$ for symmetry reason, see Refs.~\cite%
{ner03,ner07}). Obviously this effect is pronounced by increasing
transversal velocity as shown in Fig.~\ref{fig:4}. The similar feature is
observed also for electron--electron interactions shown in the right panels
of Figs.~\ref{fig:6} and \ref{fig:7}.

For ion--electron collisions we also compare the cross-sections for
attractive ($Z>0$) and repulsive ($Z<0$) interactions in Fig.~\ref%
{fig:at_vs_rep}. The solid curves represent the second order energy transfer
which is quadratic in $Z$. For small velocities the cross--section $%
\sigma_{i}$ for repulsive interaction considerably exceeds the
cross--section for attractive interaction. This is because of the
backscattering events at $Z<0$ with large energy--velocity transfers.
However, with increasing relative velocity the agreement between second
order theory and CTMC and between two CTMC with positive and negative $Z$
continuously improves and is almost perfect over the full range of the
velocity at $v_{r\parallel}/v_{s} \gtrsim 1$.

As has been emphasized in the preceding sections there are similarities
between electron--electron and ion--electron (with $\mathbf{v}_{i}
=v_{i\parallel}\mathbf{b}$) interactions but the energy transfers are not
quite the same. In particular, the approximate relation $\sigma_{i} \simeq
2\sigma_{\parallel}$ for the negative ions with $Z<0$ may be violated at
small velocities where the electron--electron cross--section $\overline{%
\sigma}_{\parallel}(\varphi)$ is not in general isotropic with respect to $%
\varphi$. In Figs.~\ref{fig:5} and \ref{fig:6} the cyclotron radius of the
second electron vanishes $a_{2}=0$ and thus $a=a_{1}$ as well as $\overline{%
\sigma}_{\parallel}(\varphi)$ are isotropic. In this case the relation
between the cross--sections is exact $\sigma_{i}= 2\sigma_{\parallel}$. In
Fig.~\ref{fig:7} we show an anisotropic case where $a_{2}=a_{1}/2$ and $a_{1}
$ varies from a small (left panel) to a large value (right panel). Another
feature which is absent in ion--electron collisions is the cm cyclotron
motion for two--electron collisions which causes the transversal energy
transfer $\sigma_{\perp}$ shown in Figs.~\ref{fig:8} and \ref{fig:9}. For
symmetry reason this quantity vanishes at $a_{1} =a_{2}$ both in CTMC
simulations (within the unavoidable numerical fluctuations) and the second
order theory, see Eq.~(\ref{eq:a63x}). In the examples shown in Figs.~\ref%
{fig:8} and \ref{fig:9} the cyclotron radius $a_{2}$ of the second electron
is smaller $a_{1}$, $a_{2}<a_{1}$. Since the cross--section $\sigma_{\perp}$
is positive the transverse energy of the first electron with $a_{1}>a_{2}$
transfers to the energy of parallel and transverse motion of the second
electron.

\section{Conclusion}
\label{sec:disc}

In this paper we have investigated the binary collisions (BC) of two charged
particles in the presence of constant magnetic field employing second order
perturbation theory and classical trajectory Monte--Carlo (CTMC)
simulations. Two distinct cases with symmetric and strongly asymmetric
charges and masses of the particles have been considered in detail: (i) BC
between two identical particles (e.g. electrons) and (ii) between an
electron and heavy ion which moves with rectilinear trajectory (no cyclotron
motion) along the magnetic field. Physically these two distinct cases are
similar except the time--dependent center of mass cyclotron motion in the
case of two identical particles. The second-order energy transfers for
two--particles collision is calculated with the help of an improved BC
treatment which is valid for any strength of the magnetic field and thus
involves all cyclotron harmonics of the particles motion. For further
applications (e.g., in cooling of ion beams, transport phenomena in
magnetized plasmas) the actual calculations of the energy transfers have
been done with a screened interaction potential which is regularized at the
origin. The use of that potential can be viewed as an alternative to the
standard cutoff procedure. For a repulsive ion--electron collisions we
furthermore presented an exact solution for the energy transfer in the
presence of an infinitely strong magnetic field. Here backscattering events
may occur when the relative velocity transfer is independent on the strength
of the Coulomb (regularized and screened) interaction $\Delta
v_{\parallel}=-2v_{r\parallel}$, where $v_{r\parallel}$ is the initial
relative velocity of the guiding centers of the particles. The second order
perturbative treatment is clearly invalid in this case. It has been shown
that for repulsive interaction of the particles in a strong but finite
magnetic field two scattering regimes must be distinguished with and without
backscattering events with large energy--velocity transfers. This also
suggests an improved perturbative treatment for repulsive interaction and in
the case of strong magnetic field.

For checking the validity of the perturbative approach and also for
applications beyond the perturbative regime we have employed numerical CTMC
simulations. These CTMC calculations have been performed in a wide range of
parameters (magnetic field and the relative velocities of the particles) and
for a small regularization parameter, that is, for an interaction which is
rather close to Coulomb at short distances. Within the second order
treatment we have introduced a dynamical cutoff parameter which
substantially improves the agreement of the theory with CTMC simulations.
From a comparison with the non-perturbative CTMC simulations we have found
as a quite general rule which is widely independent of the magnetic field
strength that the predictions of the second order perturbative treatment are
very accurate for $v_{r\parallel}/v_{s}\gtrsim 4$ for all studied parameters
and cases, with the characteristic velocity $v_{s}$ given by Eq.~(\ref%
{eq:a70}). In contrast, for low relative velocities $v_{r\parallel}/v_{s}%
\lesssim 1$ the results obtained from perturbation theory strongly deviate
from the CTMC simulations. Moreover in this regime the CTMC calculations
display a large difference between positive and negative ions which
disappears for high velocities $v_{r\parallel}$. Such a difference is
completely absent in the second order perturbative treatment where the
energy transfers is quadratic in the ion charge $Ze$. We have also tested
the exact analytical model Eq.~(\ref{eq:apa3}) derived for repulsive
interaction and an infinitely strong magnetic field by comparing it in Figs.~%
\ref{fig:3} and \ref{fig:4} with the CTMC simulations and found that the
agreement is rather satisfactory even for finite (but strong) magnetic
fields.

We believe that our theoretical findings will be useful for the
interpretation of experimental investigations. Here, it is of particular
interest to study some macroscopic physical quantities on the basis of the
presented theoretical model such as cooling forces in storage rings and
traps, stopping power of ion beams as well as transport coefficients in
strongly magnetized plasmas. These studies require an average of the energy
or velocity transfers with respect to the velocity distribution of the
electrons. The cooling forces obtained by the perturbative approach are
expected to be quite accurate if the low velocity regime only slightly
contributes to the $\mathbf{v}_{r}$-average over $\langle\Delta E\rangle$.
That is, if the typical $v_{r\parallel}$, given by the maximum of the
thermal electron velocity and the ion velocity, are large compared to $v_{s}$%
, as it is usually the case for e.g.~electron cooling in storage rings.

Another interesting issue not considered here is the interaction of the
magnetized electrons with light ions (in particular with positrons, protons
and antiprotons) when the magnetic field is so strong that the cyclotron
motion of the ion cannot be neglected anymore. In this case the relative and
center of mass motions are coupled to each other and the center of mass
velocity of the particles cannot be represented in the form of simple
cyclotron motion, Eq.~(\ref{eq:a6-1}). A detailed comparison with cooling
force measurements and a study of other aspects are in progress and the
results will be reported elsewhere.

\begin{acknowledgments}
One of the authors, H.B.N., is grateful for the support of the Alexander von
Humboldt Foundation, Germany, and National Academy of Sciences of Armenia.
This work was supported by the Bundesministerium f\"{u}r Bildung und
Forschung (BMBF) under contract 06ER145 and by the Gesellschaft f\"{u}r
Schwerionenforschung (GSI, ER/TOE). Discussions with C. Toepffer are
gratefully acknowledged.
\end{acknowledgments}

\appendix

\section{Derivation of the Coulomb logarithm $\Lambda (u)$}
\label{sec:ap1}

In this Appendix we evaluate the effective cross section in \ the case of
vanishing relative cyclotron radius. Using the definition of $\overline{%
\sigma }\left( \varphi \right) $, Eq.~(\ref{eq:a43}), and the energy
transfer Eq.~(\ref{eq:a59a}) we obtain
\begin{equation}
\overline{\sigma }\left( \varphi \right) =-\frac{4q^{4}e\!\!\!/^{4}V_{0%
\parallel }}{mv_{r\parallel }^{3}}\Lambda (u).  \label{eq:ap1}
\end{equation}%
Here $u=\chi _{1}/\kappa _{1}$ and we have introduced the generalized
Coulomb logarithm
\begin{equation}
\Lambda (u)=\int_{0}^{\infty }\left[ K_{1}(s)-uK_{1}(us)\right] ^{2}sds.
\label{eq:ap2}
\end{equation}%
The integration in Eq.~(\ref{eq:ap2}) can be done using the indefinite
integrals of the functions $sK_{1}^{2}(s)$, $sK_{1}^{2}(us)$ and $%
sK_{1}(s)K_{1}(us)$ \cite{gra80}. This yields
\begin{eqnarray}
\Lambda \left( u\right) &=&\left\{ -\frac{s^{2}}{2}K_{1}^{2}\left( s\right) -%
\frac{u^{2}s^{2}}{2}K_{1}^{2}\left( us\right) +K_{0}\left( us\right) \left[
\frac{u^{2}s^{2}}{2}K_{2}\left( us\right) -\frac{2u^{2}sK_{1}\left( s\right)
}{u^{2}-1}\right] \right.  \notag \\
&&\left. +K_{0}\left( s\right) \left[ \frac{s^{2}}{2}K_{2}\left( s\right) +%
\frac{2usK_{1}\left( us\right) }{u^{2}-1}\right] \right\} _{s\rightarrow 0}
\label{eq:ap3} \\
&=&\left\{ K_{0}\left( us\right) \left[ \frac{u^{2}s^{2}}{2}K_{2}\left(
us\right) -1+\frac{2u^{2}\left( 1-sK_{1}\left( s\right) \right) }{u^{2}-1}%
\right] \right.  \notag \\
&&\left. +K_{0}\left( s\right) \left[ \frac{s^{2}}{2}K_{2}\left( s\right)
-1-2\frac{1-usK_{1}\left( us\right) }{u^{2}-1}\right] \right\}
_{s\rightarrow 0}+\frac{u^{2}+1}{u^{2}-1}\left[ K_{0}\left( s\right)
-K_{0}\left( us\right) \right] _{s\rightarrow 0}-1.  \notag
\end{eqnarray}%
Since at $s\rightarrow 0$ the modified Bessel functions behave as (see,
e.g., \cite{gra80}) $K_{2}\left( s\right) =2/s^{2}+\mathrm{O}(1)$, $%
K_{1}\left( s\right) =1/s+\mathrm{O}(s\ln s)$, $K_{0}\left( s\right) =\ln
(2/s)+\mathrm{O}(1)$, the expression in the large brackets in the last line
of Eq.~(\ref{eq:ap3}) vanishes as $s^{2}\ln ^{2}s\rightarrow 0$. There
remains
\begin{equation}
\Lambda \left( u\right) =\frac{u^{2}+1}{u^{2}-1}\left[ K_{0}\left( s\right)
-K_{0}\left( us\right) \right] _{s\rightarrow 0}-1.  \label{eq:ap4}
\end{equation}%
Since $K_{0}\left( s\right) =\ln (2/s)+\mathrm{O}(1)$ at $s\rightarrow 0$
from Eq.~(\ref{eq:ap4}) we finally obtain
\begin{equation}
\Lambda \left( u\right) =\frac{u^{2}+1}{u^{2}-1}\ln u-1.  \label{eq:ap5}
\end{equation}

\section{The energy transfer in a small velocity limit}
\label{sec:ap2}

For the second order BC treatment the most critical situation is the small
velocity regime where we expect some deviations from the non-perturbative
CTMC simulations. For the improvement of the theoretical approach it is
therefore imperative to investigate the energy transfer in the small
velocity limit, $\vert v_{r\parallel }\vert \ll v_{0\bot }$ or alternatively
$\delta \ll a$. In principle this limit can be evaluated using the integral
representation of the cross-sections, Eqs.~(\ref{eq:a63}), (\ref{eq:a63x}), (%
\ref{eq:rel8}) and (\ref{eq:rel9}). However, while these expressions are
very convenient to calculate the high velocity limit of the energy transfers
(see Sec.~\ref{sec:s3}) they are not adopted for the evaluation of the small
velocity limit due to the oscillatory nature of the function $R(t)$ at $%
v_{r\parallel} \rightarrow 0$. In this Appendix we consider instead an
alternative but equivalent expressions for the effective cross-sections. For
the axially symmetric interaction potential the effective cross-sections can
be evaluated using Eqs.~(\ref{eq:a42a}), (\ref{eq:a42c}), (\ref{eq:a42e})
and (\ref{eq:rel2}). We refer the reader to Refs.~\cite{ner07,ner03} for
details. The integration of these expressions with respect to the impact
parameter $s$ yields the two-dimensional $\delta$-function, $\delta(\mathbf{Q%
})=\delta(\mathbf{k}_{\bot}+\mathbf{k}^{\prime}_{\bot})$. Combining this
function with the one-dimensional $\delta$-function $\delta(k_{%
\parallel}+k^{\prime}_{\parallel})$ in Eqs.~(\ref{eq:a42a}), (\ref{eq:a42c})
and (\ref{eq:a42e}) yields a three-dimensional $\delta$-function $\delta(%
\mathbf{k}+\mathbf{k}^{\prime})$ and the $\mathbf{k}^{\prime}$-integration
in the energy transfers can be performed exactly. Furthermore it can be
shown that only the imaginary part $\mathrm{Im} [G_{n}(\mathbf{k},-\mathbf{k}%
)]$ contributes to the cross-sections. From Eq.~\eqref{eq:a41} follows that
the imaginary part of $G_{n}(\mathbf{k},-\mathbf{k})$ is expressed by the
Dirac functions $\delta (\zeta_{n}(\mathbf{k}))$ which allows to perform the
$k_{\parallel}$-integration. The final result reads

\begin{eqnarray}
\overline{\sigma }\left( \varphi \right) &=&-\frac{4q^{4}e\!\!\!/^{4}}{%
mv_{r\parallel }^{3}}\sum_{n=0}^{\infty }\eta _{n}\left\{ \frac{2n^{2}}{%
\delta ^{2}}\overline{V}\left( \varphi \right) \left[ 3\Phi _{n;1}\left(
k_{\parallel },a\right) +k_{\parallel }\frac{\partial }{\partial
k_{\parallel }}\Phi _{n;1}\left( k_{\parallel },a\right) +\frac{\delta ^{2}}{%
a}\frac{\partial }{\partial a}\Phi _{n;1}\left( k_{\parallel },a\right) %
\right] \right.  \label{eq:apb1} \\
&&\left. +\frac{v_{r\parallel }}{2}f(\varphi )\left[ \frac{2n^{2}}{\delta
^{2}}\Phi _{n;1}\left( k_{\parallel },a\right) +\Phi _{n-1;3}\left(
k_{\parallel },a\right) +\Phi _{n+1;3}\left( k_{\parallel },a\right) \right]
\right\} _{k_{\parallel }=n/\delta },  \notag
\end{eqnarray}%
\begin{equation}
\overline{\sigma }_{r\bot }\left( \varphi \right) =\frac{8q^{4}e\!\!\!/^{4}%
\omega _{c}^{2}}{mv_{r\parallel }^{4}}\sum_{n=1}^{\infty }n^{2}\left\{ 2\Phi
_{n;1}\left( k_{\parallel },a\right) +k_{\parallel }\frac{\partial }{%
\partial k_{\parallel }}\Phi _{n;1}\left( k_{\parallel },a\right) +\frac{%
\delta ^{2}}{a}\frac{\partial }{\partial a}\Phi _{n;1}\left( k_{\parallel
},a\right) \right\} _{k_{\parallel }=n/\delta },  \label{eq:apb3}
\end{equation}%
\begin{equation}
\overline{\sigma }_{r1}\left( \varphi \right) =\frac{8q^{4}e\!\!\!/^{4}}{%
\delta ^{2}}\sum_{n=1}^{\infty }\left[ n^{2}\Phi _{n;1}\left( k_{\parallel
},a\right) \right] _{k_{\parallel }=n/\delta },  \label{eq:apb2}
\end{equation}%
where the function $\Phi _{n;m}(k_{\parallel },a) $ is defined as
\begin{equation}
\Phi _{n;m}\left( k_{\parallel },a\right) =\frac{\left( 2\pi \right) ^{4}}{4}%
\int_{0}^{\infty }U^{2}\left( k_{\parallel },k_{\perp }\right)
J_{n}^{2}\left( k_{\bot }a\right) k_{\perp }^{m}dk_{\perp } .
\label{eq:apb5}
\end{equation}%
In Eq.~(\ref{eq:apb1}) we have introduced the notations $\eta _{n}=1-\frac{1%
}{2}\delta _{n0}$, $\overline{V}(\varphi )=V_{0\parallel }-(v_{r\parallel
}/2)f(\varphi )$ with $f(\varphi )=(a_{1}^{2}-a_{2}^{2})/a^{2}(\varphi )$.

Equations (\ref{eq:apb1})-(\ref{eq:apb2}) are equivalent to the integral
representations (\ref{eq:cross})-(\ref{eq:a43bb}), (\ref{eq:rel6}) and (\ref%
{eq:rel7}) of the cross-sections, respectively. The function $\Phi
_{n;m}(k_{\parallel },a)$ is taken at $k_{\parallel }=n/\delta $ which in
the limit of small velocities becomes very large (except the term with $n=0$
in Eq.~\eqref{eq:apb1}). For the regularized interaction potential (\ref%
{eq:a48}) from Eq.~(\ref{eq:apb5}) at $k_{\parallel }a\gg 1$ we obtain
\begin{equation}
\Phi _{n;1}\left( k_{\parallel },a\right) \simeq \frac{5\lambda ^{3}}{32a}%
\frac{\left( \varkappa ^{2}-1\right) ^{2}}{\left( k_{\parallel }\lambda
\right) ^{7}},\quad \Phi _{n;3}\left( k_{\parallel },a\right) \simeq \frac{%
\lambda }{32a}\frac{\left( \varkappa ^{2}-1\right) ^{2}}{\left( k_{\parallel
}\lambda \right) ^{5}} .  \label{eq:apb6}
\end{equation}%
Substituting these expressions into Eqs.~(\ref{eq:apb1})-(\ref{eq:apb3}) in
the lowest order with respect to $v_{r\parallel }$ (or $\delta $) we arrive
at
\begin{equation}
\overline{\sigma }\left( \varphi \right) \simeq \frac{q^{4}e\!\!\!/^{4}}{m}%
\left[ \frac{\left\vert v_{r\parallel }\right\vert }{v_{r\parallel }}\frac{%
5\zeta \left( 5\right) V_{0\parallel }}{\left( \omega _{c}\lambda \right)
^{3}}\left( \varkappa ^{2}-1\right) ^{2}\left( \frac{\delta }{\lambda }%
\right) ^{2}\frac{\lambda }{a}-\frac{2}{v_{r\parallel }^{2}}f(\varphi )\psi
\left( \frac{a}{\lambda }\right) \right] ,  \label{eq:apb7}
\end{equation}%
\begin{equation}
\overline{\sigma }_{r\bot }\left( \varphi \right) \simeq -\frac{25\zeta
\left( 5\right) q^{4}e\!\!\!/^{4}}{4m\omega _{c}^{2}\lambda ^{2}}\left(
\varkappa ^{2}-1\right) ^{2}\left( \frac{\delta }{\lambda }\right) ^{3}\frac{%
\lambda }{a} ,  \label{eq:apb9}
\end{equation}%
\begin{equation}
\overline{\sigma }_{r1}\left( \varphi \right) \simeq \frac{5\zeta \left(
5\right) q^{4}e\!\!\!/^{4}}{4}\left( \varkappa ^{2}-1\right) ^{2}\left(
\frac{\delta }{\lambda }\right) ^{5}\frac{\lambda }{a} .  \label{eq:apb8}
\end{equation}%
Here $\zeta (z) $ is the Riemann function with $\zeta (5)\simeq 1.0369$. The
function $\psi (u)$ is expressed by the modified Bessel functions
\begin{eqnarray}
\psi \left( u\right) &=&\frac{1}{2u}\frac{\partial }{\partial u}\left\{ u^{2}%
\left[ I_{1}\left( u\right) K_{1}\left( u\right) +I_{1}\left( \varkappa
u\right) K_{1}\left( \varkappa u\right) \right] \right\}  \label{eq:apb10} \\
&&+\frac{2}{\varkappa ^{2}-1}\left[ I_{1}\left( u\right) K_{1}\left(
u\right) -\varkappa ^{2}I_{1}\left( \varkappa u\right) K_{1}\left( \varkappa
u\right) \right] .  \notag
\end{eqnarray}%
Note two limiting cases of this function. At $u\ll 1$ (vanishing cyclotron
radius) and $\varkappa \rightarrow 1$ from Eq.~(\ref{eq:apb10}) we obtain
\begin{equation}
\psi \left( u\right) \simeq \frac{u^{2}}{8}\left( \varkappa ^{2}+1-\frac{%
4\varkappa ^{2}}{\varkappa ^{2}-1}\ln \varkappa \right) ,  \label{eq:apb11}
\end{equation}%
\begin{equation}
\psi \left( u\right) \simeq \left( \varkappa ^{2}-1\right) ^{2}\frac{u^{3}}{%
48}\frac{\partial }{\partial u}\frac{1}{u}\frac{\partial }{\partial u}\frac{1%
}{u}\frac{\partial }{\partial u}\left[ u^{2} I_{1}\left( u\right)
K_{1}\left( u\right) \right] ,  \label{eq:apb12}
\end{equation}%
respectively. The case of ion--electron BC is easily recovered from the
obtained expressions as described at the end of Sec.~\ref{sec:s3}. Note that
the cm cyclotron motion is absent here and in Eqs.~(\ref{eq:apb1}) and (\ref%
{eq:apb7}) the terms proportional to the function $f(\varphi )$ must be
neglected.

From Eqs.~\eqref{eq:apb7}-\eqref{eq:apb8} we then obtain that the relative
transversal cross-sections at small relative velocities behave as $\overline{%
\sigma}_{r\bot}(\varphi) \sim v^{3}_{r\parallel}$, $\overline{\sigma}
_{r1}(\varphi) \sim v^{5}_{r\parallel}$. The first term in the cross-section
$\overline{\sigma} (\varphi)$ vanishes as $\sim v^{2}_{r\parallel}$ while
the second one which corresponds to the zero harmonic with $n=0$ in Eq.~%
\eqref{eq:apb1} diverges as $\sim v_{r\parallel }^{-2}$. This later term
vanishes at $a_{1} =a_{2}$ and predicts an infinitely large energy transfer
at $a_{1} >a_{2}$ (or energy gain at $a_{1} <a_{2}$). As expected the
cross-sections (\ref{eq:apb7})-(\ref{eq:apb8}) are strongly anisotropic with
respect to the phase $\varphi $.

Finally, we consider the small velocity limit of the cross-sections $%
\overline{\sigma }\left( \varphi \right) $ and $\overline{\sigma }_{r\bot
}\left( \varphi \right) $ at vanishing cyclotron radius, $a=0$ (see Eq.~(\ref%
{eq:a69})),
\begin{equation}
\overline{\sigma }\left( \varphi \right) =-\frac{V_{0\parallel }}{%
v_{r\parallel }}\overline{\sigma }_{r\bot }\left( \varphi \right) \simeq -%
\frac{q^{4}e\!\!\!/^{4}V_{0\parallel }v_{r\parallel }}{3m\omega
_{c}^{4}\lambda ^{4}}\left( \varkappa ^{2}-1\right) ^{2}.  \label{eq:apb16}
\end{equation}%
In this case the cross-sections behave as $\overline{\sigma }(\varphi ) \sim
v_{r\parallel }$ and $\overline{\sigma }_{r\bot }(\varphi ) \sim
v_{r\parallel }^{2}$ at small relative velocity $v_{r\parallel } $ (cf. with
Eqs.~(\ref{eq:apb7}) and (\ref{eq:apb9})).

\end{document}